\documentclass[aps, 12pt, showkeys, nofootinbib]{revtex4}

\newcommand{\be}{\begin{eqnarray}}
\newcommand{\ee}{\end{eqnarray}}
\newcommand{\ba}{\begin{array}}
\newcommand{\ea}{\end{array}}
\newcommand{\ds}{\displaystyle}

\newcommand{\pa}[1]{\left(#1\right)}
\newcommand{\paq}[1]{\left[#1\right]}
\newcommand{\pag}[1]{\left\{#1\right\}}

\newcommand{\K}{\mathbf{k}}
\newcommand{\Pp}{\mathbf{p}}

\usepackage[english]{babel}
\usepackage{amsmath}
\usepackage{amssymb}
\usepackage{graphicx,color}
\usepackage{tikz}
\usetikzlibrary{decorations.pathmorphing}

\begin{document}

\title{Effects of short-distance modifications to general relativity in
    spinning binary systems}

\author{Aline Nascimento Lins}
\affiliation{Departamento de F\'\i sica Te\'orica e Experimental, Universidade Federal do Rio Grande do Norte, Av. Sen. Salgado Filho, Natal-RN 59078-970, Brazil}

\author{Riccardo Sturani}
\affiliation{International Institute of Physics, Universidade Federal do Rio Grande do Norte, Campus Universit\'ario, Lagoa Nova CP:1613, Natal-RN 59078-970, Brazil}

\email{alinelins@fisica.ufrn.br, riccardo@iip.ufrn.br}

\begin{abstract}
  We investigate the possibility of testing short distance modifications to
  General Relativity via higher curvature terms in the fundamental gravity
  Lagrangian by analysing their impact on observations of spinning astronomical
  binary systems.
  By using effective field theory methods applied to the 2-body problem,
  generic lower bounds on the short-distance scale accompanying high
  curvature terms can be set. In particular we extend known results
  by deriving spin-dependent effects in binding energy, radiation emission
  process,
    and spin precession equations in binary systems,
    which are the fundamental ingredients to observe
  spin-dependent effects in gravitational wave detections from
  compact binary coalescences and spin precession in double binary
  pulsars.

\end{abstract}


\maketitle

\section{Introduction}
\label{sec:intro}
The recent Gravitational Wave (GW) detections
\cite{LIGOScientific:2018mvr,Abbott:2020niy} by the LIGO \cite{TheLIGOScientific:2014jea} and Virgo \cite{TheVirgo:2014hva} large interferometers have opened
a new chapter in the history of physics and astronomy, with countless new
investigation which are now made possible.
The output of GW detectors is
processed via \emph{matched-filtering} \cite{Allen:2005fk}, which is particularly
sensitive to the phase of the GW signals, bearing the imprint of both
the astrophysical parameter of the source, like masses and spins, and of the
details of the gravitational theory, well in its non-linear regime, ruling the
motion of the 2-body system sourcing GWs.

In the present work we admit the possibility that General Relativity (GR) may
not be the ultimate theory of gravity but may be completed at
\emph{short distances} (UV henceforth) by higher curvature terms.
We adopt the framework introduced by \cite{Endlich:2017tqa}, i.e. we add
\emph{quartic} curvature terms of the type Riemann to the fourth power to the
(gauge-fixed)
Einstein-Hilbert Lagrangian and compute their lowest order effects,
in both the spinning and non-spinning case, to the 2-body
dynamics by using the effective field theory methods for gravity pioneered in
\cite{Goldberger:2004jt}, later applied also to spinning sources \cite{Porto:2006bt,Porto:2016pyg}, also known as Non-Relativistic GR (NRGR).

The introduction of higher curvature terms than in GR introduces
new phenomenological constants (with dimensions) parameterising the strength of
the GR modifications, playing the role of the UV cutoff of the effective theory.

We consider in this work astronomical observations of \emph{spinning} binary
systems, focusing on both detection of GWs from compact binary coalescences and
observation of double binary pulsars, which also allows a measure of
individual pulsar spin precession.

We extend the work of \cite{Endlich:2017tqa} by systematically computing
  the linear-in-spin processes in two-body potential and two-body radiation
  emission derived from the GR-modified model proposed there (from which
  we find a minor qualitative discrepancy in the radiative sector, see discussion in sec.~\ref{ssec:rad_cc}).
  While energy and luminosity functions are necessary ingredients to construct
  GW-form template to analyse data collected by GW detectors,
  spin-precession as observed in binary pulsars or by Gravity Probe B \cite{PhysRevLett.106.221101} can also provide
  GR precision tests. In view of the application to spinning systems we derive
  linear-in-spin phenomenological effects by computing the
  \emph{geodetic precession} of spins in a binary systems within the same
  GR-modified model, which is also an original contribution of this work.

The paper is organised as follows: in sec.~\ref{sec:method} we summarise
the GR UV completion introduced in \cite{Endlich:2017tqa}
and briefly review the effective field theory method description of gravity for
non-relativistic (i.e for small velocity) spinning systems \cite{Porto:2016pyg,Levi:2015msa}.
In sec.~\ref{sec:results} we present the computation
of the effects that higher-curvature terms have for both spinning and
non spinning binary systems on the energy and luminosity function which determine
the GW-phasing,
and on the spin precession in binary systems, the treatment of the spinning case
being the main original part of present work.
A discussion of the results and comparisons with current observational limits
are presented in sec.~\ref{sec:conclusions}.

\section{Method}
\label{sec:method}
Considering a generic parameterisation of possible UV completions of GR, we adopt
the effective Lagrangian proposed in \cite{Endlich:2017tqa}
\be
\label{eq:fund_lag}
{\cal S}_{eff}=\frac 1{16\pi G_N}\int d^4x\sqrt{-g}\pa{R-\frac 12\Gamma^\mu\Gamma_\mu+\frac{{\cal C}^2}{\Lambda^6}+\frac{{\cal C}\tilde{\cal C}}{\Lambda_-^6}
  +\frac{\tilde{\cal C}^2}{\tilde\Lambda^6}}\,,
\ee
with $R$ the Ricci scalar, $\Gamma^\mu\equiv g^{\nu\rho}\Gamma^\mu_{\nu\rho}$ enters the gauge fixing term
and
\be
\ba{rcl}
\ds{\cal C}&\equiv&\ds R_{\alpha\beta\gamma\delta}R^{\alpha\beta\gamma\delta}\,,\\
\ds\tilde{\cal C}&\equiv&\ds R_{\alpha\beta\gamma\delta}\epsilon^{\alpha\beta\mu\nu}R_{\mu\nu\rho\sigma}g^{\gamma\rho}g^{\delta\sigma}\,.
\ea
\ee
Eq.~(\ref{eq:fund_lag}) adds to the Einstein-Hilbert Lagrangian terms at \emph{fourth}
order of the curvature, where $\Lambda,\Lambda_-, \tilde \Lambda$ are constant
with unit of inverse length. Terms quadratic in curvature tensors do not
contribute
to the equations of motions as they can be written in terms of total
derivatives plus terms vanishing on the equations of motion, and cubic terms
are forbidden by the causality argument presented in \cite{Camanho:2014apa}.

For non- or mildly-relativistic binary systems it is natural to expand perturbatively
the dynamics according to the post-Newtonian (PN) approximation of GR, i.e terms
at $n$-th PN order are of the type $G_N^{n-j+1}v^{2j}$, with $0\leq j\leq n$, the case
$n=j=0$ corresponding to the leading order (LO) Newtonian potential, being $v$
the relative velocity of binary components.

\begin{figure}[t]
  \begin{center}
    \begin{tikzpicture}
      \draw (-1.5,0) -- (2.5,0);
      \draw (-1.5,3) -- (2.5,3);
      \draw [dashed] (0.5,0) -- (0.5,3);
      \coordinate [label=center:$m_1$] (m) at (0.5,3.3);
      \coordinate [label=center:$m_2$] (m) at (0.5,-0.3);
      \draw (4.,0) -- (8.,0);
      \draw (4.,3) -- (8.,3);
      \draw [dashed] (4.5,0) -- (7.5,3);
      \draw [dashed] (7.5,0) -- (4.5,3);
      \coordinate [label=center:$m_1$] (m) at (4.5,3.3);
      \coordinate [label=center:$m_1$] (m) at (7.5,3.3);
      \coordinate [label=center:$m_2$] (m) at (4.5,-0.3);
      \coordinate [label=center:$m_2$] (m) at (7.5,-0.3);
      \draw (9.5,0) -- (13.5,0);
      \draw (9.5,3) -- (13.5,3);
      \draw [dashed] (11.5,3) -- (11.5,1.5);
      \draw [dashed] (11.5,1.5) -- (10.,0);
      \draw [dashed] (11.5,1.5) -- (11.5,0);
      \draw [dashed] (11.5,1.5) -- (13.,0);
      \coordinate [label=center:$m_1$] (m) at (11.5,3.3);
      \coordinate [label=center:$m_2$] (m) at (10.,-0.3);
      \coordinate [label=center:$m_2$] (m) at (11.5,-0.3);
      \coordinate [label=center:$m_2$] (m) at (13.,-0.3);
    \end{tikzpicture} 
    \caption{Diagrams representing the Newtonian potential (first on the left)
      and the leading corrections from the Riemann$^4$ terms in the fundamental
      Lagrangian in eq.~(\ref{eq:fund_lag}). Last diagram must be supplemented
      with its mirror image under $1\leftrightarrow 2$.} 
    \label{fig:fund_diags}
  \end{center}
\end{figure}
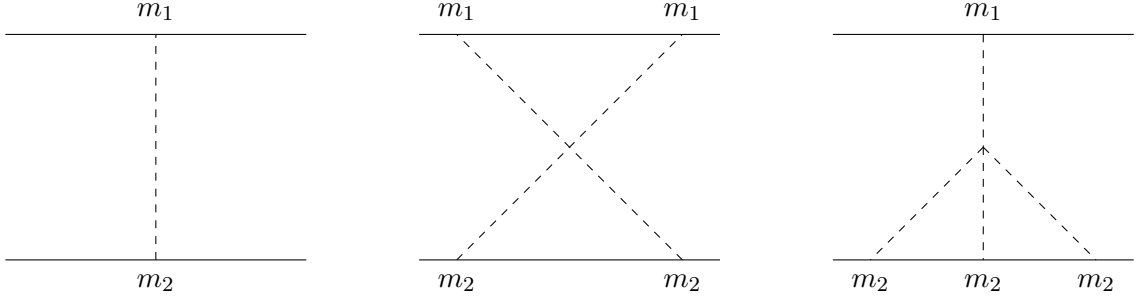

The extra terms introduce bulk interactions of quartic and higher orders
affecting the equation of motions and the radiation emission (note that to derive
the GW phase dynamical evolution one needs both the energy of bound orbits and
the luminosity function) for both non-spinning and spinning sources.

Defining $M\equiv m_1+m_2$ as the sum of rest masses of binary constituents
$m_{1,2}$, and introducing for later use the
reduced mass $\mu\equiv m_1m_2/M$ and the symmetric mass ratio $\eta\equiv \mu/M$, straightforward
dimensional argument shows that the quartic interactions introduced above
add to the two-body potential terms of the order of $(G M/r)^3/(\Lambda r)^6$,
leading to corrections of the order 2PN  $\times 1/(\Lambda r)^6$,
and analogously for the spin equations of motion, with
exception for the parity violating term
${\mathcal C}\tilde{\mathcal C}$ that will be discussed later.

To make contact with the $v$ expansion of the PN approximation,
we find convenient to express the metric via the Kaluza-Klein (KK)
parameterisation \cite{Kol:2007bc}
\be
\label{eq:met_KK}
g_{\mu\nu}=e^{2\phi/m_{Pl}}\pa{
  \ba{cc}
  -1 & \dfrac{A_i}{m_{Pl}}\\
  \dfrac{A_j}{m_{Pl}} &\, e^{-c_d\phi/m_{Pl}}\pa{\delta_{ij}+\dfrac{\sigma_{ij}}{m_{Pl}}}-
  \dfrac{A_iA_j}{m^2_{Pl}}
  \ea}\,,
\ee
with $m_{Pl}\equiv \pa{32\pi G_N}^{-1/2}$, $c_d\equiv 2(d-1)/(d-2)$,
where $d$ is the number of space dimension, $d=3$ in the rest of this work,
and Latin indices $i,j\ldots$ run over pure space dimensions.
The quadratic terms of the gravity bulk Lagrangian, unaffected by the
${\cal C}^2,{\cal C}\tilde {\cal C},\tilde{\cal C}^2$ terms, are then given by
\be
\label{eq:sEH_KK}
\ba{rcl}
\ds {\mathcal S}_{eff} &=&\ds \int {\rm d}^{d+1}x\sqrt{-\gamma}
\left\{\frac 14\paq{\pa{\nabla_k\sigma}^2-2\pa{\nabla_k\sigma_{ij}}^2-
  \pa{\dot\sigma^2-2\pa{\dot\sigma_{ij}}^2}}\right.\\
&&\ds\left. -c_d\paq{\pa{\nabla_k\phi}^2-\dot\phi^2}
+\paq{\frac 12F_{ij}^2+\pa{\nabla_kA_i}^2-\dot A_i^2}\right\}\,,
\ea
\ee
with $\sigma\equiv\sigma_{ij}\delta^{ij}$,
where time and space-derivatives have been split to make manifest the
scaling in $v$ of the potential gravitational modes
as their derivatives scale differently
with $v$: $\frac {\rm d}{{\rm d}t}\sim v^i\partial_i$.

The parameterisation in eq.~(\ref{eq:met_KK}) has the advantage that expanding
around the Minkowski metric $\eta_{\mu\nu}\equiv \rm{diag}(-1,1,1,1)$ it returns
diagonal Feynman propagators for the fields $\phi,A_i,\sigma_{ij}$
\cite{Foffa:2011ub,Foffa:2012rn}:
\be
\label{eq:propsKK}
\left.
\ba{rcl}
\ds P[\phi,\phi]&=&\ds-\frac 18\\
\ds P[A_i,A_j]&=&\ds \frac{\delta_{ij}}2\\
\ds P[\sigma_{ij},\sigma_{kl}]&=&\ds -\frac 12\pa{\delta_{ik}\delta_{jl}+\delta_{il}\delta_{jk}-2\delta_{ij}\delta_{kl}}
\ea
\right\}\times \frac i{k^2-k_0^2-i\epsilon}\,,
\ee
where $k\equiv |\vec k|$ is the modulus of the three-momentum.

\subsection{Spin-less case}

\subsubsection{Two-body potential in GR}
The standard coupling to gravity of a spin-less particle of mass $m_a$
with trajectory $x_a$ is given by the world-line
\be
\label{eq:wl_KK}
\left.{\cal S}_{pp-wl}\right|_{\vec S=0} =-m_a\int d_{x_a}\tau\supset \frac {m_a}{m_{Pl}}\int_{x_a} {\rm d}t\paq{
  -\phi+A_iv_i+\frac 12\sigma_{ij}v^iv^j+\ldots}\,,
\ee
where dots stand for non-linear coupling with gravity fields
and higher order in velocity expansion.
The lowest order potential is given by the first diagram in fig.~\ref{fig:fund_diags} \footnote{We adopt the notation $\int_\K\equiv\int\frac{d^dk}{(2\pi)^d}$
and $d=3$ throughout this paper.}
\be
V_N=-\frac{m_1m_2}{8m^2_{Pl}}\int_\K
\frac{e^{i\vec k\cdot \vec r}}{k^2}=-\frac{G_Nm_1m_2}r\,,
\ee
which is the standard Newton potential, obtained by taking the static limit of
both the world-line action eq.~(\ref{eq:wl_KK}) and of the
propagators (\ref{eq:propsKK}). The additional diagrams in fig.~\ref{fig:fund_diags}
are the LO ones modifying the Newtonian potential
due to Riemann$^4$ terms in eq.~(\ref{eq:fund_lag}).

\subsubsection{Radiation emission in GR}

\begin{figure}[t]
  \begin{center}
    \begin{tikzpicture}
      \draw (-1.5,0) -- (2.5,0);
      \draw (-1.5,3) -- (2.5,3);
      \draw [decorate, decoration=snake, green] (0.5,3.) -- (2.5,4.5);
      \draw (4.,0.) -- (8.,0);
      \draw (4.,3.) -- (8.,3);
      \draw [dotted, blue, thick] (6.,0) -- (6.,3);
      \draw [decorate, decoration=snake, green] (6.,1.5) -- (8.,1.5);
    \end{tikzpicture}
  \end{center}
  \caption{Diagrams describing LO radiative process from a spin-less binary
    system. The first one must be supplemented by its mirror image under
    $1\leftrightarrow 2$. Green wavy lines represent radiation, blue dotted
  line are $\phi$ longitudinal modes.}
  \label{fig:rad_quad_GR}
\end{figure}
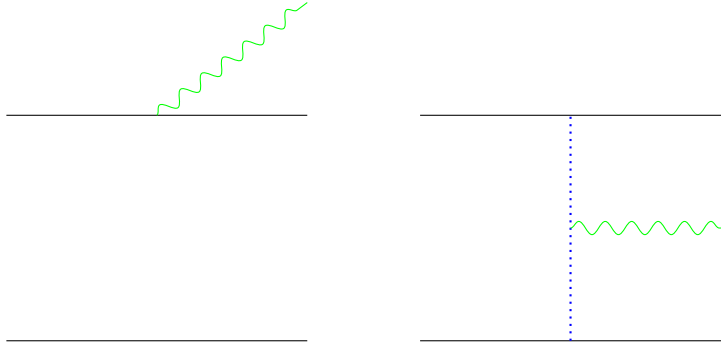

The coupling of a binary system to radiative gravitational modes
in GR can be computed via the diagrams in fig.~\ref{fig:rad_quad_GR}
\cite{Goldberger:2004jt}, i.e. by analysing the emission of a
(trace-less) $\sigma_{ij}$ mode, which gives the
following LO effective Lagrangian:
\be
\label{eq:rad_LO}
\ds{\cal L}_{rad-I}&=&\ds\frac 12 T^{ij}\frac{\sigma_{ij}}{m_{Pl}}\\
\label{eq:rad_LO2}
&=&\ds \frac{m_1}2\pa{v_1^iv_1^j-\frac{G_Nm_2r^ir^j}{2r^3}}
\frac{\sigma_{ij}}{m_{Pl}}+(1\leftrightarrow 2)\,.
\ee

By using the (spin-less) Newtonian equation of motion
\be
\label{eq:Newt}
\left.\vec a_1\right|_{\vec S_{1,2}=0}=-\frac{G_Nm_2\vec r}{r^3}\,,
\ee
one can recast eq.~(\ref{eq:rad_LO2}) into the following standard form
\be
\label{eq:rad_ele_GR}
{\cal L}_{rad-I}=-\frac 12I^{ij}R_{0i0j}\,,
\ee
where $I^{ij}$ is the radiative electric quadrupole which at LO
equals the (trace-less) mass quadrupole
$Q^{ij}=\sum_{a=1}^2 m_a\pa{x_a^ix_a^j-\frac 13\delta^{ij}x_a^2}$ and $R_{0i0j}$ is
the \emph{electric} part of the Riemann tensor, whose explicit expression at
linear order in the metric perturbation reads:
\be
\label{eq:riem_ele_tt}
m_{Pl}R_{0i0j}&\simeq& -\frac 12\pa{\ddot\sigma_{ij}-\dot A_{i,j}-\dot A_{j,i}}
+\phi_{,ij}+\frac{\delta_{ij}}{d-2}\ddot\phi\,.
\ee

More generally, the radiative coupling can be equivalently expressed either
in terms of moments of space-space components the energy-momentum tensor
like in (\ref{eq:rad_LO}), or in
terms of mass and momentum multipole moments like in (\ref{eq:rad_ele_GR}).
The derivation of the equivalence is a standard GR-course exercise which uses
the conservation of the energy-momentum tensor
\be
\label{eq:eomcons}
T^{\mu\nu}_{\ \ ,\nu}=0\,,
\ee
and we report here the radiative
multipolar coupling up to NLO in terms of the electric octupole $O^{ijk}$ and
the magnetic quadrupole $J^{ij}$, see app.~\ref{app:rad_mult} for details,
\be
\label{eq:mult_rad_GR}
L_{rad-GR}=-\int {\rm d}t \pa{\frac 12I^{ij}{\cal E}_{ij}+\frac 16O^{ijk}{\cal E}_{ij,k}-
  \frac 23J^{ij}{\cal B}_{ij}}+{\rm higher\ multipoles}\,,
\ee
where ${\cal E}_{ij}\equiv R_{0i0j}$ and ${\cal B}_{ij}\equiv\frac 12\epsilon_{ikl}R_{0jkl}$
is the magnetic part of the Riemann tensor.
From (\ref{eq:mult_rad_GR}) one can derive the standard GR flux formula
which is given, writing explicitly only the electric and magnetic quadrupole
contributions, by \cite{Thorne:1980ru}
\be
\label{eq:flux_GR}
F=G_N\pa{\frac 15\dddot I_{ij}^2+\frac{16}{45}\dddot J_{ij}^2+\ldots}\,,
\ee
from which the electric quadrupole contribution gives the LO formula for emission
of GWs from circular orbits $F_{LOcirc}=\frac{32}{5G_N}\eta^2 v^{10}$.

\subsection{Spin degrees of freedom in GR}
Spin degrees of freedom in GR require the introduction of additional degrees
of freedom than just position and velocity, embodied by two tetrads:
$e^\mu_a$, relating the metric into the locally free-falling frame
\cite{Hanson:1974qy,Porto:2005ac}
\be
g_{\mu\nu}e^\mu_a e^\nu_b=\eta_{ab}\,,
\ee
and $e^{\mu}_A$, co-rotating with the spinning body, and related to the
former by a local Lorentz transformation $e^{\mu}_A=\Lambda^a_A e^{\mu}_a$ (we use
$a,b,c,d$ to denote flat space-time Lorentz indices with their capitalised
version transforming under the residual Lorentz invariance).

The transport of $e^{\mu}_A$ along the world-line of a reference point chosen
inside the extended body defines the generalised angular velocity
$\Omega^{\mu\nu}$
\be
\frac{{\rm d}e^{A\mu}}{{\rm d}\tau}\equiv u^\rho e^{A\mu}_{;\rho}=\Omega_{\nu}^{\mu} e^{A\nu}\Longrightarrow\Omega^{\mu\nu}=
e^\mu_A\frac{{\rm d}e^{A\nu}}{{\rm d}\tau}=-\Omega^{\nu\mu}\,,
\ee
where $u^\rho$ is the four-velocity of the object's world-line.
Local coordinate, Lorentz and parameterisation invariances require the
Lagrangian to be made of invariant contractions of $\Omega^{\mu\nu}$, $u^\rho$ and
eventually of the local curvature tensors, but do not unambiguously fix its
form even in the case of flat space-time.
However it turns out that if one neglects finite-size effects
the variation of any possible Lagrangians w.r.t. to the spinning body local position and tetrad, when expressed in terms of the conjugate momenta
$p^\mu=\frac{\delta{\cal L}}{\delta u_\mu}$ and
$S^{\mu\nu}=\frac{\delta{\cal L}}{\delta \Omega_{\mu\nu}}$, gives the
\emph{Mathisson-Papapetrou} equations of motion
\cite{Mathisson:1937zz,Papapetrou:1951pa,Dixon:1970zza}:
\renewcommand{\arraystretch}{1.6}
\be
\ba{rcl}
\ds\frac{{\rm d}p^\mu}{{\rm d}\tau}&=&\ds -\frac{1}{2}R_{\mu\nu\rho\sigma}u^\nu S^{\rho\sigma}\,,\\
\ds\frac{{\rm d}S^{\mu\nu}}{{\rm d}\tau}&=&\ds p^\mu u^\nu-p^\nu u^\mu\,.
\ea
\ee
\renewcommand{\arraystretch}{1.}

Since the physical spin variables are related to the conjugate momentum $S^{\mu\nu}$
rather than to the fundamental tetrad variables, it can be convenient
to work with a functional that behaves as an Hamiltonian with respect to the
spin, while remaining a Lagrangian with respect to the body position $x^\mu$.
Such hybrid functional is called a Routhian \cite{goldstein}, defined as
the Legendre transform of the Lagrangian $\cal L$, which gives the explicit
form (valid up to linear order in the spin)
\be
\label{routhian}
{\cal R}=m\sqrt{u^2}+\frac 12\omega_\mu^{ab}S_{ab}u^\mu\,,
\ee
being $\omega_{\mu}^{ab}\equiv e^{b\nu} e^a_{\nu;\mu}$ the spin connection.
One then recovers the Mathisson-Papapetrou equations via
\be
\label{eqs:eom}
\frac{\delta}{\delta x^\mu}\int{\rm d}t\, {\cal R}=0\,,\quad \frac{{\rm d}S^{ab}}{{\rm d}\tau}=\left\{{\cal R},S^{ab}\right\}\,,
\ee
once the following Poisson bracket is taken into account\footnote{Note the
  difference in sign with respect to eq.~(8.25) of \cite{Porto:2016pyg}, where
  a mostly minus metric signature convention is used.}:
\be
\label{eq:spin_Poi}
\left\{S^{ab},S^{cd}\right\}=\eta^{ad}S^{bc}+\eta^{bc}S^{ad}-\eta^{ac}S^{bd}-\eta^{bd}S^{ac}\,.
\ee

The anti-symmetric tensor $S^{\mu\nu}$ (which appears above through its locally
flat-frame components $S^{ab}\equiv S^{\mu\nu}e_\mu^a e_\nu^b$) is the generalised,
relativistically-covariant spin of the body, however it contains redundant
degrees of freedom.
The redundancy corresponds to the ambiguity related the choice of a reference
world-line inside the body. One can reduce from 6 to the 3 degrees of freedom
needed to describe an ordinary spin vector by imposing the
\emph{Spin Supplementary Condition} (SSC), which relates the $3$-vector $S^{0i}$
to the physical spin components $S^{i}\equiv\frac 12\varepsilon^{ijk} S_{jk}$.
There is not a unique way to impose such condition, e.g. one can use
the \emph{covariant} SSC $S^{\mu\nu}p_\nu=0$ condition \cite{Hanson:1974qy}, the
\emph{baryonic} $S^{i0}=\frac 12 S^{ij}u_j$ \cite{Barker:1975ae}, among others,
both of which can be described at LO by
\be
\label{eq:ssc_kappa}
S^{i0}=\kappa S^{ij}v_j\,,
\ee
with $\kappa=1,\frac 12$.
The requirement of SSC conservation along the world line
shifts the momentum of the particle by a quantity \emph{quadratic} in the spin
\cite{Hanson:1974qy}, and it will be neglected in this work that is restricted
to liner-in-spin effects.

Being an algebraic constraint, as far as the orbital equation of motions are
concerned, the SSC can be imposed by direct replacement of
$S^{i0}$ indifferently at the level of the fundamental Routhian, in the
effective potential or in the equations of motion:
we will adopt here the second option (substitute into the potential).
However when deriving the spin equations of motion the SSC constraint has
to be imposed at the level of the equation of motions, i.e. after applying
the Poisson bracket in eq.~(\ref{eqs:eom}).

Note that adopting the covariant SSC, i.e. $\kappa=1$ in eq.~(\ref{eq:ssc_kappa}),
the Poisson brackets become \emph{non-canonical}:
brackets between coordinates and between coordinates and spin
do not vanish \cite{Hanson:1974qy}.
Instead of dealing with a non-canonical algebra one can equivalently
shift the world-line of each particle according to \cite{Porto:2005ac}
\be
\label{eq:cssc_shift}
\vec x_{1,2}\to \vec x_{1,2}-\frac 1{2m_{1,2}}\vec S_{1,2}\times \vec v_{1,2}\,,
\ee
and use a canonical algebra when deriving the
equation of motions for coordinates and spin, or equivalently one can adopt
the baryonic SSC, $\kappa=1/2$ in eq.~(\ref{eq:ssc_kappa}), and use canonical
Poisson brackets \cite{Porto:2005ac}.

The spin-dependent part of the world-line action in terms of the KK
parameterisation, for a particle with mass $m$ and
velocity $v$ at linear order in spin is (for $d=3$)\footnote{Note that the point-particle Routhian has the same sign as the potential, hence opposite sign with respect to the Lagrangian.} \cite{Foffa:2013qca}
\be
\label{eq:lag_KK}
\ba{rcl}
\ds {\cal S}_{pp-wl} &\supset&\ds \frac 1{m_{Pl}}\int_{x_a}{\rm d}t\left[
  S^{ij}\pa{-\frac{1}{4}F_{ij}+\frac 12\sigma_{ik,j}v_a^k+\phi_{,j}v_{ai}+\frac 14F_{jk}\sigma^k_{\ i}+\frac 12\sigma_{ik}\pa{\phi_{,j}v_a^k+\phi^{,k}v_{aj}}}\right.\\
  &&\ds\left. \qquad+S^{0i}\pa{-\phi_{,i}+\frac 12\dot\sigma_{ij}v^j_a
    -\frac 12\phi_{,j}\sigma_{ij}}\right]\,,
\ea
\ee
where all field are understood to be evaluated on the world-line of the
source-particle and we displayed only terms that will be needed in the rest of
this work.

For power counting spin $|\vec S_a|\sim G_Nm_a^2$, angular momentum
$|\vec L|\sim \eta G_NM^2/v$ hence terms linear in spin, i.e. of the type
$\vec S\cdot \vec L$ appear at lower order than terms $\sim \vec S^2$ in
the PN expansion.

\subsubsection{Spin terms in two-body potential in GR}

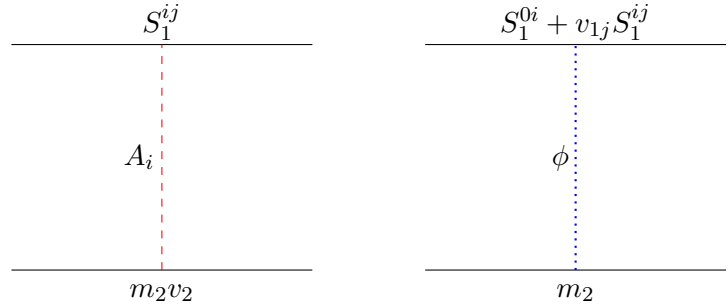
\begin{figure}[ht]
  \begin{center}
    \begin{tikzpicture}
      \draw (-0.5,0) -- (3.5,0);
      \draw (-0.5,3) -- (3.5,3);
      \draw [dashed, red] (1.5,0) -- (1.5,3);
      \coordinate [label=center:$S_1^{ij}$] (m) at (1.5,3.3);
      \coordinate [label=center:$m_2 v_2$] (m) at (1.5,-0.3);
      \coordinate [label=center:$A_i$] (m) at (1.2,1.5);
      \draw (5.,0) -- (9.,0);
      \draw (5.,3) -- (9.,3);
      \draw [dotted, blue, thick] (7.,0) -- (7.,3);
      \coordinate [label=center:$S_1^{0i}+v_{1j}S_1^{ij}$] (m) at (7.,3.3);
      \coordinate [label=center:$m_2$] (m) at (7.,-0.3);
      \coordinate [label=center:$\phi$] (m) at (6.8,1.5);
    \end{tikzpicture} 
    \caption{Diagrams representing the LO spin-orbit potential
      in GR. Diagrams must be supplemented with their $1\leftrightarrow 2$
      mirror image. Dashed red line represent a $A_i$-polarised longitudinal
      mode, blue dotted lines are $\phi$-modes.} 
    \label{fig:spin_orbit}
  \end{center}
\end{figure}

The lowest order spin contribution to the two body potential in GR occurs at 1.5PN order,
i.e. $v^3$ order with respect to the leading,
and is due to the sum of the two processes represented in
fig.~\ref{fig:spin_orbit} (and their mirror images under $1\leftrightarrow 2$).

The sum of the exchange of gravitational modes with $A_i$ and $\phi$
polarisations reported in fig.~\ref{fig:spin_orbit}, with vertices given by
the Lagrangian (\ref{eq:lag_KK}) gives ($\vec r\equiv \vec x_1-\vec x_2$)
for the LO spin-dependent (\emph{spin-orbit}) potential
\be
\label{eq:so_l}
V_{SO}^{(LO)}=2\frac{G_N m_2}{r^3}\vec S_1\cdot\pa{\vec r\times \vec v}+\pa{\kappa-1}\frac{G_Nm_2}{r^3}\vec S_1\cdot\pa{\vec r\times \vec v_1}+1\leftrightarrow 2\,.
\ee
After substituting the covariant SSC (\ref{eq:ssc_kappa}) with $\kappa=1$
one obtains a different result than the classical one
\cite{Barker:1970zr,Barker:1975ae,Thorne:1984mz} which implicitly uses $\kappa=1/2$,
however things reconcile at the level of the equation of motions.
Indeed using the covariant SSC (i.e. $\kappa=1$), the spin equations of motion,
which are first order,
are obtained by using the second of the eqs.~(\ref{eqs:eom}), \emph{then}
applying the SSC, giving:
\be
\label{eq:spin_eom}
\dot S^i_1=\frac 12\epsilon^{ijk}\pag{V_{SO}^{(L)},S_{1jk}}=\frac{G_N m_2}{r^2}\paq{
\vec S_1\times\pa{\hat n\times\pa{2\vec v_2-\vec v_1}}
+\hat n\times\pa{\vec v_1\times\vec S_1}}^i\,,
\ee
which implies that the norm of $\vec S_1$ is not conserved, since
$\dot{\vec S}_1$ is not perpendicular to $\vec S_1$.
The previous result can be recast into the standard one by performing the $O(v^2)$ shift of the spin variable according to
\be
\label{eq:spin_redef}
\vec S_a\to \pa{1-\frac{v_a^2}2}\vec S_a+
\frac{\vec v_a}2\pa{\vec v_a\cdot \vec S_a}\,,
\ee
for $a=1,2$. The shift (\ref{eq:spin_redef}) (together with
the shift (\ref{eq:cssc_shift}) to be used for $\kappa=1$ only) recast
the non-canonical Poisson brackets into canonical ones, finally
obtaining a spin vector with constant norm (neglecting absorption effects
\cite{Poisson:1994yf}),
and whose derivative at LO, after substituting the center of mass
relationships
\be
\label{eq:com}
\vec v_1=\ds\frac{m_2}M \vec v\,,\qquad
\vec v_2=\ds -\frac{m_1}M \vec v\,,
\ee
is given by the standard form \cite{Kidder:1992fr,Kidder:1995zr}
\be
\label{eq:sdotLO}
\frac{{\rm d}{\vec S}_1}{{\rm d}t}= \frac{G_N}{r^3}\paq{2+\frac{3m_2}{2m_1}}
\vec L\times \vec S_1\,,
\ee
being $\vec L\equiv \mu\, \vec r\times \vec v$ the Newtonian orbital angular momentum.
The result in eq.~(\ref{eq:sdotLO}) could have been obtained straightforwardly
by substituting relations
(\ref{eq:com}) and $\kappa=1/2$ into the potential (\ref{eq:so_l}) and
then applying canonical Poisson brackets, without the shifts
(\ref{eq:cssc_shift}) and (\ref{eq:spin_redef}) \cite{Barker:1975ae}.

Eq.~(\ref{eq:sdotLO}) is responsible for the precession of the spin
around the orbital angular momentum\footnote{
The orbital angular momentum is usually larger than individual spins, unless $\eta\ll 1$ and $v\lesssim 1$.} with angular velocity $\Omega_S\sim v^3/r$, i.e.
the precession time scale is longer by a factor $v^{-2}$ than the orbital scale
$r/v$ but shorter
than the typical dissipation time-scale $\eta^{-1} r/v^6$.

Finally, from the potential (\ref{eq:so_l}) and the world-line coordinate shift
(\ref{eq:cssc_shift}) one can derive the Newtonian equation of
motion including effects linear in spin for the relative acceleration of two
point-particles at LO in $v$ \cite{Kidder:1992fr}:
\be
\label{eq:eom_GR_SO}
{\vec a}=-\frac{G_NM {\vec r}}{r^3}+2\frac{G_N}{r^3}\paq{
  2\pa{\vec S_m\times \vec v}
  +3\frac{\vec r\cdot\vec v}{r^2}\pa{\vec r\times\vec S_m}
  +3\frac{\vec r}{r^2}\pa{\vec S_m\cdot\pa{\vec r\times \vec v}}}\,,
\ee
where $\vec S_m\equiv \pa{1+\frac{3m_2}{4m_1}}\vec S_1+\pa{1+\frac{3m_1}{4m_2}}\vec S_2$. Also in this case the equation of motion (\ref{eq:eom_GR_SO}) can be
obtained straightforwardly from the potential (\ref{eq:so_l}) with $\kappa=1/2$.

\subsubsection{Spin terms in radiation emission in GR}

\begin{figure}[t]
  \begin{center}
    \begin{tikzpicture}
      \draw (-1.5,4) -- (2.5,4);
      \draw (-1.5,7) -- (2.5,7);
      \draw [decorate, decoration=snake, green] (.5,7.) -- (2.5,8.);
      \filldraw[black] (0.5,7.) circle (2pt) node[anchor=south] {$S_1$};
      \draw (4.,4) -- (8.,4);
      \draw (4.,7) -- (8.,7);
      \draw [dotted, blue, thick] (6.,4.) -- (6.,7.);
      \draw [decorate, decoration=snake, green] (6.,7.) -- (8.,8.);
      \filldraw[black] (6.,7.) circle (2pt) node[anchor=south] {$S_1$};
      \draw (9.5,4) -- (13.5,4);
      \draw (9.5,7) -- (13.5,7);
      \draw [dashed, red] (11.5,4.) -- (11.5,7.);
      \draw [decorate, decoration=snake, green] (11.5,7.) -- (13.5,8.);
      \filldraw[black] (11.5,7.) circle (2pt) node[anchor=south] {$S_1$};
      \draw (-1.5,0) -- (2.5,0);
      \draw (-1.5,3) -- (2.5,3);
      \draw [dashed, red] (.5,0.) -- (.5,3.);
      \draw [decorate, decoration=snake, green] (.5,1.5) -- (2.5,1.5);
      \filldraw[black] (.5,3.) circle (2pt) node[anchor=south] {$S_1$};
      \draw (4.,0) -- (8.,0);
      \draw (4.,3) -- (8.,3);
      \draw [dashed, red] (6.,3.) -- (6.,1.5);
      \draw [dotted, blue, thick] (6.,0.) -- (6.,1.5);
      \draw [decorate, decoration=snake, green] (6.,1.5) -- (8.,1.5);
      \filldraw[black] (6.,3.) circle (2pt) node[anchor=south] {$S_1$};
      \draw (9.5,0) -- (13.5,0);
      \draw (9.5,3) -- (13.5,3);
      \draw [dotted, blue, thick] (11.5,0.) -- (11.5,3.);
      \draw [decorate, decoration=snake, green] (11.5,1.5) -- (13.5,1.5);
      \filldraw[black] (11.5,3.) circle (2pt) node[anchor=south] {$S_1$};
    \end{tikzpicture}
  \end{center}
  \caption{Diagrams determining (part of) the LO spin terms in gravitational
    radiation emission. The first one contributes at LO to the
    magnetic quadrupole, the remaining ones to the electric quadrupole.
    Additional contribution comes from the expression of the radiative multipoles
    in terms of the energy-momentum tensor, see app.~\ref{app:rad_mult}.}
  \label{fig:rad_spin_GR}
\end{figure}
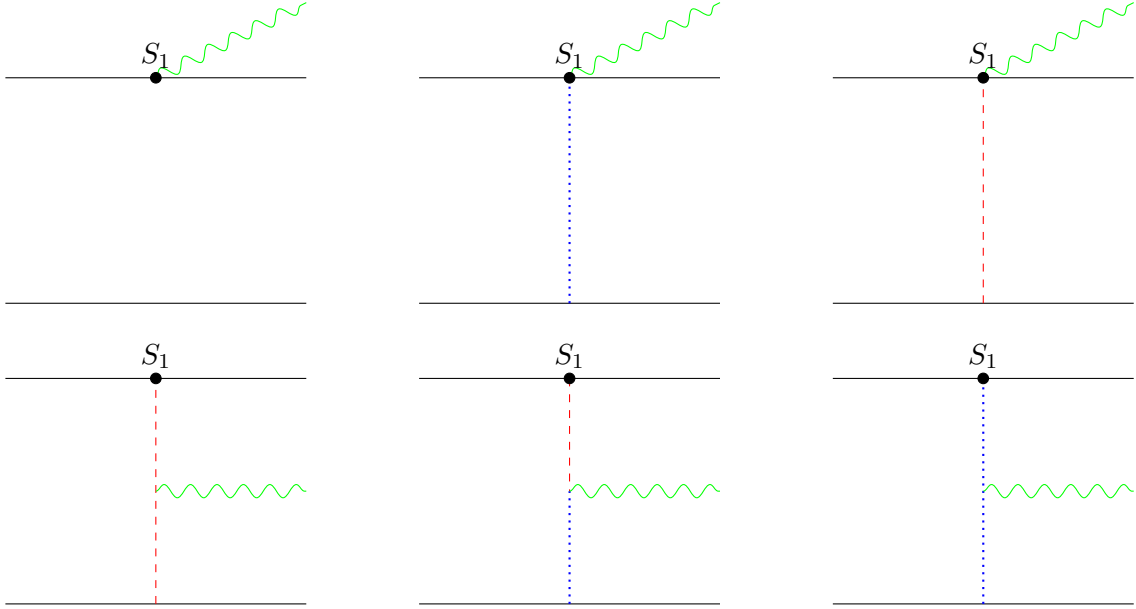

The lowest order spin-dependent source coupling to $\sigma_{ij}$
is given by the diagrams in fig.~\ref{fig:rad_spin_GR}.
The first diagram does not involve any field propagator, hence it can be
directly read from the Lagrangian, and its leading contribution from the
$\sim S^{ij}\sigma_{ik,j}v^j$ term in eq.~(\ref{eq:lag_KK}) is of \emph{magnetic}
quadrupole type
\be
\label{eq:rad_JS}
{\cal L}_{rad-JS}= \frac 12S_{1ij}\frac{\sigma_{jk,i}}{m_{Pl}}v^k_1+1\leftrightarrow 2=
\frac 14\pa{S^i_1x^j_1+S^j_1x^i_1+1\leftrightarrow 2}\frac{\epsilon^{ikl}}{2m_{Pl}}
\pa{\dot\sigma_{ik,l}-\dot\sigma_{il,k}}\,,
\ee
implying by comparison with eq.~(\ref{eq:mult_rad_GR}) that the spinning part of
the magnetic quadrupole $J^{ij}$ is
\be
J^{ij}_S=\frac 34\sum_a\pa{S_a^ix_a^j+S^j_ax^i_a}_{TF}\,,
\ee
where ``$TF$'' stands for trace-free part.

The first diagram in fig.~\ref{fig:rad_spin_GR} also includes the
$\sim S^{0i}\dot\sigma_{ij}v^j$ interaction from the Lagrangian (\ref{eq:lag_KK})
which contributes to the electric quadrupole coupling as
\be
\label{eq:rad_QSI_GR}
\ba{rcl}
\ds{\cal L}_{rad-QSI}&=&\ds
\paq{\frac 12\frac{\rm d}{{\rm d}t}\pa{S^{i0}_1v_1^j+S^{j0}_1v_1^i}+
  1\leftrightarrow 2}\frac{\sigma_{ij}}{2m_{Pl}}\\
&=&\ds -\frac\kappa 2 \frac{G_Nm_2^2}{Mr^3}\paq{\pa{\vec v\times \vec S_1}^ir^j
  +\pa{\vec r\times \vec S_1}^iv^j+i\leftrightarrow j}\frac{\sigma_{ij}}{2m_{Pl}}
+1\leftrightarrow 2\,,
\ea
\ee
where in the second line the equation of motion (\ref{eq:Newt}) and the center
of mass relationships (\ref{eq:com}) have been used.

The 5 remaining diagrams in fig.~\ref{fig:rad_spin_GR} contribute to the electric
quadrupole, the leading contribution is of order $v$ with respect to
(\ref{eq:rad_JS}) giving in total:
\be
\label{eq:rad_QSII_GR}
\ba{rcl}
\ds{\cal L}_{rad-QSII}&=&\ds\frac{G_Nm_2}{r^3}\left\{
-\paq{\pa{\frac{\kappa+3}2\vec v_1-2\vec v_2}\times S_1}^ir^j-\frac 12
\pa{\vec r\times \vec S_1}^iv_1^j\right.\\
&&\ds\left.+3r^i\pa{\vec r\cdot \vec v}\pa{\vec r\times \vec S_1}^j+
\frac 32r^ir^j\paq{\pa{\vec r\times\pa{\pa{1+\kappa}\vec v_1-2\vec v_2}}
  \cdot\vec S_1}\right\}\frac{\sigma_{ij}}{m_{Pl}}+1\leftrightarrow 2\\
&=&\ds\frac{G_Nm_2}{Mr^3}\left\{
-\pa{\frac{\kappa+3}2m_2+2m_1}\pa{\vec v\times S_1}^ir^j-\frac 12
\pa{\vec r\times \vec S_1}^iv^j\right.\\
&&\ds\left.+3Mr^i\pa{\vec r\cdot \vec v}\pa{\vec r\times \vec S_1}^j+
\frac 32r^ir^j\pa{\pa{1+\kappa}m_2+2m_1}\pa{\vec r\times\vec v}\cdot\vec S_1
\right\}\frac{\sigma_{ij}}{m_{Pl}}+1\leftrightarrow 2\,,
\ea
\ee
where in the last passage the center of mass relationships (\ref{eq:com}) have
been inserted.

Eqs.~(\ref{eq:rad_QSI_GR}) and (\ref{eq:rad_QSII_GR}) give the explicit form of the linear
coupling of the sources to $\sigma_{ij}$, i.e. of $T^{ij}$, as per
eq.~(\ref{eq:rad_LO}). To check that this result coincides with the standard one
eq.~(\ref{eq:rad_ele_GR}), one has to use the conservation of the
energy momentum tensor (equivalent to the source equations of motion)
eq.~(\ref{eq:eomcons}) to reproduce the quadrupole moment from $T_{ij}$
via the standard GR textbook trick reported in app.~\ref{app:rad_mult}.

Indeed the linear-in-spin mass quadrupole
\be
Q^{ij}_S\equiv\int_V \left.T^{00}\right|_Sx^ix^j=\pa{S^{i0}_1x_1^j+S^{j0}_1x_1^i}+1\leftrightarrow 2\,,
\ee
satisfies $\ddot Q^{ij}=2\int_VT^{ij}$ on the equations of motion with
$T^{ij}$ obtained from ${\cal L}_{rad-QSI+II}=\frac 12T^{ij}\sigma_{ij}/m_{Pl}$, i.e. from the sum of eqs.~(\ref{eq:rad_QSI_GR}) and
(\ref{eq:rad_QSII_GR}), considering the contribution of the equation of
motion (\ref{eq:eom_GR_SO}) that needs to be used to cast  (\ref{eq:rad_LO2})
into (\ref{eq:rad_ele_GR}), using $\kappa=1/2$.~\footnote{We use here the baryonic SSC for simplicity, to avoid the complication of a
  non-standard Poisson algebra, see eq.~(\ref{eq:rad_cnstnc}) for details.}

For the spin-dependent part however, the radiative quadrupole $I^{ij}$ coupling to the
Riemann as per eq.~(\ref{eq:rad_ele_GR}) does not coincide
with the mass quadrupole even at LO, as it is rather given by
\cite{Thorne:1984mz}
\be
\ba{rcl}
\left.\ds I^{ij}\right|_S&=&\ds\left.\int_V \pa{T^{00}+T^{ll}-\frac 43 \dot T^{0l}x^l}\pa{x^ix^j}_{TF}\right|_S\\
&=&\ds \pag{\pa{\kappa+1}\paq{\pa{\vec v_1\times \vec S_1}^ir^j}
  -\frac 23\paq{\pa{\vec v_1\times \vec S_1}^i\vec r_1^j+\pa{\vec r_1\times \vec S_1}^i\vec v_1^j}+i\leftrightarrow j}_{TF}+1\leftrightarrow 2\,,
\ea
\ee
where for $T^{0l}$, $T^{ij}$ we inserted their LO expressions that can be read
directly from the Lagrangian (\ref{eq:wl_KK}): $\frac 14S_a^{ij}F_{ji}$ and
$\frac 12S_a^{ik}v^j_a\sigma_{ij,k}$.

Had not we used the GR equations of motions, we would have obtained the coupling
\be
\label{eq:rad_ele}
   {\cal L}_{rad}=\frac{\sigma_{ij}}{2m_{Pl}}\int_V\paq{T^{ij}+\frac 17\frac{{\rm d}^2}{{\rm d}t^2}
     \pa{\frac 23T^{ll}x^ix^j+\frac{11}6T^{ij}r^2-T^{il}x^lx^j-T^{jl}x^lx^i}}_{TF}\,,
   \ee
which is equivalent to (\ref{eq:rad_ele_GR}) in GR,
see app.~\ref{app:rad_mult} for details and derivation.

In total one has the electric quadrupole coupling, for $\kappa=1/2$ and in the
center of mass:\footnote{Note that
  using the LO expression for the linear in spin $T^{ij}$
  the term $\frac{{\rm d}^2}{{\rm d}t^2}\pa{T^{il}x^lx^j+T^{jl}x^lx^i}$ in
  eq.~(\ref{eq:rad_ele}) vanishes.}
\be
\label{eq:rad_IS}
      {\cal L}_{rad-QS}=-\frac 1{6}\frac{G_Nm_2^2}{M r^3}
      \paq{\pa{\vec r\times \vec S_1}^iv^j+\pa{\vec v\times\vec S_1}^ir^j
        -\frac 32\frac{\pa{\vec r\cdot\vec v}}{r^2}\pa{\vec S_1\times \vec r}^ir^j}
      \frac{\sigma_{ij}}{m_{Pl}}+1\leftrightarrow 2\,.
\ee

\section{Results for Riemann$^4$ terms}
\label{sec:results}

To systematically show the results when short-scale deviations from GR are
present,
we present separately the effects of the terms ${\cal C}^2$, ${\cal C}\tilde{\cal C}$
and $\tilde{\cal C}^2$ of eq.~(\ref{eq:fund_lag}).

\subsection{$\mathcal{C}^2$}
\label{ssec:cc}
The Riemann to the fourth power addition to the GR Lagrangian introduce 4-point
interactions that change both the potential and the emission formulae.
The spin-less LO corrections depending on the $\mathcal{C}^2$ terms can be derived from the contributions of
diagrams in fig.~\ref{fig:pot_cc} for the potential,
and fig.~\ref{fig:rad_cc} for the emission, which gives $G_N^2$
(or equivalently $v^4$) corrections with respect to the LO energy and
emitted radiation flux.

Considering effects linear in spin with the lowest number of velocity factors
(i.e. the lowest PN order), one has to compute the four-point vertex involving
the lowest number of time derivative and velocities (and neglecting terms vanishing on the equations of motion), hence one can approximate
\be
\ba{rcl}
\ds {\cal C}&\simeq&\ds \frac 8{m_{Pl}^2}
\left\{\pa{\partial_i\partial_j\phi}\pa{\partial^i\partial^j\phi}+
  \pa{\partial_i\dot A_j\pa{\partial^i\partial^j\phi}-\pa{\partial_i\partial_j A^j}\partial_i\dot\phi}
  -\frac 12\pa{\ddot\sigma_{ij}+\nabla^2\sigma_{ij}}\pa{\partial^i\partial^j\phi}
  \right.\\
  &&\ds \qquad 
  +\pa{\partial_i\partial_j\phi}\partial^i\partial_k\pa{\sigma^{jk}-\frac{\delta^{jk}}2\sigma}
  +\frac 12\partial_i\pa{\partial_kA_j-\partial_jA_k}\partial_k\dot \sigma_{ij}\\
  &&\ds\left.\qquad
  +\frac 14\pa{\partial_i\partial_jA_k}\partial_i\pa{\partial_jA_k-\partial_kA_j}
  +\frac 18\pa{\partial_i\partial_j\sigma_{kl}}\paq{
    \pa{\partial_i\partial_j\sigma_{kl}}+\pa{\partial_k\partial_l\sigma_{ij}}
    -2\pa{\partial_i\partial_k\sigma_{jl}}}
  +\ldots\right\}\,,
  \ea
  \ee
  \be
\label{eq:CC_KK}
\ba{rcl}
\ds {\mathcal C}^2&\simeq&\ds \frac{64}{m_{Pl}^4}\pa{\partial_m\partial_n\phi}\pa{\partial^m\partial^n\phi}\times
\left\{\pa{\partial_i\partial_j\phi}\pa{\partial_i\partial_j\phi}+
2\paq{\partial_i\dot A_j\pa{\partial^i\partial^j\phi}-\pa{\partial_i\partial_j A^j}\partial_i\dot\phi}\right.\\
&&\ds \
-\pa{\ddot\sigma_{ij}+\nabla^2\sigma_{ij}}\pa{\partial^i\partial^j\phi}
+2\pa{\partial_i\partial_j\phi}\partial^i\partial_k\pa{\sigma^{jk}-\frac{\delta^{jk}}2\sigma}+
\partial_i\pa{\partial_kA_j-\partial_jA_k}\partial_k\dot \sigma_{ij}\\
&&\ds\
+\frac 14\pa{\partial_i\partial_j\sigma_{kl}}\paq{
\pa{\partial_i\partial_j\sigma_{kl}}+\pa{\partial_k\partial_l\sigma_{ij}}
  -2\pa{\partial_i\partial_k\sigma_{jl}}}+\ldots\Big\}\,.
\ea
\ee
Note that in the spin-less case the potential gravitational mode $\phi,A_i,\sigma_{ij}$ couples to world-line with respectively 0,1,2 powers of velocity,
in the spinning case $A_i$ couples without time derivative or power of velocity,
whereas $\phi$ and $\sigma_{ij}$ require one power of velocity or a time
derivative, see eq.~(\ref{eq:lag_KK}).

\subsubsection{Potential}
\label{ssec:pot_cc}

\begin{figure}[ht]
  \begin{center}
    \begin{tikzpicture}
      \draw (4.,8) -- (8.,8);
      \draw (4.,11) -- (8.,11);
      \draw [dotted, blue, thick] (6,9.5) -- (4.5,8.);
      \draw [dotted, blue, thick] (6,11.) -- (6,8.);
      \draw [dotted, blue, thick] (6,9.5) -- (7.5,8.);
      \coordinate [label=center:$a$] (m) at (6.,7.5);
      \draw (9.5,8) -- (13.5,8);
      \draw (9.5,11) -- (13.5,11);
      \draw [dotted, blue, thick] (13.,8.) -- (10.,11);
      \draw [dotted, blue, thick] (13.,11.) -- (10.,8.);
      \coordinate [label=center:$b$] (m) at (11.5,7.5);
    \end{tikzpicture}
    \caption{Diagrams representing the leading corrections to the non-spinning
      potential from $\mathcal{C}^2$ interactions (blue dotted lines represent
      $\phi$ propagators). The cross diagram vanishes. The $a$ diagram must be
      complemented by its mirror image under particle exchange.} 
    \label{fig:pot_cc}
  \end{center}
\end{figure}
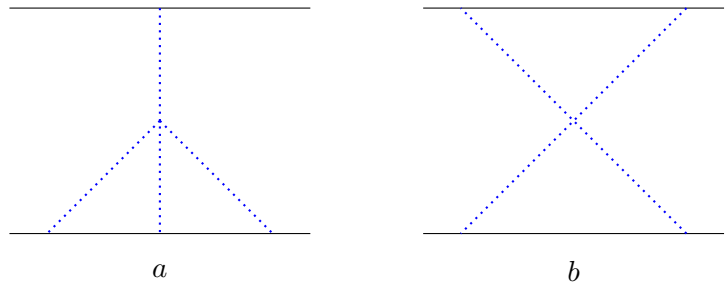

\noindent{\bf No spin}\\
The lowest order spin-dependent contributions to the potential
due to the ${\cal C}^2$ term are given by the
processes represented in fig.~\ref{fig:pot_cc}, plus their mirror image under
$1\leftrightarrow 2$ exchange. They correspond to the
the ``cross'' and ``peace/log'' diagrams as they were named in \cite{Endlich:2017tqa}.

The LO contribution to the potential is, see app.~\ref{ssapp:pot_cc} for
details, first computed in eq.~(5.9) of \cite{Endlich:2017tqa}
\be
\label{eq:pot_cc}
V_\Lambda=-V_N\frac{512}{\pa{\Lambda r}^6}\pa{\frac{G_Nm_2}r}^2+1\leftrightarrow 2\,,
\ee
where $V_N$ is the standard Newtonian potential, modifying the spin-independent
part of the equation of motion to 
\be
\label{eq:eom_cc}
{\vec a}_{\Lambda}=\vec a_{1\Lambda}-\vec a_{2\Lambda}=-\frac{G_NM {\vec r}}{r^3}
\paq{1-\frac{4608}{\pa{\Lambda r}^6}\frac{G_N^2(m_1^2+m_2^2)}{r^2}}\,.
\ee

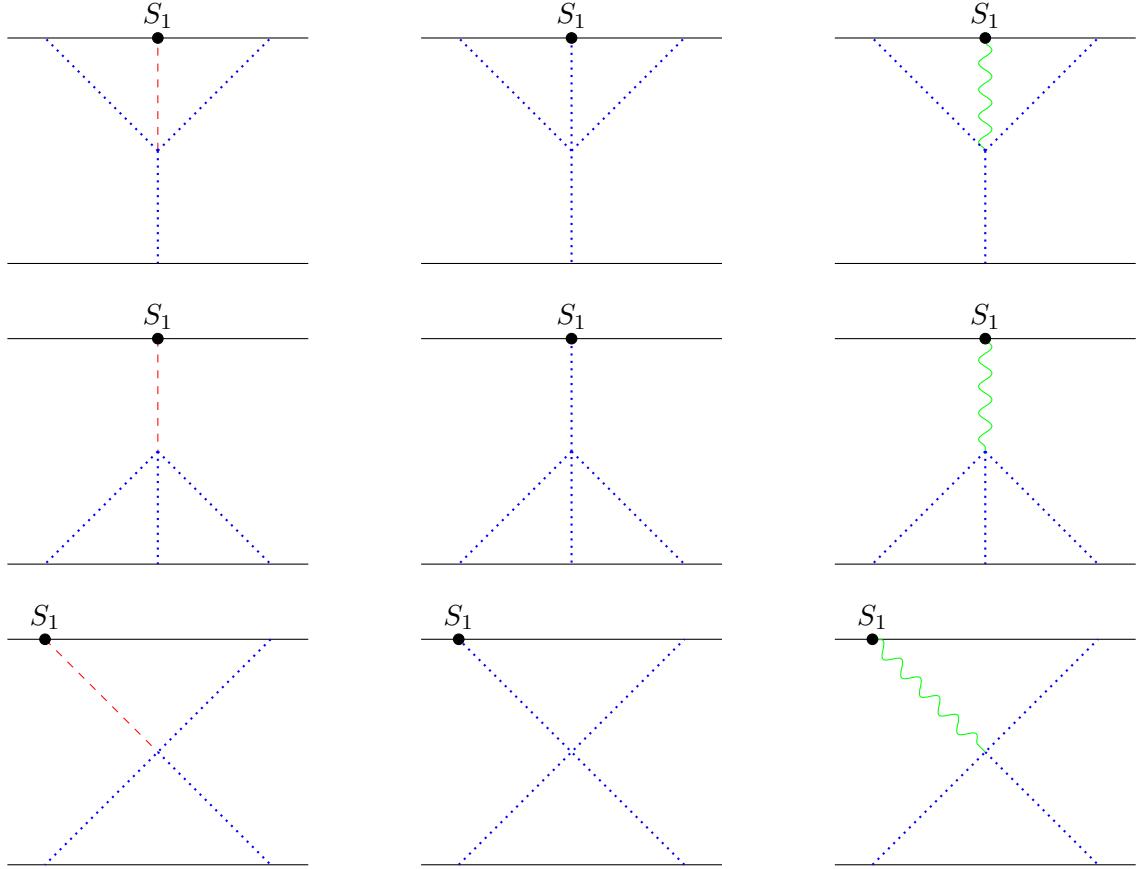
\begin{figure}[ht]
  \begin{center}
    \begin{tikzpicture}
      \draw (-1.5,0) -- (2.5,0);
      \draw (-1.5,3) -- (2.5,3);
      \draw [dashed, red] (-1.,3) -- (0.5,1.5);
      \draw [dotted, blue, thick] (0.5,1.5) -- (2.,0);
      \draw [dotted, blue, thick] (2.,3.) -- (-1.,0.);
      \filldraw[black] (-1.,3.) circle (2pt) node[anchor=south] {$S_1$};
      \draw (4.,0) -- (8.,0);
      \draw (4.,3) -- (8.,3);
      \draw [dotted, blue, thick] (4.5,0) -- (7.5,3);
      \draw [dotted, blue, thick] (7.5,0) -- (4.5,3);
      \filldraw[black] (4.5,3.) circle (2pt) node[anchor=south] {$S_1$};
      \draw (9.5,0) -- (13.5,0);
      \draw (9.5,3) -- (13.5,3);
      \draw [decorate, decoration=snake, green] (10,3) -- (11.5,1.5);
      \draw [dotted, blue, thick] (11.5,1.5) -- (13.,0.);
      \draw [dotted, blue, thick] (10,0) -- (13.,3.);
      \filldraw[black] (10.,3.) circle (2pt) node[anchor=south] {$S_1$};
      \draw (-1.5,4) -- (2.5,4);
      \draw (-1.5,7) -- (2.5,7);
      \draw [dashed, red] (.5,7) -- (0.5,5.5);
      \draw [dotted, blue, thick] (0.5,5.5) -- (-1.,4);
      \draw [dotted, blue, thick] (0.5,5.5) -- (.5,4);
      \draw [dotted, blue, thick] (0.5,5.5) -- (2.,4);
      \filldraw[black] (.5,7.) circle (2pt) node[anchor=south] {$S_1$};
      \draw (4.,4) -- (8.,4);
      \draw (4.,7) -- (8.,7);
      \draw [dotted, blue, thick] (6,5.5) -- (4.5,4);
      \draw [dotted, blue, thick] (6,7.) -- (6,4.);
      \draw [dotted, blue, thick] (6,5.5) -- (7.5,4);
      \filldraw[black] (6,7.) circle (2pt) node[anchor=south] {$S_1$};
      \draw (9.5,4) -- (13.5,4);
      \draw (9.5,7) -- (13.5,7);
      \draw [decorate, decoration=snake, green] (11.5,7.) -- (11.5,5.5);
      \draw [dotted, blue, thick] (11.5,5.5) -- (13.,4.);
      \draw [dotted, blue, thick] (11.5,5.5) -- (11.5,4.);
      \draw [dotted, blue, thick] (11.5,5.5) -- (10.,4.);
      \filldraw[black] (11.5,7.) circle (2pt) node[anchor=south] {$S_1$};
      \draw (-1.5,8) -- (2.5,8);
      \draw (-1.5,11) -- (2.5,11);
      \draw [dotted, blue, thick] (0.5,9.5) -- (0.5,8);
      \draw [dashed, red] (0.5,9.5) -- (0.5,11);
      \draw [dotted, blue, thick] (0.5,9.5) -- (-1.,11);
      \draw [dotted, blue, thick] (0.5,9.5) -- (2.,11);
      \filldraw[black] (0.5,11.) circle (2pt) node[anchor=south] {$S_1$};
      \draw (4.,8) -- (8.,8);
      \draw (4.,11) -- (8.,11);
      \draw [dotted, blue, thick] (6,9.5) -- (4.5,11);
      \draw [dotted, blue, thick] (6,11.) -- (6,8.);
      \draw [dotted, blue, thick] (6,9.5) -- (7.5,11);
      \filldraw[black] (6,11) circle (2pt) node[anchor=south] {$S_1$};
      \draw (9.5,8) -- (13.5,8);
      \draw (9.5,11) -- (13.5,11);
      \draw [dotted, blue, thick] (11.5,9.5) -- (11.5,8.);
      \draw [decorate, decoration=snake, green] (11.5,9.5) -- (11.5,11);
      \draw [dotted, blue, thick] (11.5,9.5) -- (13.,11);
      \draw [dotted, blue, thick] (11.5,9.5) -- (10.,11);
      \filldraw[black] (11.5,11) circle (2pt) node[anchor=south] {$S_1$};
    \end{tikzpicture} 
    \caption{Diagrams representing the leading corrections to the linear-in-spin potential
      from $\mathcal{C}^2$ interactions. Blue dotted lines represent $\phi$
      propagators, red dashed for $A$, green wavy for $\sigma$.
      All cross diagrams vanish, as well as the first
      (diagram with the field $A$ coupling to $S_1$) and third
      ($\sigma$ coupling to $S_1$) diagram on the second line.
      Diagrams obtained by exchanging particles 1 and 2 should be added.} 
    \label{fig:pot_spin_cc}
  \end{center}
\end{figure}

\noindent {\bf Spin}\\
The LO, linear-in-spin contribution to the potential can be derived
by computing the diagrams in fig.~\ref{fig:pot_spin_cc}, see
app.~\ref{ssapp:pot_spin_cc} for details.
They are all diagrams with 1 power of the spin
and 1 power of the velocity of any of the two particles (there are no
spin-dependent contributions for $\vec v_1=\vec v_2=0$).

Overall they give the following potential
\be
\label{eq:DV_spin_cc}
V_{\Lambda S_1}=-\pa{1+\kappa}\frac{4608}{\pa{\Lambda r}^6} \vec S_1\cdot\pa{\vec r\times v_1}\frac{G_N^3m_2\pa{m_1^2+m_ 2^2}}{r^5}\,,
\ee
leading to a modification of the equation of motion
\be
\label{eq:a_lambdaS}
\vec a_{\Lambda S}=-6912\frac{G_N^3\pa{1-2\eta}M^2}{\Lambda^6r^{11}}
\paq{2\vec S_{\cal C}\times \vec v+11\frac{\vec r\cdot \vec v}{r^2}\pa{\vec r\times \vec S_{\cal C}}
  +11\frac{\vec r}{r^2}\pa{\vec S_{\cal C}\cdot\pa{\vec r\times\vec v}}}\,,
\ee
which is independent of $\kappa$ and where the definition
$\vec S_{\cal C}\equiv \frac 1M\pa{\frac{m_2^2}{m_1}\vec S_1+\frac{m_1^2}{m_2}\vec S_2}$ has been used.

The spin equation of motion with the contribution of the $\mathcal{C}^2$
interaction are obtained following the same procedure as the one used in
sec.~\ref{sec:method} to derive the LO spin-orbit coupling,
to obtain the ${\cal C}^2$ correction
\be
\label{eq:ds_cc}
\left.\frac{{\rm d}\vec S_1}{{\rm d}t}\right|_\Lambda=
-\frac{6912}{(\Lambda r)^6}\frac{G_N^3\pa{m_1^2+m_2^2}}{r^5}
\pa{\frac{m_2}{m_1}}\vec L\times \vec S_1\,,
\ee
which is independent of the SSC chosen (i.e. $\kappa=0$ or $\kappa=1/2$).

\subsubsection{Radiation}
\label{ssec:rad_cc}
\begin{figure}
  \begin{center}
    \begin{tikzpicture}  
      \draw (-1.5,0) -- (2.5,0);
      \draw (-1.5,3) -- (2.5,3);
      \draw [dotted, blue, thick] (0.5,1.5) -- (0.5,3);
      \draw [dotted, blue, thick] (0.5,1.5) -- (2.,0);
      \draw [dotted, blue, thick] (0.5,1.5) -- (-1.,0.);
      \draw [decorate, decoration=snake, green] (.5,1.5) -- (2.5,1.5);
    \end{tikzpicture}
  \end{center}
  \caption{Leading ${\cal C}^2$ correction to the emission process for non-spinning sources. Blue dotted lines are longitudinal modes exchanged between the sources, green wavy line the radiative gravitational mode.}
  \label{fig:rad_cc}
\end{figure}
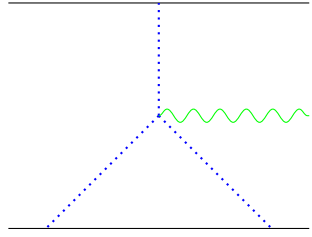

\noindent{\bf No spin}\\
The relevant diagram for the ${\cal C}^2$ contribution to the emission from
non-spinning sources is given in fig.~\ref{fig:rad_cc}, first computed in
eq.~(6.6) of \cite{Endlich:2017tqa}, and it is worth
\be
\label{eq:rad_cc}
L_{fig.\ref{fig:rad_cc}}=1344\frac{G_N^2m_1^2m_2}{\Lambda^6r^8}r^ir^j\frac{\ddot\sigma_{ij}}{2m_{Pl}}+1\leftrightarrow 2=1344\frac{G_N^2\eta M^3}{\pa{\Lambda r}^6}
n^in^j\frac{\ddot\sigma_{ij}}{2m_{Pl}}\,,
\ee
with $n^i\equiv \frac{r^i}r$,
where to simplify the calculations we work in the Transverse Traceless (TT)
gauge, which is valid only for the radiative field (i.e. on-shell and in vacuum),
enabling to write
$\ddot \sigma_{ij}=\nabla^2\sigma_{ij}=-2R_{0i0j}m_{Pl}$ and $\dot\sigma_{ij,k}-\dot\sigma_{ik,j}=2R_{0ijk}m_{Pl}$, see app.~\ref{ssapp:rad_cc} for details.

To express the leading radiative coupling in terms of the quadrupole,
one needs to combine eq.~(\ref{eq:rad_cc}) with the contribution
from the modified equation of motion (\ref{eq:eom_cc}), that enters the
game when expressing the radiative coupling (\ref{eq:rad_LO2}) in terms of
quadrupole derivatives, see eq.~(\ref{eq:rad_cnstnc}) for details.
In total one has
\be
\label{eq:rad_cc_tot}
L_{rad-\Lambda}=\paq{\frac{{\rm d}^2}{{\rm d}t^2}\pa{
    \frac{Q^{ij}}2+1344 \frac{G_N^2\eta M^3}{\pa{\Lambda r}^6}n^in^j}
-4608\frac{G^3_NM^4\eta(1-2\eta)}{\Lambda^6r^9}n^in^j}
\frac{\sigma_{ij}}{2m_{Pl}}\,.
\ee
Beside the shift in the quadrupole coupling given
by (\ref{eq:rad_cc}), we see that the non-GR structure of our effective theory,
via non-conservation of the energy-momentum tensor, introduces terms
in radiative interaction which do not have a multipolar structure, as the last
term in square bracket in eq.~(\ref{eq:rad_cc_tot}).
In terms of PN scaling, the $\Lambda$ dependent terms in (\ref{eq:rad_cc_tot})
are of the same order (remember that $\frac{{\rm d}}{{\rm d}t}\sim \frac vr$):
2PN$\times (\Lambda r)^{-6}$ with respect to the leading term (\ref{eq:rad_LO2}).

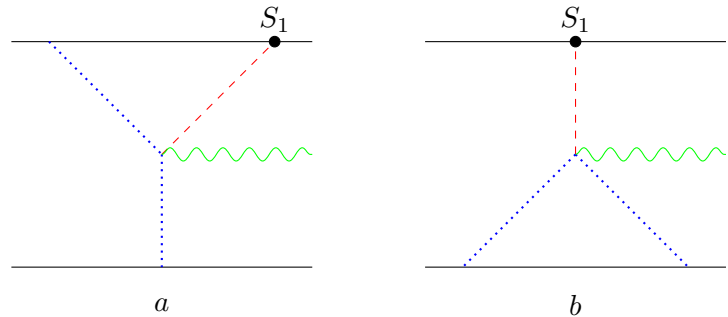
\begin{figure}[htb]
  \begin{center}
    \begin{tikzpicture}
      \draw (-1.5,0) -- (2.5,0);
      \draw (-1.5,3) -- (2.5,3);
      \draw [dotted, blue, thick] (-1.,3) -- (0.5,1.5);
      \draw [dashed, red] (0.5,1.5) -- (2.,3.);
      \draw [dotted, blue, thick] (.5,1.5) -- (.5,0.);
      \draw [decorate, decoration=snake, green] (.5,1.5) -- (2.5,1.5);
      \filldraw[black] (2.,3.) circle (2pt) node[anchor=south] {$S_1$};
      \coordinate [label=center:$a$] (m) at (0.5,-0.5);
      \draw (4.,0) -- (8.,0);
      \draw (4.,3.) -- (8.,3.);
      \draw [dashed, red] (6.,1.5) -- (6.,3);
      \draw [dotted, blue, thick] (6.,1.5) -- (4.5,0.);
      \draw [dotted, blue, thick] (6.,1.5) -- (7.5,0);
      \draw [decorate, decoration=snake, green] (6.,1.5) -- (8.,1.5);
      \filldraw[black] (6.,3.) circle (2pt) node[anchor=south] {$S_1$};
      \coordinate [label=center:$b$] (m) at (6.,-0.5);
    \end{tikzpicture}
    \caption{Diagrams representing the leading \emph{magnetic} contribution to 
      radiation emission process from $\mathcal{C}^2$ interaction, generated by
      bulk terms $\sim (\partial_m\partial_n\phi)^2\times \partial_i\partial_j A_k\, R^{0ijk}$.
      Blue dotted lines represent $\phi$ propagators, red dashed for $A_i$,
      green wavy for $\sigma_{ij}$.}
    \label{fig:rad_spin_cc_mag}
  \end{center}
\end{figure}

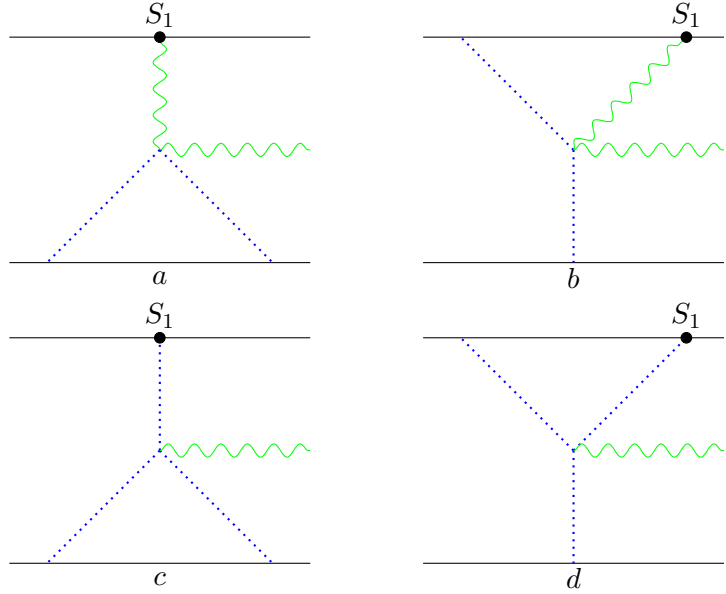
\begin{figure}[htb]
  \begin{center}
    \begin{tikzpicture}
      \draw (-1.5,0) -- (2.5,0);
      \draw (-1.5,3) -- (2.5,3);
      \draw [dotted, blue, thick] (0.5,1.5) -- (0.5,3.);
      \draw [dotted, blue, thick] (0.5,1.5) -- (-1.,0.);
      \draw [dotted, blue, thick] (0.5,1.5) -- (2.,0.);
      \draw [decorate, decoration=snake, green] (0.5,1.5) -- (2.5,1.5);
      \filldraw[black] (.5,3.) circle (2pt) node[anchor=south] {$S_1$};
      \coordinate [label=center:$c$] (m) at (0.5,-0.2);
      \draw (4.,0) -- (8.,0);
      \draw (4.,3) -- (8.,3);
      \draw [dotted, blue, thick] (6.,1.5) -- (4.5,3.);
      \draw [dotted, blue, thick] (6.,1.5) -- (7.5,3.);
      \draw [dotted, blue, thick] (6.,1.5) -- (6.,0.);
      \draw [decorate, decoration=snake, green] (6.,1.5) -- (8.,1.5);
      \filldraw[black] (7.5,3.) circle (2pt) node[anchor=south] {$S_1$};
      \coordinate [label=center:$d$] (m) at (6.,-0.2);
      \draw (-1.5,4) -- (2.5,4);
      \draw (-1.5,7) -- (2.5,7);
      \draw [decorate, decoration=snake, green] (0.5,5.5) -- (0.5,7.);
      \draw [dotted, blue, thick] (0.5,5.5) -- (-1.,4.);
      \draw [dotted, blue, thick] (0.5,5.5) -- (2.,4.);
      \draw [decorate, decoration=snake, green] (0.5,5.5) -- (2.5,5.5);
      \filldraw[black] (.5,7.) circle (2pt) node[anchor=south] {$S_1$};
      \coordinate [label=center:$a$] (m) at (0.5,3.8);
      \draw (4.,4) -- (8.,4);
      \draw (4.,7) -- (8.,7);
      \draw [dotted, blue, thick] (6.,5.5) -- (4.5,7.);
      \draw [decorate, decoration=snake, green] (6.,5.5) -- (7.5,7.);
      \draw [dotted, blue, thick] (6.,5.5) -- (6.,4.);
      \draw [decorate, decoration=snake, green] (6.,5.5) -- (8.,5.5);
      \filldraw[black] (7.5,7.) circle (2pt) node[anchor=south] {$S_1$};
      \coordinate [label=center:$b$] (m) at (6.,3.8);
    \end{tikzpicture}
    \caption{Diagrams representing the leading \emph{electric} contribution to
      radiation emission process from $\mathcal{C}^2$ interaction, generated by
      bulk terms $\sim (\partial_m\partial_n\phi)^2 \partial_i\partial_j\phi\pa{\ R_{0i0j}+\delta^{kl}R_{ikjl}}$.
      Blue dotted lines represent $\phi$ propagators, green wavy for $\sigma$.}
    \label{fig:rad_spin_cc_ele}
  \end{center}
\end{figure}

\noindent {\bf Spin}\\
The contribution to radiation emission processes introduced by the
quartic interaction due to the $\mathcal{C}^2$ term
can be distinguish in magnetic and electric ones.
Contrarily from the non-spinning case, the spinning LO emission process
is of \emph{magnetic} type, and described by the
processes reported in fig.~\ref{fig:rad_spin_cc_mag}, the ones of electric type are
reported in fig.~\ref{fig:rad_spin_cc_ele} and are $O(v)$ with respect to the former.

Indeed, as it can be seen from the scaling of the world-line (\ref{eq:lag_KK})
and bulk (\ref{eq:CC_KK}) couplings, the processes in fig.~\ref{fig:rad_spin_cc_mag}
give a contribution to the \emph{magnetic} quadrupole which is 2PN$\times (\Lambda r)^{-6}$ order
with respect to the leading GR magnetic quadrupole, whereas the electric
emission processes in fig.~\ref{fig:rad_spin_cc_ele} contribute at order
2.5PN$\times(\Lambda r)^{-6}$ with
respect to the leading electric quadrupole in GR, see fig.~\ref{fig:rad_quad_GR}:
as a result the contribution from ${\mathcal C}^2$ terms to the radiation flux $F$ in eq.~(\ref{eq:flux_GR})
from the magnetic and the electric quadrupole is of the same order, since for the spin-less part $J_{ij}\sim v\times Q_{ij}$.

The diagrams in fig.~\ref{fig:rad_spin_cc_mag} represents the linear-in-spin
${\mathcal C}^2$ contributions to the magnetic quadrupole, giving, see
app.~\ref{ssapp:rad_spin_cc} for details,
\be
\label{eq:rad_smag_cc}
      {\cal L}_{fig.\ref{fig:rad_spin_cc_mag}}=\frac{384}{\pa{\Lambda r}^6}\frac{G_N^2m_2}{r^2}
      \paq{\pa{4m_1+5m_2}\pa{S_1^ir^j+S_1^jr^i}
        -2\pa{5m_1+8m_2}n^in^j\pa{\vec r\cdot\vec S_1}}{\cal B}_{ij}\,,
\ee
which corrects the magnetic quadrupole according to
\be
\ba{rcl}
\left.J^{ij}\right|_{LO,\Lambda S_1}&=&\ds\dfrac 34\pa{S^i_1r^j+S_1^jr^i}_{TF}\times\paq{1+
  512\frac{G_N^2m_2}{\Lambda^6r^8}\pa{4m_1+5m_2}}\\
&&\ds-768\pa{n^in^j}_{TF}\pa{\vec r\cdot\vec S_1}\frac{G_N^2m_2}{\Lambda^6 r^8}\pa{5m_1+8m_2}
+1\leftrightarrow 2\,,
\ea
\ee
which is again a 2PN$\times (\Lambda r)^{-6}$ correction to the LO.

From fig.~\ref{fig:rad_spin_cc_ele} one gets the linear-in-spin ${\cal C}^2$
corrections to the electric quadrupole, see app.~\ref{ssapp:rad_spin_cc}
for detailed computations,
\be
\label{eq:rad_ele_cc_spin}
\ba{rcl}
\ds{\cal L}_{fig.\ref{fig:rad_spin_cc_ele}}&=&\ds\frac{192}{\pa{\Lambda r}^6}
   \frac{G_N^2m_2}{r^2}\Big\{
     \pa{5m_1+8m_2}n^i\pa{\hat n\cdot \vec v_1}\pa{\vec S_1\times \vec r}^j
     -\pa{m_1+4m_2}v_1^i\pa{\vec S_1\times \vec r}^j\\
     &&\ds\qquad\qquad-\pa{15+14\kappa}\pa{m_1+m_2}r^i\pa{\vec S_1\times\vec v_1}^j\\
     &&\ds\qquad\qquad -56\pa{1+\kappa}\pa{m_1+m_2}n^in^j\paq{\vec S_1\cdot\pa{\vec r\times \vec v_1}}
   \Big\}R_{0i0j}\,.
\ea
\ee
For the linear-in-spin part, like in the non-spinning case
eq.~(\ref{eq:rad_cc_tot}), the radiative coupling receives contributions from
the modification of the equation of motion eq.~(\ref{eq:a_lambdaS})
as per eq.~(\ref{eq:rad_cnstnc})
\be
\ba{rcl}
L_{rad-\Lambda S}&=&\ds-\frac{I^{ij}}2{\cal E}_{ij}+
6912\frac{G_N^3M^3\eta(1-2\eta)}{\Lambda^6 r^{11}}r^i\Big\{
2\pa{\vec S_{\cal C}\times \vec v}^j\\
&&\ds\quad +11\frac{\vec r\cdot \vec v}{r^2}\pa{\vec r\times \vec S_{\cal C}}^j
+11\frac{r^j}{r^2}\paq{\vec S_{\cal C} \cdot \pa{\vec r\times\vec v}}
+i\leftrightarrow j\Big\}\frac{\sigma_{ij}}{2m_{Pl}}\,,
\ea
\ee
and the electric quadrupole $I_{ij}$ is modified to:
\be
\ba{rcl}
\ds\left. I_{ij}\right|_{S_1}&=&\ds\frac{m_2^2}{M^2}\left\{
\pa{\vec v\times \vec S_1}^ir^j\paq{2\kappa+\frac 23+\frac{192}{\pa{\Lambda r}^6}\frac{G_N^2M^2}{r^2}\pa{15+14\kappa}}\right.\\
&&\ds\qquad+\pa{\vec r\times\vec S_1}^iv^j\pa{-\frac 43+\frac{192}{\pa{\Lambda r}^6}\frac{G_N^2M\pa{m_1+4m_2}}{r^2}}\\
&&\ds\qquad +\frac{192}{\pa{\Lambda r}^6}\frac{G_N^2M}{r^2}\left[
    -56M\pa{1+\kappa}n^in^j\pa{\vec S_1\cdot\pa{\vec r\times \vec v}}\right.\\
    &&\ds\left.\qquad\quad\left.
    +\pa{5m_1+8m_2}n^i\pa{\hat n \cdot \vec v}\pa{\vec S_1\times \vec r}^j\right]
+1\leftrightarrow 2\right\}_{STF}\,.
\ea
\ee
where ``$STF$'' stand for symmetric (in $i,j$) and trace-free part.

\subsection{$\mathcal{C\tilde C}$}

The $\mathcal{\tilde C}$ term violates both parity and time-reversal invariance,
the explicit expression of $\tilde{\mathcal C}$ and $\mathcal{C\tilde C}$,
limited to the terms with lower number of time derivatives of interest for this
work, are
\be
\label{eq:tildec}
\ba{rcl}
\ds\mathcal{\tilde C}&\simeq&\ds 4\frac{\epsilon_{jkl}}{m_{Pl}^2}
\Big[-4\pa{\partial_i\partial_j\phi}\pa{\partial_i\partial_kA_l}
  +\ddot\sigma_{ij}\pa{\partial_i\partial_kA_l}+
  \partial_k\partial_n\sigma_{ml}\partial_j\pa{\partial_nA_m-\partial_nA_m}\\
  &&\ds \qquad -4\pa{\partial_i\partial_j\phi}\partial_l\dot\sigma_{ik}+
\partial_n\dot\sigma_{jm}\partial_k\pa{\partial_m\sigma_{ln}-\partial_n\sigma_{lm}
  }\Big]\,,
\ea
\ee
\be
\label{eq:ctildec}
\ba{rcl}
\ds\mathcal{C\tilde C}&\simeq&\ds 32\frac{\epsilon_{jkl}}{m_{Pl}^4}
\left[\pa{\partial_m\partial_n\phi}\pa{\partial_m\partial_n\phi}+
  \frac 12\epsilon_{amn}\epsilon_{ars}\pa{\partial_c\partial_{m}A_n}
  \partial_{r}\dot\sigma_{cs}+\frac 14\pa{\partial_m\partial_nA_c}
  \partial_m\pa{\partial_nA_c-\partial_cA_n} \right.\\
  &&\ds\qquad\left.+\pa{\partial_m\dot A_n}\pa{\partial_m\partial_n\phi}
  -\pa{\partial_m\partial_nA_n}\pa{\partial_m\dot\phi}
  +\pa{\partial_c\partial_m\phi}\partial_c\partial_n
  \pa{\sigma_{mn}-\frac 12\delta_{mn}\sigma}\right]\\
&&\ds\times
\Big[-4\pa{\partial_i\partial_j\phi}\pa{\partial_i\partial_kA_l}
  +\ddot\sigma_{ij}\pa{\partial_i\partial_kA_l}+
  \partial_k\partial_n\sigma_{ml}\partial_j\pa{\partial_nA_m-\partial_mA_n}\\
  &&\ds\qquad -4\pa{\partial_i\partial_j\phi}\partial_l\dot\sigma_{ik}
  + \pa{\partial_l\dot\sigma_{ik}}\epsilon_{imn}\epsilon_{jrs}\partial_n\partial_s\sigma_{mr} +\ldots\Big]\,,
\ea
\ee

\subsubsection{Potential}
\label{ssec:pot_cctilde}

\noindent{\bf No spin}\\
The ${\cal C} \tilde{\cal C}$ contribution to the conservative potential
vanishes at all orders in the spin-less case, as this interaction involves a
Levi-Civita tensor which has to be contracted with three linearly independent
vectors to generate a scalar potential.
However for a non-spinning binary system all vectors available lie in the orbital plane, hence the $\mathcal{C\tilde C}$ contribution to the potential
vanishes.

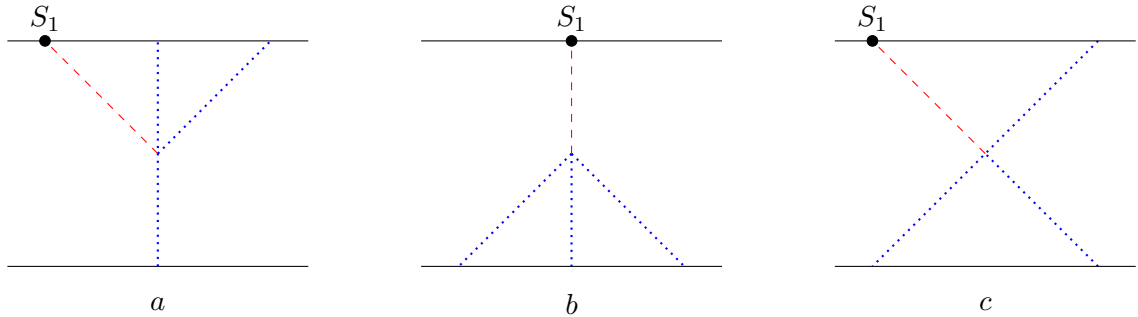
\begin{figure}[ht]
  \begin{center}
    \begin{tikzpicture}
      \draw (-1.5,8) -- (2.5,8);
      \draw (-1.5,11) -- (2.5,11);
      \draw [dashed, red] (.5,9.5) -- (-1.,11.);
      \draw [dotted, blue, thick] (.5,8.) -- (.5,11.);
      \draw [dotted, blue, thick] (.5,9.5) -- (2.,11.);
      \filldraw[black] (-1.,11.) circle (2pt) node[anchor=south] {$S_1$};
      \coordinate [label=center:$a$] (m) at (0.5,7.5);
      \draw (4.,8) -- (8.,8);
      \draw (4.,11) -- (8.,11);
      \draw [dotted, blue, thick] (6,9.5) -- (4.5,8.);
      \draw [dotted, blue, thick] (6,9.5) -- (6,8.);
      \draw [dotted, blue, thick] (6,9.5) -- (7.5,8.);
      \draw [dashed, red] (6,9.5) -- (6.,11.);
      \filldraw[black] (6.,11.) circle (2pt) node[anchor=south] {$S_1$};
      \coordinate [label=center:$b$] (m) at (6.,7.5);
      \draw (9.5,8) -- (13.5,8);
      \draw (9.5,11) -- (13.5,11);
      \draw [dashed, red] (11.5,9.5) -- (10.,11);
      \draw [dotted, blue, thick] (13.,8.) -- (11.5,9.5);
      \draw [dotted, blue, thick] (13.,11.) -- (10.,8.);
      \filldraw[black] (10.,11.) circle (2pt) node[anchor=south] {$S_1$};
      \coordinate [label=center:$c$] (m) at (11.5,7.5);
    \end{tikzpicture}
    \caption{Diagrams representing the  leading corrections to the spinning
      potential from $\mathcal{C\tilde C}$ interactions (blue dotted lines represent
      $\phi$ propagators, red dashed $A_i$ ones) at 2PN order with respect to leading GR behaviour. Only the first diagram is non-vanishing.}
    \label{fig:pot_spin_cctilde}
  \end{center}
\end{figure}

\noindent{\bf Spin}\\
In the spinning case  diagrams in fig.~\ref{fig:pot_spin_cctilde} give, see
app.~\ref{ssapp:pot_spin_cctilde} for details,
\be
\label{eq:pot_spin_cctilde}
V_{\Lambda_-S_1}=-4608\frac{G_N^3m_1^2m_2}
       {r^5\pa{\Lambda_- r}^6}\vec r\cdot \vec S_1\,,
\ee
which is a $1.5$PN$\times (\Lambda_- r)^{-6}$ correction with respect to the
leading spin term in the potential eq.~(\ref{eq:so_l}), the half-integer
relative order due to parity violation nature of the term, modifying the spin
equation of motion per
\be
\label{eq:ds_cctilde}
\left.\frac{{\rm d}\vec{S}_1}{{\rm d}t}\right|_{\Lambda_-}=-4608\frac{G_N^3m_1^2m_2}{r^5\pa{\Lambda_- r}^6}\, \vec r\times \vec S_1\,,
\ee
which induces a perturbation into spin equation of motion
with characteristic frequency given by the orbital frequency.
This is a distinctive effect generated by the parity violating Lagrangian which
is qualitative different from other modifications.

\subsubsection{Emission}
\label{ssec:rad_cctilde}

\begin{figure}[htb]
  \begin{center}
    \begin{tikzpicture}
      \draw (-1.5,0) -- (2.5,0);
      \draw (-1.5,3) -- (2.5,3);
      \draw [dotted, blue, thick] (0.5,1.5) -- (0.5,3.);
      \draw [dotted, blue, thick] (0.5,1.5) -- (-1.,0.);
      \draw [dotted, blue, thick] (0.5,1.5) -- (2.,0.);
      \draw [decorate, decoration=snake, green] (0.5,1.5) -- (2.5,1.5);
      \coordinate [label=center:$a$] (m) at (0.5,-0.5);
      \draw (4.,0) -- (8.,0);
      \draw (4.,3) -- (8.,3);
      \draw [dotted, blue, thick] (6.,1.5) -- (6.,3.);
      \draw [dotted, blue, thick] (6.,1.5) -- (4.5,0.);
      \draw [dashed, red] (6.,1.5) -- (7.5,0.);
      \draw [decorate, decoration=snake, green] (6.,1.5) -- (8.,1.5);
      \coordinate [label=center:$b$] (m) at (6.,-0.5);
      \draw (9.5,0) -- (13.5,0);
      \draw (9.5,3) -- (13.5,3);
      \draw [dashed, red] (11.5,1.5) -- (11.5,3.);
      \draw [dotted, blue, thick] (11.5,1.5) -- (10.,0.);
      \draw [dotted, blue, thick] (11.5,1.5) -- (13.,0.);
      \draw [decorate, decoration=snake, green] (11.5,1.5) -- (13.5,1.5);
      \coordinate [label=center:$c$] (m) at (11.5,-0.5);
      \end{tikzpicture}
  \end{center}
  \caption{Diagram giving the LO effect of $\mathcal{C\tilde C}$ on
    the radiation emission for the non-spinning case. At leading order
    diagram $a$ contributes to the magnetic coupling, $b$ and $c$
    to the electric part.}
  \label{fig:rad_cctilde}
\end{figure}
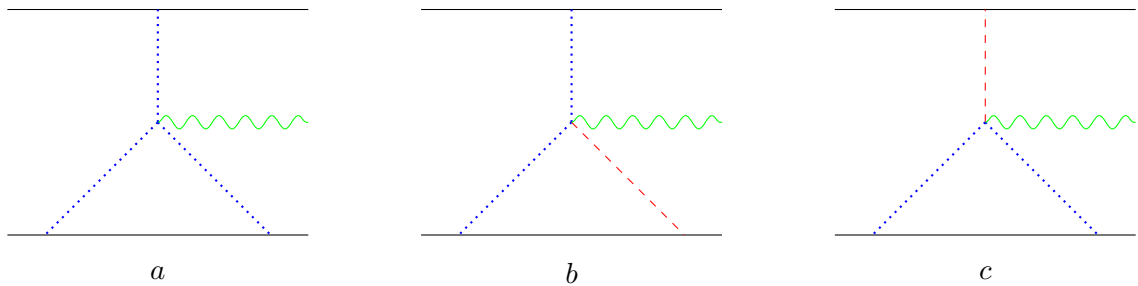

\noindent{\bf No spin}\\
The parity violating term under consideration couples at lowest order to the
magnetic part of the Riemann via an electric quadrupole
in fig.~\ref{fig:rad_cctilde}a giving a coupling to radiative field, see app.~\ref{app:rad_cctilde} for details:
\be
\label{eq:rad_mag_cctilde}
L_{fig.~\ref{fig:rad_cctilde}a}=-\frac{1344}{\pa{\Lambda_- r}^6}
\frac{G_N^2m_1m_2^2}{r^2}r^i r^j\frac 14\epsilon_{ikl}
\pa{\dot\sigma_{jk,l}-\dot\sigma_{jl,k}}+1\leftrightarrow 2\,.
\ee

Note that since at leading order ${\mathcal B}_{ij}$ couples to
$J^{ij}=\frac 12 \mu v^kx^l\epsilon_{mkl}\pa{\delta^{mi}x^j+\delta^{mj}x^i}$,
eq.~(\ref{eq:rad_mag_cctilde}) represents a 1.5PN electric correction to the magnetic quadrupole. In this case there is no contribution from
modified equations of motion, see eq.~(\ref{eq:rad_cnstnc}), as the interaction
$\cal{C}\tilde{\cal C}$ does not
affect the conservative dynamics in the non-spinning sector, giving
\be
L_{rad-\Lambda_-J}=-\paq{\frac 13\pa{r^iL^j+r^jL^i}+
  1344\frac{\eta G_N^2M^3}{r^2\pa{\Lambda_- r}^6}r^ir^j}{\cal B}_{ij}\,,
\ee
which agrees with eq.~(6.21) of \cite{Endlich:2017tqa}.

The correction to the emission process coupling to the electric part of the
Riemann is of 2.5 PN order with respect to the to leading
radiative electric coupling, and as at leading order
$J_{ij}\sim v\times I_{ij}$ the electric and magnetic terms involving
the ${\cal C}\tilde{\cal C}$ interaction contribute to the flux
(\ref{eq:flux_GR}) at the same order.

The contributions to the electric emission processes in
fig.~\ref{fig:rad_cctilde}b,c give
\be
\label{eq:rad_ele_cctilde}
L_{fig.~\ref{fig:rad_cctilde}bc}=1536\frac{G_N^2m_1m_2^2}{\Lambda_-^6r^8}r^i
\paq{\vec r\times \pa{\vec v_1+\frac 34\vec v_2}}^j
\frac{\ddot \sigma_{ij}}{2m_{Pl}}+1\leftrightarrow 2\,,
\ee
which together with the leading order gives an electric radiative coupling
\be
L_{rad-\Lambda_-I}=-\mu\pa{\frac 12r^ir^j+
  768 \frac{G_N^2M^2\pa{2-7\eta}}{r^2\pa{\Lambda_-r}^6}r^iL^j}_{STF}{\cal E}_{ij}\,.
\ee

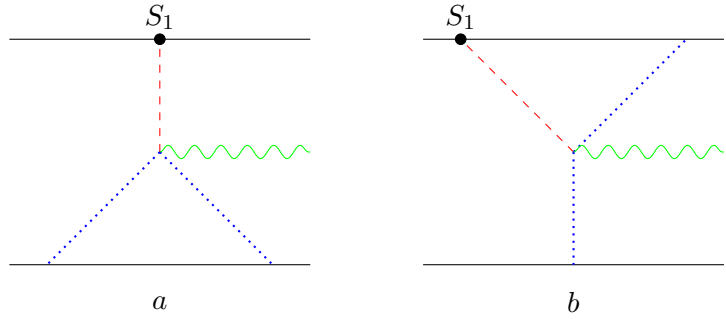
\begin{figure}[htb]
  \begin{center}
    \begin{tikzpicture}
      \draw (-1.5,0) -- (2.5,0);
      \draw (-1.5,3) -- (2.5,3);
      \draw [dashed, red] (0.5,1.5) -- (0.5,3.);
      \draw [dotted, blue, thick] (0.5,1.5) -- (-1.,0.);
      \draw [dotted, blue, thick] (0.5,1.5) -- (2.,0.);
      \draw [decorate, decoration=snake, green] (0.5,1.5) -- (2.5,1.5);
      \filldraw[black] (.5,3.) circle (2pt) node[anchor=south] {$S_1$};
      \coordinate [label=center:$a$] (m) at (.5,-0.5);
      \draw (4.,0) -- (8.,0);
      \draw (4.,3) -- (8.,3);
      \draw [dotted, blue, thick] (6.,1.5) -- (6.,0.);
      \draw [dashed, red] (6.,1.5) -- (4.5,3.);
      \draw [dotted, blue, thick] (6.,1.5) -- (7.5,3.);
      \draw [decorate, decoration=snake, green] (6.,1.5) -- (8.,1.5);
      \filldraw[black] (4.5,3.) circle (2pt) node[anchor=south] {$S_1$};
      \coordinate [label=center:$b$] (m) at (6.,-0.5);
    \end{tikzpicture}
  \end{center}
  \caption{Diagram giving the LO effect of $\mathcal{C\tilde C}$ on
    the radiation emission for the electric spinning case. They contribute also
    to the magnetic coupling.}
  \label{fig:rad_spin_cctilde_ele}
\end{figure}

\noindent {\bf Spin}\\
The diagrams giving the leading $\mathcal{C\tilde C}$ electric contribution to radiation
in the spinning case are reported in fig.~\ref{fig:rad_spin_cctilde_ele}, see app.~\ref{ssapp:rad_spin_cctilde} for details.
The electric type radiation process gives the following Lagrangian coupling
\be
\label{eq:rad_ele_cct_spin_ele}
\ba{rcl}
L_{fig.~\ref{fig:rad_spin_cctilde_ele}{\cal E}}&=&\ds 384\frac{G_N^2m_2}{r^2\pa{\Lambda_- r}^6}
\paq{\pa{7m_1+15m_2}r^iS_1^j-4\pa{7m_1+6m_2}n^in^j\pa{\vec r\cdot \vec S_1}}
   {\mathcal E}_{ij}\,,
\ea
\ee
representing a 1.5PN$\times \pa{\Lambda_- r}^{-6}$ magnetic
type correction to the spin-dependent electric quadrupole which
can be read from eq.~(\ref{eq:rad_IS}).

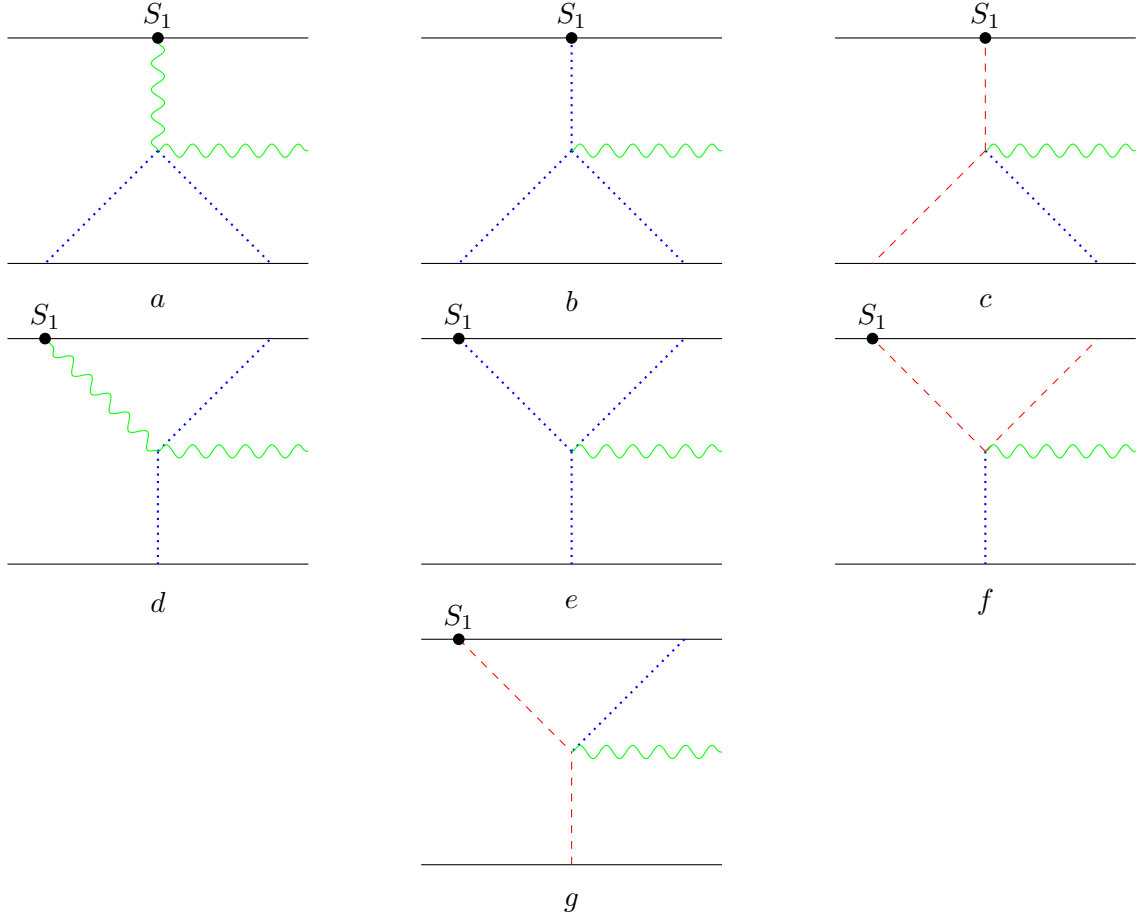
\begin{figure}[htb]
  \begin{center}
    \begin{tikzpicture}
      \draw (4.,4) -- (8.,4);
      \draw (4.,7) -- (8.,7);
      \draw [decorate, decoration=snake, green] (6.,5.5) -- (6.,7.);
      \draw [dotted, blue, thick] (6.,5.5) -- (4.5,4.);
      \draw [dotted, blue, thick] (6.,5.5) -- (7.5,4.);
      \draw [decorate, decoration=snake, green] (6.,5.5) -- (8.,5.5);
      \filldraw[black] (6.,7.) circle (2pt) node[anchor=south] {$S_1$};
      \coordinate [label=center:$a$] (m) at (6.,3.5);
      \draw (9.5,4) -- (13.5,4);
      \draw (9.5,7) -- (13.5,7);
      \draw [dotted, blue, thick] (11.5,5.5) -- (11.5,7.);
      \draw [dotted, blue, thick] (11.5,5.5) -- (10.,4.);
      \draw [dotted, blue, thick] (11.5,5.5) -- (13.,4.);
      \draw [decorate, decoration=snake, green] (11.5,5.5) -- (13.5,5.5);
      \filldraw[black] (11.5,7.) circle (2pt) node[anchor=south] {$S_1$};
      \coordinate [label=center:$b$] (m) at (11.5,3.5);
      \draw (15,4) -- (19,4);
      \draw (15,7) -- (19,7);
      \draw [dashed, red] (17.,5.5) -- (17.,7.);
      \draw [dashed, red] (17.,5.5) -- (15.5,4.);
      \draw [dotted, blue, thick] (17.,5.5) -- (18.5,4.);
      \draw [decorate, decoration=snake, green] (17.,5.5) -- (19.,5.5);
      \filldraw[black] (17.,7.) circle (2pt) node[anchor=south] {$S_1$};
      \coordinate [label=center:$c$] (m) at (17.,3.5);
      \draw (4.,0) -- (8.,0);
      \draw (4.,3) -- (8.,3);
      \draw [dotted, blue, thick] (6.,1.5) -- (6.,0.);
      \draw [decorate, decoration=snake, green] (6.,1.5) -- (4.5,3.);
      \draw [dotted, blue, thick] (6.,1.5) -- (7.5,3.);
      \draw [decorate, decoration=snake, green] (6.,1.5) -- (8.,1.5);
      \filldraw[black] (4.5,3.) circle (2pt) node[anchor=south] {$S_1$};
      \coordinate [label=center:$d$] (m) at (6.,-0.5);
      \draw (9.5,0) -- (13.5,0);
      \draw (9.5,3) -- (13.5,3);
      \draw [dotted, blue, thick] (11.5,1.5) -- (11.5,0.);
      \draw [dotted, blue, thick] (11.5,1.5) -- (10.,3.);
      \draw [dotted, blue, thick] (11.5,1.5) -- (13.,3.);
      \draw [decorate, decoration=snake, green] (11.5,1.5) -- (13.5,1.5);
      \filldraw[black] (10.,3.) circle (2pt) node[anchor=south] {$S_1$};
      \coordinate [label=center:$e$] (m) at (11.5,-0.5);
      \draw (15.,0) -- (19.,0);
      \draw (15.,3) -- (19.,3);
      \draw [red, dashed] (17.,1.5) -- (15.5,3.);
      \draw [red, dashed] (17.,1.5) -- (18.5,3.);
      \draw [dotted, blue, thick] (17.,1.5) -- (17.,0.);
      \draw [decorate, decoration=snake, green] (17.,1.5) -- (19.,1.5);
      \filldraw[black] (15.5,3.) circle (2pt) node[anchor=south] {$S_1$};
      \coordinate [label=center:$f$] (m) at (17.,-0.5);
      \draw (9.5,-4) -- (13.5,-4);
      \draw (9.5,-1) -- (13.5,-1);
      \draw [red, dashed] (10.,-1.) -- (11.5,-2.5);
      \draw [dotted, blue, thick] (13.,-1.) -- (11.5,-2.5);
      \draw [red, dashed] (11.5,-4.) -- (11.5,-2.5);
      \draw [decorate, decoration=snake, green] (11.5,-2.5) -- (13.5,-2.5);
      \filldraw[black] (10,-1.) circle (2pt) node[anchor=south] {$S_1$};
      \coordinate [label=center:$g$] (m) at (11.5,-4.5);
    \end{tikzpicture}
  \end{center}
  \caption{Additional diagrams contributing via $\mathcal{C\tilde C}$ interaction to the LO radiation emission for the magnetic spinning case.}
  \label{fig:rad_spin_cctilde_mag}
\end{figure}

The linear-in-spin magnetic type processes give 2.5PN $\times (\Lambda_-r)^{-6}$
corrections with respect to the leading order magnetic emission terms, which
then induce sub-leading effects in the emission amplitude with respect to
(\ref{eq:rad_ele_cct_spin_ele}).
However the linear-in-spin term of the electric quadrupole is
subleading by a factor of $v$ with respect to the analog magnetic quadrupole
term,
hence electric and magnetic ${\cal C}\tilde{\cal C}$ linear-in-spin terms
give same order contributions to the flux.
In addition to the diagrams in
fig.~\ref{fig:rad_spin_cctilde_ele}, in the magnetic case one has also the
diagrams of fig.~\ref{fig:rad_spin_cctilde_mag}. Overall the magnetic radiative
coupling induced by the ${\cal C}\tilde{\cal C}$ term is,
see app.~\ref{ssapp:rad_spin_cctilde} for details,
\be
\label{eq:rad_ele_cct_spin_mag}
\ba{rcl}
L_{fig.~(\ref{fig:rad_spin_cctilde_ele}+\ref{fig:rad_spin_cctilde_mag}){\cal B}}&=&\ds 24\frac{G_N^2m_2}{r^2\pa{\Lambda_- r}^6}
\left\{\pa{\vec r\times\vec S_1}^i\paq{2v_1^j\pa{41 m_2+32m_1}-16v_2^j\pa{7m_2+6m_1}}+\right.\\
&&\ds\pa{\vec r\times\vec S_1}^i\frac{r^jr^m}{r^2}
\paq{-2v_{1m}\pa{130 m_2+61m_1}+64v_{2m}\pa{8 m_1+9 m_2}}\\
&&\ds
-n^in^jS_1^m\paq{\pa{\vec r\times\vec v_1}_m\pa{346m_2+605m_1}+\pa{\vec r\times\vec v_2}_m32\pa{20 m_2+23m_1}}\\
&&\ds\left.+r^i\paq{2\pa{\vec S_1\times\vec v_1}^j\pa{27m_2+77m_1}
  +224\pa{\vec S_1\times\vec v_2}^j\pa{m_2+2m_1}}\right\}{\mathcal B}_{ij}\,,
\ea
\ee
where $\kappa=1/2$ has been used.

\subsection{$\mathcal{\tilde C}^2$}

While the $\tilde{\mathcal C}$ term violates parity, the $\tilde{\mathcal C}^2$
does not and its expression is given by the square of (\ref{eq:tildec}).

\subsubsection{Potential}
\label{ssec:pot_ctct}

\noindent{\bf No spin}\\
The potential interaction mediated by $\tilde{\mathcal C}^2$ vanish in the
static limit and it is $v^2$ with respect to the one mediated by the
$\mathcal C^2$, i.e. a 3PN correction to the Newtonian potential and it will
not be computed here.

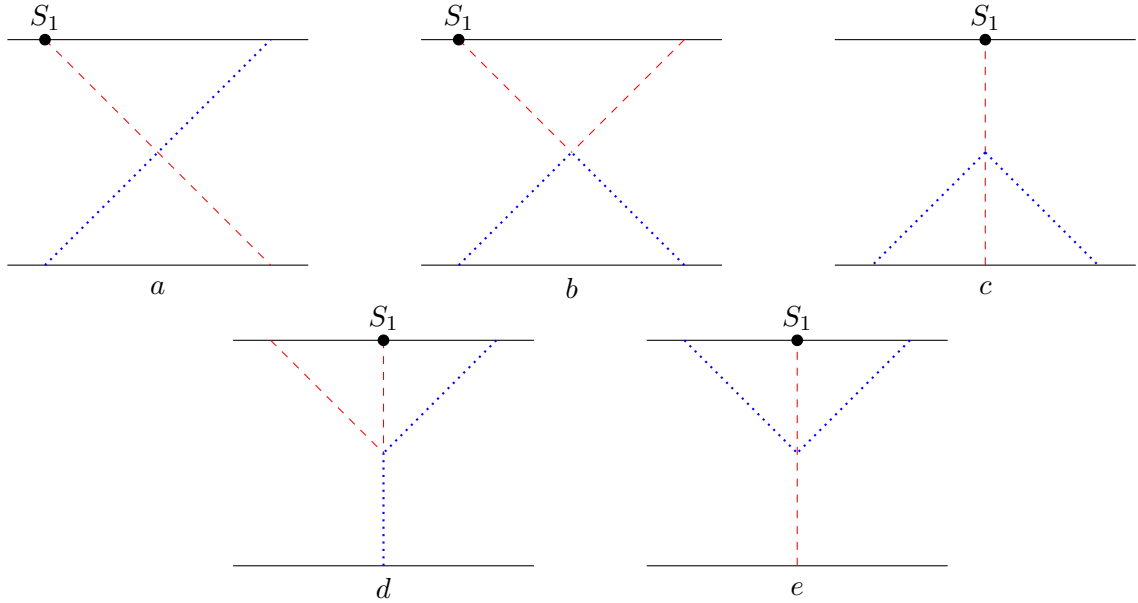
\begin{figure}[ht]
  \begin{center}
    \begin{tikzpicture}
      \draw (-1.5,0) -- (2.5,0);
      \draw (-1.5,3) -- (2.5,3);
      \draw [dashed, red] (-1.,3) -- (2.,0);
      \draw [dotted, blue, thick] (-1.0,0) -- (2.,3);
      \filldraw[black] (-1.,3.) circle (2pt) node[anchor=south] {$S_1$};
      \coordinate [label=center:$a$] (m) at (0.5,-0.3);
      \draw (4.,0) -- (8.,0);
      \draw (4.,3) -- (8.,3);
      \draw [dashed, red] (4.5,3) -- (6.,1.5);
      \draw [dashed, red] (7.5,3) -- (6.,1.5);
      \draw [dotted, blue, thick] (4.5,0) -- (6.,1.5);
      \draw [dotted, blue, thick] (7.5,0) -- (6.,1.5);
      \filldraw[black] (4.5,3.) circle (2pt) node[anchor=south] {$S_1$};
      \coordinate [label=center:$b$] (m) at (6.,-0.3);
      \draw (9.5,0) -- (13.5,0);
      \draw (9.5,3) -- (13.5,3);
      \draw [dashed, red] (11.5,0) -- (11.5,3);
      \draw [dotted, blue, thick] (11.5,1.5) -- (13.,0.);
      \draw [dotted, blue, thick] (11.5,1.5) -- (10.,0.);
      \filldraw[black] (11.5,3.) circle (2pt) node[anchor=south] {$S_1$};
      \coordinate [label=center:$c$] (m) at (11.5,-0.3);
      \draw (1.5,-1) -- (5.5,-1);
      \draw (1.5,-4) -- (5.5,-4);
      \draw [dashed, red] (2.,-1) -- (3.5,-2.5);
      \draw [dashed, red] (3.5,-1) -- (3.5,-2.5);
      \draw [dotted, blue, thick] (3.5,-4) -- (3.5,-2.5);
      \draw [dotted, blue, thick] (5.,-1) -- (3.5,-2.5);
      \filldraw[black] (3.5,-1.) circle (2pt) node[anchor=south] {$S_1$};
      \coordinate [label=center:$d$] (m) at (3.5,-4.3);
      \draw (7.,-1) -- (11.,-1);
      \draw (7.,-4) -- (11.,-4);
      \draw [dashed, red] (9.,-4) -- (9.,-1);
      \draw [dotted, blue, thick] (7.5,-1.) -- (9.,-2.5);
      \draw [dotted, blue, thick] (10.5,-1.) -- (9.,-2.5);
      \filldraw[black] (9.,-1.) circle (2pt) node[anchor=south] {$S_1$};
      \coordinate [label=center:$e$] (m) at (9.,-4.3);
    \end{tikzpicture} 
    \caption{Diagrams representing the correction to the spinning potential
      mediated by the $\tilde{\mathcal C}^2$ bulk interaction.
      Diagrams must be supplemented with their mirror images under $1\leftrightarrow 2$. The three diagrams in the first line vanish.} 
    \label{fig:pot_spin_ctct}
  \end{center}
\end{figure}

\noindent {\bf Spin}\\
The LO potential mediated by the $\tilde{\mathcal C}^2$ interaction in the spinning case is given by the processes in fig.~\ref{fig:pot_spin_ctct}
with the total result, see app.~\ref{ssapp:pot_spin_ctct} for details,
\be
\label{eq:pot_ctct_spin}
L_{pot-\tilde\Lambda S_1}=\frac{55296}{11}\frac{G_N^3m_1^2m_2}{r^5(\tilde\Lambda r)^6}
\paq{\vec r\times\pa{\vec v_1-\vec v_2}}\cdot \vec S_1\,,
\ee
which agrees with eq.~(5.18) of \cite{Endlich:2017tqa}.

This contribution is analogous to the one of the ${\mathcal C}^2$ term and it
introduces a correction to the spin equation of motion
\be
\label{eq:ds_ctct}
\left.\frac{{\rm d}\vec S_1}{{\rm d}t}\right|_{\tilde \Lambda}=\frac{55296}{11}
\frac{G_N^3Mm_2}{r^5(\tilde\Lambda r)^6}\vec L\times \vec S_1\,,
\ee
analogous to the ${\cal C}^2$ case eq.~(\ref{eq:ds_cc}).

\subsubsection{Emission}
\label{ssec:rad_ctct}

\noindent{\bf No spin}\\
The radiative interaction due to the $\tilde{\mathcal C}^2$ term
are $v^2$ with respect to the one mediated by the $\mathcal C^2$, i.e. a 3PN
correction to the quadrupole formula and it will not be computed here.

\begin{figure}[ht]
  \begin{center}
    \begin{tikzpicture}
      \draw (-1.5,0) -- (2.5,0);
      \draw (-1.5,3) -- (2.5,3);
      \draw [dashed, red] (0.5,3) -- (0.5,1.5);
      \draw [dashed, red] (-1.,0) -- (.5,1.5);
      \draw [dotted, blue, thick] (2.,0) -- (.5,1.5);
      \filldraw[black] (0.5,3.) circle (2pt) node[anchor=south] {$S_1$};
      \draw [decorate, decoration=snake, green] (.5,1.5) -- (2.5,1.5);
      \coordinate [label=center:$a$] (m) at (.5,-0.5);
      \draw (4.,0) -- (8.,0);
      \draw (4.,3) -- (8.,3);
      \draw [dashed, red] (4.5,3) -- (6.,1.5);
      \draw [dashed, red] (7.5,3) -- (6.,1.5);
      \draw [dotted, blue, thick] (6.,0) -- (6.,1.5);
      \filldraw[black] (4.5,3.) circle (2pt) node[anchor=south] {$S_1$};
      \coordinate [label=center:$b$] (m) at (6.,-0.5);
      \draw [decorate, decoration=snake, green] (6.,1.5) -- (8.,1.5);
      \draw (9.5,0) -- (13.5,0);
      \draw (9.5,3) -- (13.5,3);
      \draw [dashed, red] (10.,3) -- (11.5,1.5);
      \draw [dotted, blue, thick] (13.,3) -- (11.5,1.5);
      \draw [dashed, red] (11.5,0) -- (11.5,1.5);
      \filldraw[black] (10.,3.) circle (2pt) node[anchor=south] {$S_1$};
      \draw [decorate, decoration=snake, green] (11.5,1.5) -- (13.5,1.5);
      \coordinate [label=center:$c$] (m) at (11.5,-0.5);
    \end{tikzpicture} 
    \caption{Diagrams representing the correction to the linear-in-spin
      \emph{electric} radiation emission mediated by the $\tilde{\mathcal C}^2$ bulk interaction.
      Diagrams must be supplemented with their mirror images under $1\leftrightarrow 2$.}
    \label{fig:rad_spin_ctct_ele}
  \end{center}
\end{figure}
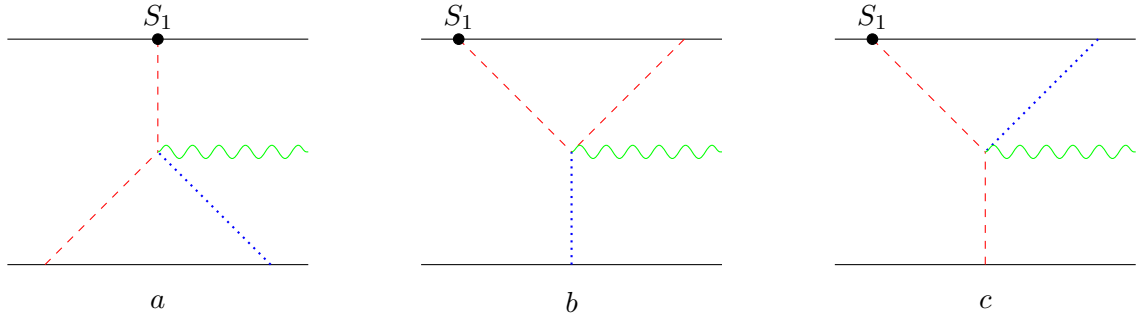

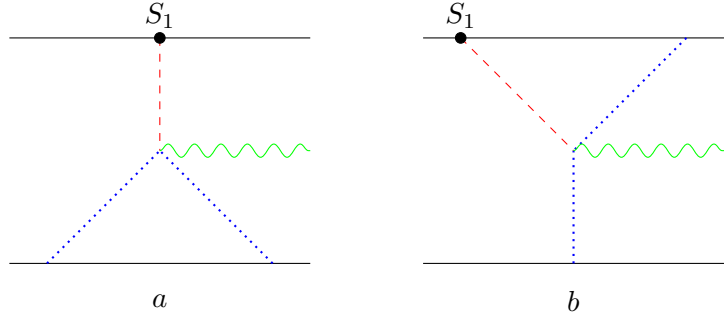
\begin{figure}[ht]
  \begin{center}
    \begin{tikzpicture}
      \draw (-1.5,0) -- (2.5,0);
      \draw (-1.5,3) -- (2.5,3);
      \draw [dashed, red] (0.5,3) -- (0.5,1.5);
      \draw [dotted, blue, thick] (-1.,0) -- (.5,1.5);
      \draw [dotted, blue, thick] (2.,0) -- (.5,1.5);
      \filldraw[black] (0.5,3.) circle (2pt) node[anchor=south] {$S_1$};
      \draw [decorate, decoration=snake, green] (.5,1.5) -- (2.5,1.5);
      \coordinate [label=center:$a$] (m) at (.5,-0.5);
      \draw (4.,0) -- (8.,0);
      \draw (4.,3) -- (8.,3);
      \draw [dashed, red] (4.5,3) -- (6.,1.5);
      \draw [dotted, blue, thick] (7.5,3) -- (6.,1.5);
      \draw [dotted, blue, thick] (6.,0) -- (6.,1.5);
      \filldraw[black] (4.5,3.) circle (2pt) node[anchor=south] {$S_1$};
      \draw [decorate, decoration=snake, green] (6.,1.5) -- (8.,1.5);
      \coordinate [label=center:$b$] (m) at (6.,-0.5);
    \end{tikzpicture} 
    \caption{Diagrams representing the correction to the linear-in-spin
      \emph{magnetic} radiation emission mediated by the $\tilde{\mathcal C}^2$ bulk interaction.
      Diagrams must be supplemented with their mirror images under $1\leftrightarrow 2$.}
    \label{fig:rad_spin_ctct_mag}
  \end{center}
\end{figure}

\noindent {\bf Spin}\\
The radiative, linear-in-spin processes contribute to both the electric and
magnetic coupling, each being 2PN order $\times (\tilde\Lambda r)^{-6}$ the
respective leading order one (\ref{eq:rad_IS}) and (\ref{eq:rad_JS}).
From fig.~\ref{fig:rad_spin_ctct_ele}, see app.~\ref{ssapp:rad_spin_ctct} for
details, one gets the Lagrangian contributions to the electric-type radiative
couplings
\be
\label{eq:rad_ele_ctct_spin}
\ba{rcl}
\ds L_{fig.\ref{fig:rad_spin_ctct_ele}}&=&\ds \frac{384}{(\tilde\Lambda r)^6}
\frac{G_N^2\mu}{r^2}
\left\{-2\pa{94 m_1+43m_2}n^in^j\paq{\vec S_1\cdot\pa{\vec r\times\vec v}}\right.\\
&&\ds\qquad+4\pa{42m_1+19m_2}n^i\pa{\hat n\cdot \vec v}\pa{\vec S_1\times\vec r}^j\\
&&\ds\qquad\left.-4\pa{32m_1+7m_2}r^i\pa{\vec S_1\times \vec v}^j-2
\pa{12m_1+13m_2}v^i\pa{\vec S_1\times \vec r}^j\right\} {\mathcal E}_{ij}\,.
\ea
\ee
For the radiative processes with magnetic parity one has
\be
\label{eq:rad_mag_ctct_spin}
\ba{rcl}
\ds L_{fig.\ref{fig:rad_spin_ctct_mag}}&=&\ds
768\frac{G_N^2 m_2}{r^2(\tilde\Lambda r)^6}
\paq{-\pa{3m_1+5m_2}S_1^ir^j+2\pa{9m_1+4m_2}n^in^j\pa{\vec S_1\cdot \vec r}}
   {\mathcal B}_{ij}\,.
\ea
\ee

\section{Discussion}
\label{sec:conclusions}

Following the parameterisation of short-distance gravity modification introduced
in \cite{Endlich:2017tqa}, we applied effective field theory methods of
non-relativistic General
Relativity \cite{Goldberger:2004jt} to preliminarly re-derive in this modified gravity model
the energy and luminosity functions of a compact two-body system for non-spinning systems,
which are the ingredients to compute the
phase of gravitational waves observed routinely since the year 2015 by the
gravitational wave observatories LIGO and Virgo.
Moreover, applying standard treatment of spin degrees of freedom, we have extended these
results to the linear-in-spin case, with the new result of deriving
linear-in-spin potentials, radiative terms and 
spin precessing equations from the same modifications to General
Relativity, hence adding a new test of General Relativity.
Another new finding of the present work is given by corrections to the quadrupole formula that must be included into emission processes and luminosity functions
when dealing with General Relativity modifications altering the equation
of motion of the sources, as explained in sec.~\ref{ssec:rad_cc}.

However, irrespectively of physical quantities being considered (2-body potential, luminosity,
gravitational wave phasing, spin equation of motion),
the corrections due to the quartic interactions introduced by the extra terms
of the type Riemann to the fourth power
can be classified as 2PN order in the post-Newtonian approximation
($n$PN meaning $v^{2n}$ corrections with respect to leading order)
times a factor $(\Lambda r)^{-6}$
(being $\Lambda$ the generic short-distance scale introduced by the quartic
curvature terms and $r$ the typical source size)
with respect to leading order
for parity (and time-reversal) preserving interactions,
and 1.5 PN (again times $(\Lambda_- r)^{-6}$)
for parity-violating corrections for both spinning and non-spinning systems.

For spins however, we have to consider that in General Relativity the leading
spin-dependent term in the gravitational wave phasing formula appears at 1.5PN
order, and the quartic curvature
interactions considered in this work give a 1.5PN
(as it happens in the parity-violating case, or 2PN in the parity preserving one)
correction with respect to the leading spin order,
hence 3PN (3.5PN) order in absolute terms,
times the usual $(\Lambda_- r)^{-6}$ or $(\Lambda r)^{-6}$ factor.

Current limits set by LIGO/Virgo detections on 2PN and 3PN phasing terms
\cite{LIGOScientific:2019fpa,Abbott:2020jks} require them to be within $\sim$
10\% of the value predicted by General Relativity,
giving a mild bound on $\Lambda\gtrsim 1/r$, with typical $r$ for
LIGO/Virgo detections of the order of $\sim 10-100$ km, which translated
to an energy scale is $\Lambda \gtrsim 10^{-12,-13}$ eV, see \cite{Sennett:2019bpc}
for a Bayesian model selection approach used on two real gravitational wave events
and \cite{Cardoso:2018ptl} for the geometry of single black hole solutions.
Note that while the bound that can be set with this type of analysis are
rather loose, we have provided in this work a first \emph{ab initio} computation of spin
effects in two body potential, emission and spin precession equations.

Other tests are possible via \emph{geodetic precession}, i.e.
relativistic spin precession which in General Relativity is caused at
leading order by the spin-orbit coupling, that has been observed in the
spin of double pulsar PSR J0737-3039
\cite{Breton:2008xy,Perera:2010sp,Perrodin:2017bxr} and it is
currently in agreement with the value predicted by General Relativity at $13\%$ level.

For the gravitational model under study here, parity-preserving
terms give rise to corrections of the order
$v^4\times(\Lambda r)^{-6}$ to the spin precessing equations, as reported in eqs.~(\ref{eq:ds_cc}) and (\ref{eq:ds_ctct}).
Using the value of orbital velocity of the double pulsar $v\sim 10^{-3}$, geodetic precession
gives a weaker constraint on
$\Lambda,\tilde \Lambda$ ($\Lambda \gtrsim 10^{-17}$ eV)
than the one from GW phasing because of the larger size
of the binary pulsars systems ($r\sim 10^6$ km)
with respect to typical GW coalescing binary sources.
Considering the bounds from Gravity Probe B \cite{PhysRevLett.106.221101},
which constrained the geodetic precession with $1\%$ precession, the
relevant velocity and system size being $v \sim 3\times 10^{-5}$ and
$r\sim 4000$ km, one obtains $\Lambda\gtrsim 10^{-16}$eV.

Note however the presence in the spin precession equation (\ref{eq:ds_cctilde})
of a term $\sim \vec r\times \vec S_1$, being $\vec r$ the orbital radius vector,
at 1.5PN $\times \pa{\Lambda_-r}^{-6}$
level in the parity violating case.
Such term induces an oscillation at the orbital frequency $\sim v/r$ on the
top of the leading precession effect which has frequency $v^3/r$, see eq.~(\ref{eq:sdotLO}).
In General Relativity the spin equations of motion at next-to-leading order
have a term quadratic in the $\vec r$ vector which can produce nutation of the spin
at twice the orbital frequency \cite{Racine:2008kj}, but the presence of a
modulation at the orbital frequency in the spin precession equation
would be a smoking gun of the parity violating term $\vec r\times \vec S_1$.
The magnitude of this effect is suppressed by $v^3\times (\Lambda_- r)^{-6}$ with respect
to the leading order precession equation, hence assuming $N$ orbital cycles of observations,
one would get the limit $\Lambda_-\gtrsim 10^{-16,-15}{\rm eV}\times N^{1/12}$
for the double pulsar and Gravity Probe B respectively, which can observe
$10^{3,4}$ orbital cycles per year.

Overall, we have extended the analysis of \cite{Endlich:2017tqa} on the effect that
high curvature terms have on binary systems to include linear-in-spin effects
in energy, radiation emission and the spin precession equations, the last being
an important test-bed of gravity for spinning systems.

\section*{Acknowledgements}
The authors wish to thank Shao Lijing for useful correspondence.
The work of ANL is financed in part by the Coordena\c{c}\~ao de
Aperfei\c{c}oamento de Pessoal de N\'\i vel Superior - Brasil (CAPES) - Finance
Code 001.
RS is partially supported by CNPq.
RS would like to thank ICTP-SAIFR FAPESP grant 2016/01343-7.

\appendix

\section{Derivation of multipolar radiative coupling in GR}
\label{app:rad_mult}
The first diagram in fig.~\ref{fig:rad_quad_GR} can be read directly from the
Lagrangian (\ref{eq:wl_KK}), the second one needs the $\phi^2\sigma$ Lagrangian
bulk term which appears in eq.~(\ref{eq:sEH_KK})
\be
L_{rad-\ref{fig:rad_quad_GR}b}=\frac{m_1m_2}{8m_{Pl}^2}\int \frac{d^3k}{(2\pi)^3}\frac{k^ik^j}{k^4}e^{i\vec k\cdot \vec r}=-\frac{G_Nm_1m_2r^i r^j}{2r^3}\frac{\sigma_{ij}}{m_{Pl}}\,,
\ee
where a term proportional to the trace of $\sigma$ has been neglected, since it
does not contribute to radiation in the TT gauge.
Using trivial identities and $m_1\ddot x_1+m_2\ddot x_2=0$, which is valid in general, not only in GR, one can rewrite the radiative coupling as
\be
\label{eq:rad_cnstnc}
\ba{rcl}
\ds L_{rad-LO}&=&\ds
\ds m_1\pa{v_1^iv_1^j-\frac{G_Nm_2}{2r^3}r^ir^j}\frac{\sigma_{ij}}{2m_{Pl}}+1\leftrightarrow 2\\
&=&\ds m_1\paq{v_1^iv_1^j+\frac 12\ddot x_1^i\pa{x_1^j-x_2^j}-
  \frac{r^j}2\pa{\frac{G_Nm_2}{r^3}r^i+\ddot x_1^i}}\frac{\sigma_{ij}}{2m_{Pl}}+1\leftrightarrow 2\\
&=&\ds
\paq{m_1v_1^iv_1^j+\frac{m_1}2\ddot x_1^ix_1^j+\frac{m_2}2\ddot x_2^i x_2^j-
  \frac{m_1}2r^i\pa{\frac{G_Nm_2}{r^3}r^j+\ddot x_1^j}}\frac{\sigma_{ij}}{2m_{Pl}}+1\leftrightarrow 2\\
&=&\ds\paq{\frac{m_1}2\frac{{\rm d}^2}{{\rm d}t^2}\pa{x_1^ix_1^j}+\frac{m_2}2
 \frac{{\rm d}^2}{{\rm d}t^2}\pa{x_2^ix_2^j}-
 r^i\pa{\frac{G_Nm_1m_2}{r^3}r^j+\frac{m_1}2\ddot x_1^j-\frac{m_2}2\ddot x_2^j}}
 \frac{\sigma_{ij}}{2m_{Pl}}\,,
\ea
\ee
which shows explicitly that the standard GR result (\ref{eq:rad_ele_GR})
is recovered on the equations of motions, but it also shows that in the
case of modified equations of motion, for which e.g.
$\ddot x_1^i=-Gm_2 r^i(1+O(\Lambda r)^{-6})$, extra terms appear in the
radiative coupling.

The usual textbook procedure to derive the source multipole
coupling to radiation is equivalent for the mass quadrupole to the
explicit caculation (\ref{eq:rad_cnstnc}) in GR, but it does not take
account GR deviations, and for an extended source
(like a binary system, extended over a volume $V$) whose size
$r\ll\lambda_{gw}$, being $\lambda_{gw}$ the GW wavelength, goes as follows:
\be
\label{eq:mult_exp}
\ba{rcl}
\ds{\cal S}_{mult}&=&\ds\frac 12\int {\rm d}t\int d^3xT^{\mu\nu}(t,\vec x)h_{\mu\nu}(t,\vec x)\simeq\\
&&\ds\frac 12\int {\rm d}t\left\{\pa{\int_VT^{00}}h_{00}+
  \paq{2\pa{\int_VT_{0i}}h_{0i}+\pa{\int_VT^{00}x^i}h_{00,i}}+\right.\\
&& \ds\paq{\pa{\int_VT^{ij}}h_{ij}+\pa{\int_VT_{0i,j}}\pa{h_{0i,j}+h_{0j,i}}+
    \frac 12\pa{\int_VT_{00}x^ix^j}h_{00,ij}}\\
  &&\left.\ds+\pa{\int_VT_{0i,j}}\pa{h_{0i,j}-h_{0j,i}}+
  \pa{\int_V T^{ij}x^k}h_{ij,k}+\ldots\right\}\,,
  \ea
  \ee
  where the notation $\int_V=\int d^3x$ has been adopted for brevity and all
  gravitational fields $h$ and their derivatives have to be computed at $\vec x=0$.
  Note that this Taylor expansion is actually an expansion in
  $r/\lambda_{gw}\sim v$ where $v$ is the source internal velocity.
  
  Using repeatedly the energy-momentum conservation in the form $\dot T^{\mu 0}=-T^{\mu i}_{\ \ ,i}$
  one can derive
  \be
  \label{eq:emt_cons}
  \ba{rcl}
\ds  \int_V T^{ij}&=&\ds\frac 12\frac{{\rm d}^2}{{\rm d}t^2}\pa{\int_V T^{00}x^ix^j}\equiv \frac 12\ddot Q^{ij}\,,\\
\ds  \int_V T^{0i}&=&\ds -\frac{\rm d}{{\rm d}t}\pa{\int_V T^{00}x^i}\,,\\
\ds  \int_V T^{ij}x^k&=&\ds\frac 13\int_V\pa{T^{ij}x^k+T^{ki}x^j+T^{jk}x^i}+\frac 13\int_V \pa{2T^{ij}x^k-T^{ik}x^j-T^{jk}x^i}\,,\\
&=&\ds\frac 16\frac{{\rm d}^2}{{\rm d}t^2}\pa{\int_V T^{00}x^ix^jx^k}
+\frac 13\frac{\rm d}{{\rm d}t}\paq{\int_V \pa{T^{0i}x^kx^j+T^{0j}x^kx^i-2T^{0k}x^ix^j}}\,,
\ea
\ee
The coupling $T^{ij}h_{ij}$ hence gives rise to the electric quadrupole coupling
$\frac 12 Q^{ij}R_{0i0j}$ term,
where the linear order expression of $R_{0i0j}$ in terms of the Kaluza-Klein fields $\phi,A_i,\sigma_{ij}$ is given in eq.~(\ref{eq:riem_ele_tt}).

Moreover defining
\be
\ba{rcl}
O^{ijk}&\equiv&\ds\int_VT^{00}x^ix^jx^k\,,\\
J^{ij}&\equiv&\ds\frac 12\int_V T^0_{\ l}x_k\pa{x^i\epsilon^{jkl}+x^j\epsilon^{ikl}}\,,
\ea
\ee
using the identity
\be
\ba{l}
\ds\pa{T^{0i}x^kx^j-T^{0k}x^ix^j}\sigma_{ij,k}=
T^{0m}x^nx^j\pa{\delta^l_m\delta^k_n-\delta^k_m\delta^l_n}\sigma_{lj,k}\\
\ds =\epsilon_{imn}T^{0m}x^nx^j\frac 12\epsilon^{ikl}\pa{\sigma_{jk,l}-\sigma_{jl,k}}
\,,
\ea
\ee
and the definition of the magnetic part of the Riemann tensor
\be
   {\cal B}_{ij}\equiv\frac 12\epsilon_{ikl}R_{0jkl}\simeq \frac 1{4m_{Pl}}\epsilon_{ikl}
   \paq{\dot \sigma_{jk,l}-\dot\sigma_{jl,k}+ A_{l,jk}-A_{k,jl}+
     \frac 1{d-2}\pa{\dot\phi_{,k}\delta_{jl}-\dot\phi_{,l}\delta_{jk}}}\,,
\ee
one finds explicitly the magnetic and electric quadrupole and electric octupole
gravitational couplings of eq.~(\ref{eq:mult_rad_GR}), with
$J_{ij}=T^{0m}x^n\pa{\epsilon_{imn}x_j+\epsilon_{jmn}x_i}/2$.

At the next order in eq.~(\ref{eq:mult_exp}), i.e electric exadecapole and
magnetic octupole, one has the following identity, see \cite{Ross:2012fc}
for an explicit Lagrangian treatment or \cite{Thorne:1980ru} for general
derivation of the multipole expansion,
\be
\label{eq:exa}
\ba{rcl}
\ds\frac 14\int_V T^{ij}x^kx^l&=&\ds\frac 14\times\frac 16\int_V\pa{
  T^{ij}x^kx^l+T^{li}x^jx^k+T^{kl}x^ix^j+T^{jk}x^lx^i+T^{ik}x^jx^l+T^{jl}x^ix^k}\\
&&\ds +\frac 14\times \frac 12\int_V \pa{T^{ij}x^kx^l-T^{kl}x^ix^j}\\
&&\ds +\frac 14\times\frac 16\int_V \pa{2T^{ij}x^kx^l+2T^{kl}x^ix^j-T^{li}x^jx^k-T^{jk}x^lx^i-T^{ik}x^jx^l-T^{jl}x^ix^k}\,,
\ea
\ee
where the first line contains the electric exadecapole $l=4$, the second line contains
the magnetic octupole $l=3$ and the third line is a contribution to the electric
quadrupole ($l=2$).
From the third line of (\ref{eq:exa}) one gets the multipole coupling
\be
\label{eq:mult_nnl2}
\ba{l}
\ds\frac 14\left.\pa{\int_VT^{ij}x^kx^l}\right|_{(\ref{eq:exa})3^{rd}line}
\sigma_{ij,kl}\\
\ds=\frac 1{12}\pa{\int_VT^{ij}x^kx^l}
\pa{\sigma_{ij,kl}+\sigma_{kl,ij}-\sigma_{il,jk}-\sigma_{jk,il}}\\
\ds=-\frac 1{12}\epsilon_{ikm}\epsilon_{jln}\pa{\int_VT^{ij}x^kx^l}\frac{\ddot\sigma^{mn}}{m_{Pl}}\\
\ds=\frac 1{12}\pa{\int_VT^{ll}x^mx^n+T^{mn}r^2-T^{ml}x^nx^l-T^{nl}x^lx^m}
\frac{\ddot\sigma_{mn}}{m_{Pl}}\,,
\ea
\ee
where the relationships
\be
\label{eq:riem_ele}
R_{ijkl}&=&-\epsilon_{ijm}\epsilon_{kln}R_{0m0n}\,,\\
R_{ijkl}&=&\frac 12\pa{\sigma_{il,jk}+\sigma_{jk,il}-\sigma_{ik,jl}-\sigma_{jl,ik}}
+O(\sigma^2)\,,
\ee
have been used, which are valid for the radiative field.
To collect all the electric quadrupole contributions from second order
multipole expansions one must add to eq.~(\ref{eq:mult_nnl2}) the terms obtained
by subtracting traces from the first and second line of (\ref{eq:exa}):
\be
\ba{rcl}
\ds\left.\frac 14\pa{\int_V T^{ij}x^kx^l}\right|_{Tr-Exa}\sigma_{ij,kl}&=&\ds\frac 14\times\frac 1{42}\int_V\pa{T^{ij}r^2+T^{ll}x^ix^j+2\pa{T^{il}x^lx^j+T^{jl}x^lx^i}}\ddot\sigma_{ij}\,,\\
\ds\left.\frac 14\pa{\int_V T^{ij}x^kx^l}\right|_{Tr-Oct}\sigma_{ij,kl}&=&\ds
-\frac 14\times\frac 16\int_V\pa{T^{ll}x^ix^j-T^{ij}r^2}\ddot\sigma_{ij}\,,
\ea
\ee
where we used that the gravitational field is on-shell and in vacuum (i.e. $\ddot \sigma_{ij}=\nabla^2\sigma_{ij}$ and $\sigma_{ij,j}=0$).

In the standard approach one then uses repeatedly the energy momentum
conservation equation to derive
\be
\label{eq:eom_ids}
\ba{rcl}
\ds \int_V\pa{T^{ij}r^2+T^{ll}x^ix^j}&=&\ds\int_V \left[
  \pa{-\frac 12\ddot T^{00}r^2+2\dot T^{0l}x^l}x^ix^j+
  \pa{\dot T^{i0}x^j+\dot T^{j0}x^i}r^2\right.\\
  &&\ds\left. \qquad -2\pa{T^{im}x^mx^j+T^{jm}x^mx^i}\right]\,,\\
\ds \int_V\pa{T^{il}x^lx^j+T^{jl}x^lx^i}&=&\ds
\int_V\paq{\pa{\dot T^{0l}x^l-T^{ll}}x^ix^j}\,,\\
\ds\int_V\pa{\dot T^{i0}x^j+\dot T^{j0}x^i}r^2&=&\ds
  \int_V\pa{\ddot T^{00}r^2x^ix^j-2\dot T^{0l}x^l x^i x^j}\,,
\ea
\ee
from which one recovers the standard form of the radiative electric quadrupole
involved in the GR electric quadrupole coupling
(\ref{eq:rad_ele_GR}) (valid up to $O(v^2)$ with respect to LO)
\be
\label{eq:ele_quad_NLO}
-\frac 12I^{ij}R_{0i0j}=-\frac 12R_{0i0j}
\int_V\pa{T^{00}+T^{ll}-\frac 43\dot T^{0l}x^l+\frac{11}{42}\ddot T^{00}r^2}x^ix^j\,.
\ee
However since we are using non-GR equation of motions to derive the main
results of this paper, in particular in secs.~\ref{ssec:rad_cc},\ref{ssec:rad_cctilde} and \ref{ssec:rad_ctct}, we will use the
GR-equivalent multipole expanded coupling (\ref{eq:mult_exp})
\be
\label{eq:rad_ele_nGR}
\ba{l}
\ds\frac 12\int_VT^{ij}(t,\vec x)\frac{\sigma_{ij}(t,\vec x)}{m_{Pl}}\\
\simeq\ds
\frac{\sigma_{ij}(t,0)}{2m_{Pl}}\int_V\paq{T^{ij}+
\frac 17\frac{{\rm d}^2}{{\rm d}t^2}\pa{\frac 23T^{ll}x^ix^j+\frac{11}6T^{ij}r^2-T^{il}x^lx^j-T^{jl}x^lx^i}}\,.
\ea
\ee
One can check explicitly that eqs.~(\ref{eq:ele_quad_NLO}) and
(\ref{eq:rad_ele_nGR}) are equivalent in GR on the equation of motions,
in particular for the spinning case
by using in eq.~(\ref{eq:rad_ele_nGR}) the energy momentum tensor $T_{ij}$
derived via the sum of eqs.~(\ref{eq:rad_QSI_GR}) and (\ref{eq:rad_QSII_GR}),
the equation of motion (\ref{eq:eom_GR_SO}) and (\ref{eq:rad_cnstnc}).
In eq.~(\ref{eq:ele_quad_NLO}) the LO part of $T^{00}$, $T^{0i}$ and electric
part of $T^{ij}$ can
be read directly from the point-particle Lagrangian (\ref{eq:lag_KK}).
In the right hand side of eq.~(\ref{eq:rad_ele_nGR}) the LO part of $T^{ij}$
does not contribute to the electric coupling in the first term
(the one without time derivatives), but it does in the terms involving two time
derivatives.

\section{Higher order curvature contributions to the conservative Lagrangian}

\subsection{No Spin}

\subsubsection{${\mathcal C}^2$}
\label{ssapp:pot_cc}
The amplitudes corresponding to the diagrams in fig.~\ref{fig:pot_cc} are
\be
\ba{rcl}
\ds A_{fig.~\ref{fig:pot_cc}a}&=&\ds i\frac{m_1m_2^3}{8m_{Pl}^6\Lambda^6}\int_{\Pp\,,\K_1\,,\K_2}
e^{i\vec p\cdot \vec r}\frac{\paq{\vec k_2\cdot\pa{\vec k_1-\vec k_2}}^2\paq{\vec p\cdot\pa{\vec p-\vec k_1}}^2}{p^2\pa{p-k_1}^2\pa{k_1-k_2}^2k_2^2}=-i512\frac{G_N^3m_1m_2^3}{\Lambda^6r^9}\,,\\
\ds A_{fig.~\ref{fig:pot_cc}b}&=&\ds i\frac{m_1^2m_2^2}{16m_{Pl}^6\Lambda^6}\int_{\Pp\,,\K_1\,,\K_2}
e^{i\vec p\cdot \vec r}\times\\
&&\ds
\frac{\paq{\vec k_1\cdot\pa{\vec p-\vec k_1}}^2\paq{\vec k_2\cdot\pa{\vec p-\vec k_2}}^2+2\pa{\vec k_1\cdot \vec k_2}^2
  \paq{\pa{\vec p-\vec k_1}\cdot\pa{\vec p-\vec k_2}}^2}{\pa{\vec p-\vec k_1}^2\vec k_1^2\pa{\vec p-\vec k_2}^2\vec k_2^2}=0\,,
\ea
\ee
giving the result of eq.~(\ref{eq:pot_cc}).
They can be computed by repeated use (and differentiations) of the standard
1-loop master integral
\be
\int_\K\frac 1{k^{2a}\pa{p-k}^{2b}}=\frac{\pa{p^2}^{d/2-2}}{\pa{4\pi}^{d/2}}
\frac{\Gamma(d/2-a)\Gamma(d/2-b)\Gamma(a+b-d/2)}
     {\Gamma(a)\Gamma(b)\Gamma(d-a-b)}\,,
\ee
and of the fundamental integral
\be
\int_\Pp \frac{e^{i\vec p\cdot \vec r}}{p^{2a}}=\frac 1{2^{2a}\pi^{d/2}}
\frac{\Gamma(d/2-a)}{\Gamma(a)}r^{2a-d}\,.
\ee

\subsection{Spin}

\subsubsection{${\mathcal C}^2$}
\label{ssapp:pot_spin_cc}

The amplitudes from diagrams in fig.~\ref{fig:pot_spin_cc}, grouped according
to the gravitational polarisation attached to the spin, are the following:
\be
\ba{rcl}
\ds A_{fig.\ref{fig:pot_spin_cc}-A}&=&\ds -\frac{m_1^2m_2}{8 m_{Pl}^6\Lambda^6}
\int_{\Pp,\K_1,\K_2}
\frac {e^{i\vec p\cdot \vec r}}{p^2\pa{p-k_1}^2\pa{k_1-k_2}^2k_2^2}\times
\pa{\vec k_2\cdot \vec v_1}\\
&&\ds\times \Big\{
\paq{\vec p\cdot \pa{\vec p-\vec k_1}}^2\pa{\vec k_1-\vec k_2}\cdot \vec k_2
\,\pa{k_1-k_2}^j\\
&&\ds\quad+\paq{\pa{\vec k_1-\vec k_2}\cdot\pa{\vec p-\vec k_1}}^2
\vec p\cdot \vec k_2\,\frac 12 p^j\Big\}k_2^iS_{1ij}\\
&&\ds=i\frac{1152}{\pa{\Lambda r}^6}\frac{G_N^3m_1^2m_2}{r^5}r_iv_{1j}S_1^{ij}\,,\\
\ds A_{fig.\ref{fig:pot_spin_cc}-\phi}&=&\ds -\frac{m_2}{8m_{Pl}^6\Lambda^6}\int_{\Pp\,\K_1,\K_2}\frac{e^{i\vec p\cdot \vec r}}{p^2\pa{p-k_1}^2\pa{k_1-k_2}^2k_2^2}\\
&&\ds \times \Big\{m_1^2\Big\{2\paq{\pa{\vec k_1-\vec k_2}\cdot \vec k_2}^2\paq{\vec p\cdot\pa{\vec p-\vec k_1}}^2\\
&&\quad +\paq{\pa{\vec k_1-\vec k_2}\cdot \pa{\vec p-\vec k_1}}^2
\paq{\vec p\cdot \vec k_2}^2\Big\}\\
&&-m_2^2\pag{\vec p\cdot\pa{\vec p-\vec k_1}}^2
\paq{\vec k_2\cdot\pa{\vec k_1-\vec k_2}}^2\Big\}p_i\pa{S^{i0}_1+v_{1j}S^{ij}_1}\\
&=&\ds i\frac{4608}{\pa{\Lambda r}^6}
\frac{G_N^3m_2}{r^5}\pa{m_1^2+m_2^2}\pa{r_iS_1^{i0}+r_iv_{1j}S_1^{ij}}\,,\\
\ds A_{fig.\ref{fig:pot_spin_cc}-\sigma}&=&\ds \frac{m_1^2m_2}{16m_{Pl}^6\Lambda^6}\int_{\Pp,\K_1,\K_2}
\frac{e^{i\vec p\cdot \vec r}}{p^2\pa{p-k_1}^2\pa{k_1-k_ 2}^2k_2^2}
\Big\{\paq{\pa{\vec k_1-\vec k_2}\cdot\pa{\vec p-\vec k_1}}^2\vec p\cdot \vec k_2p^j\\
&&\ds\quad+2\paq{\vec p\cdot \pa{\vec p-\vec k_1}}^2\pa{\vec k_1-\vec k_2}\cdot \vec k_2 \pa{k_1-k_2}^j\Big\}\vec k_2\cdot \vec v_1k_2^iS_{1ij}\\
&=&\ds -i\frac{1152}{\pa{\Lambda r}^6}
\frac{G_N^3m^2_1m_2}{r^5}r_iv_{1j}S^{ij}_1\,,
\ea
\ee
which add up to give the potential
\be
V_{\Lambda S_1}&=&\ds -\frac{4608}{\pa{\Lambda r}^6}
\frac{G_N^3m_2}{r^5}\pa{m_1^2+m_2^2}\pa{r_iS^{i0}_1+r_iv_{1j}S^{ij}_1}\,,
\ee
leading via eq.~(\ref{eq:spin_eom}) to eq.~(\ref{eq:ds_cc}).

\subsubsection{$\mathcal{C\tilde C}$}
\label{ssapp:pot_spin_cctilde}

The amplitudes from diagrams in fig.~\ref{fig:pot_spin_cctilde} are
\be
\ba{rcl}
\ds A_{fig.\ref{fig:pot_spin_cctilde}a}&=&\ds -\frac{m_1^2m_2}{4m_{Pl}^6\Lambda^6_-}\int_{\Pp,\K_1,\K_2}\frac{e^{i\vec p\cdot \vec r}\pa{p-k_1}^l}{p^2\pa{p-k_1}^2\pa{k_1-k_2}^2k_2^2}\left\{
\paq{\pa{\vec k_1-\vec k_2}\cdot \vec k_2}^2\paq{\vec p\cdot\pa{\vec p-\vec k_1}}p^j\right.\\
&&\ds\left.+
  2  \paq{\vec p\cdot \vec k_2}^2\paq{\pa{\vec p-\vec k_1}\cdot\pa{\vec k_1-\vec k_2}}\pa{k_1-k_2}^j\right\}\pa{p-k_1}^k\epsilon_{ijk}S^{il}_1\\
&=&\ds i\frac{2304}{(\Lambda_- r)^6}\frac{G_N^3m_1^2m_2}{r^5}
\epsilon_{ijk}r^iS_1^{jk}\,,\\
\ds A_{fig.\ref{fig:pot_spin_cctilde}b}&=&\ds\frac{m_2^3}{8m_{Pl}^6\Lambda^6_-}\int_{\Pp,\K_1,\K_2}\frac{e^{i\vec p\cdot \vec r}}{p^2\pa{p-k_1}^2\pa{k_1-k_2}^2k_2^2}\\
&&\ds\times
\paq{\pa{\vec k_1-\vec k_2}\cdot \vec k_2}^2\paq{\vec p\cdot\pa{\vec p-\vec k_1}}
p^lk_1^jp^k\epsilon_{ijk}S^{il}=0\,,\\
\ds A_{fig.\ref{fig:pot_spin_cctilde}c}&=&\ds\frac{m_1m_2^2}{8m_{Pl}^6\Lambda^6_-}\int_{\Pp,\K_1,\K_2}\frac{e^{i\vec p\cdot \vec r}}{\pa{p-k_1}^2k_1^2\pa{p-k_2}^2k_2^2}\\
&&\ds\times
\left\{\paq{\vec k_2\cdot\pa{\vec p- \vec k_2}}^2\paq{\pa{\vec p-\vec k_1}\cdot \vec k_1}k_1^jp^k\right.\\
&&\left. +\pa{\vec k_1\cdot \vec k_2}^2\paq{\pa{\vec p-\vec k_1}\cdot\vec k_2}
\pa{p-k_2}^j\pa{p-k_2}^k\right\}\pa{\vec p-\vec k_1}^l\epsilon_{ijk}S_1^{il}=0\,,
\ea
\ee
giving the potential in eq.~(\ref{eq:pot_spin_cctilde}).

\subsubsection{$\tilde{\mathcal{C}}^2$}
\label{ssapp:pot_spin_ctct}

The amplitudes from the non-vanishing diagrams in fig.~\ref{fig:pot_spin_ctct} are
\be
\ba{rcl}
\ds A_{fig.\ref{fig:pot_spin_ctct}d}&=&\ds -\frac{2m_1^2m_2}{m_{Pl}^6\tilde\Lambda^6}\int_{\Pp,\K_1,\K_2}
\frac{e^{i\vec p\cdot \vec r}}{p^2\pa{p-k_1}^2\pa{k_1-k_2}^2k_2^2}\\
&&\ds\quad\times
\left\{
\paq{\pa{\vec p-\vec k_1}\cdot\vec k_2}\paq{\vec p\cdot\pa{\vec k_1-\vec k_2}}k_2^kp^n+
\paq{\pa{\vec p-\vec k_1}\cdot\vec p}\paq{\vec k_2\cdot\pa{\vec k_1-\vec k_2}}p^kk_2^n\right\}\\
&&\ds\qquad\times\pa{p-k_1}^i\pa{p-k_1}^lv_1^m\pa{k_1-k_2}^r
\epsilon_{jkl}\epsilon_{mnr}S_1^{ij}\\
&=&\ds i\frac{55296}{11(\tilde \Lambda r)^6}\frac{G_N^3m_1^2m_2}{r^5}
r^iv_1^jS_{1ij}\,,\\
\ds A_{fig.\ref{fig:pot_spin_ctct}e}&=&\ds -\frac{2m_1^2m_2}{m_{Pl}^6\tilde\Lambda^6}\int_{\Pp,\K_1,\K_2}
e^{i\vec p\cdot \vec r}
\frac{\paq{\vec p\cdot\pa{\vec k_1-\vec k_2}}\paq{\pa{\vec p-\vec k_1}\cdot\vec k_2}}{p^2\pa{p-k_1}^2\pa{k_1-k_2}^2k_2^2}\\
&&\ds\quad\times \pa{p-k_1}^ik_2^k\pa{p-k_1}^lv_2^m\pa{k_1-k_2}^np^r\epsilon_{jkl}\epsilon_{mnr}S_1^{ij}\\
&=&\ds -i\frac{55296}{11(\tilde \Lambda r)^6}\frac{G_N^3m_1^2m_2}{r^5}
r^iv_2^j S_{1ij}\,,
\ea
\ee
whose sum gives eq.~(\ref{eq:pot_ctct_spin}).

\section{Higher order curvature contributions to radiative coupling}
\label{app:rad}

\subsection{No spin}

\subsubsection{$\mathcal{C}^2$}
\label{ssapp:rad_cc}
The leading correction to radiative coupling is given by the diagram in fig.~\ref{fig:rad_cc} resulting in amplitudes
\be
\ba{rcl}
\ds A_{fig.~\ref{fig:rad_cc}}&=&\ds i\frac{m_1m_2^2}{m_{Pl}^4\Lambda^6}\int_{\Pp\,,\K}e^{i\vec p\cdot \vec r}\,
\frac{p^ip^j\paq{\vec k\cdot\pa{\vec p-\vec k}}^2+2k^ik^j\paq{\vec p\cdot\pa{\vec p-\vec k}}^2}{p^2\pa{p-k}^2k^2}\times R_{0i0j}\\
&=&\ds i 1344\frac{G_N^2m_1m_2^2}{\pa{\Lambda r}^6}n^in^jR_{0i0j}\,,
\ea
\ee
to which the diagrams obtained under $1\leftrightarrow 2$ exchange must be
added, and terms $\propto \delta^{ij}R_{0i0j}$ which are vanishing on-shell have
been neglected.
This amplitude is needed to arrive at eq.~(\ref{eq:rad_cc}).

\subsubsection{$\mathcal{C\tilde C}$}
\label{app:rad_cctilde}
The diagram in figs.\ref{fig:rad_cctilde} are worth the following amplitudes
\be
\label{eq:rad_cctilde_amp}
\ba{rcl}
A_{fig.\ref{fig:rad_cctilde}a}&=&\ds -i\frac{m_1m_2^2}{m_{Pl}^4{\tilde\Lambda}^6}
\int_{\Pp,\K}\frac{e^{i\vec p\cdot\vec r}}{p^2\pa{p-k}^2k^2}
\pag{p^ip^j\paq{\vec k\cdot\pa{\vec p-\vec k}}^2+
  2k^ik^j\paq{\vec p\cdot\pa{\vec p-\vec k}}^2}{\mathcal B}_{ij}\\
&=&\ds -i\frac{1344}{(\tilde \Lambda r)^6}\frac{G_N^2m_1m_2^2}{r^2}
r^ir^j {\cal B}_{ij}\,,\\
A_{fig.\ref{fig:rad_cctilde}b}&=&\ds i\frac{2m_1m_2^2}{m_{Pl}^4{\tilde\Lambda}^6}
\int_{\Pp,\K}\frac{e^{i\vec p\cdot\vec r}}{p^2\pa{p-k}^2k^2}
\paq{\pa{\vec k\cdot p}^2+p^4-2p^2\pa{\vec p\cdot\vec k}}
  k^ik^kv_2^l\epsilon_{jkl}R_{0i0j}\\
&=&\ds -i\frac{1152}{(\tilde \Lambda r)^6}\frac{G_N^2m_1m_ 2^2}{r^2}
  r^ir^kv_2^l\epsilon_{jkl}{\cal E}_{ij}\,,\\
 A_{fig.\ref{fig:rad_cctilde}c}&=&\ds i\frac{2m_1m_2^2}{m_{Pl}^4{\tilde\Lambda}^6}
\int_{\Pp,\K}\frac{e^{i\vec p\cdot\vec r}}{p^2\pa{p-k}^2k^2}
\paq{\pa{\vec k\cdot p}^2+k^4-2k^2\pa{\vec p\cdot\vec k}}
p^ip^kv_1^l\epsilon_{jkl}R_{0i0j}\\
&=&\ds -i\frac{1536}{(\tilde \Lambda r)^6}\frac{G_N^2m_1m_ 2^2}{r^2}
  r^ir^kv_1^l\epsilon_{jkl}{\cal E}_{ij}\,,
\ea
\ee
from which eqs.(\ref{eq:rad_mag_cctilde}) and (\ref{eq:rad_ele_cctilde}) follow.

\subsection{Spin}
\subsubsection{$\mathcal{C}^2$}
\label{ssapp:rad_spin_cc}
The leading, linear-in-spin corrections to the magnetic quadrupole comes from diagrams in fig.~\ref{fig:rad_spin_cc_mag}, whose amplitudes are respectively
for the magnetic case
\be
\ba{rcl}
\ds A_{fig.\ref{fig:rad_spin_cc_mag}a}&=&\ds \frac{2m_1m_2}{m_{Pl}^4\Lambda^6}
\int_{\Pp\,,\K}
e^{i\vec p\cdot\vec r}\frac{\paq{\vec p\cdot \pa{p-k}}^2}{p^2\pa{p-k}^2k^2}k^i
\paq{S_1^jk^2-k^j\pa{\vec k\cdot \vec S_1}}{\cal B}_{ij}\\
&=&\ds i768\frac{G_N^2m_1m_2}{\Lambda^6r^8}
\paq{2r^iS_1^j-5n^in^j\pa{\vec r\cdot \vec S_1}}{\cal B}_{ij}\,,\\
\ds A_{fig.\ref{fig:rad_spin_cc_mag}b}&=&\ds i\frac{m_2^2}{m_{Pl}^4\Lambda^6}
\int_{\Pp\,,\K}e^{i\vec p\cdot\vec r}\frac{\paq{\vec k\cdot \pa{p-k}}^2}{p^2\pa{p-k}^2k^2}p^i
\paq{S_1^j\vec p\cdot\pa{\vec p-\vec k}-\pa{p-k}^j\pa{\vec p\cdot\vec S_1}}{\cal B}_{ij}\\
&=&\ds i384\frac{G_N^2m_2^2}{\Lambda^6r^8}\pa{5r^iS_1^j-8n^in^j\pa{\vec r\cdot \vec S_1}}{\cal B}_{ij}\,,
\ea
\ee
whose sum reproduces eq.~(\ref{eq:rad_smag_cc}).

The leading, linear-in-spin electric quadrupole corrections are due to the
diagrams in fig.~\ref{fig:rad_spin_cc_ele} which give:
\be
\ba{rcl}
\ds A_{fig.\ref{fig:rad_spin_cc_ele}a}&=&\ds -i\frac{m_2^2}{4m_{Pl}^4\Lambda^6}\int_{\Pp\,,\K}
e^{i\vec p\cdot\vec r}\frac{\paq{\vec k\cdot \pa{p-k}}^2}{p^2\pa{p-k}^2k^2}
\paq{\pa{\vec p\cdot \vec v_1}p^i-p^2v_1^i}p^kS_1^{kj}{\cal E}_{ij}\\
&=&\ds i192\frac{G_N^2m_2^2}{\Lambda^6r^8}\paq{8\pa{\hat n\cdot \vec v_1}n^i\pa{\vec S_1\times \vec r}^j-
  4v_1^i\pa{\vec S_1\times \vec r}^j-r^i\pa{\vec S_1\times\vec v_1}^j}
{\cal E}_{ij}\,,\\
\ds A_{fig.\ref{fig:rad_spin_cc_ele}b}&=&\ds i\frac{m_1m_2}{2m_{Pl}^4\Lambda^6}
\int_{\Pp\,,\K}
e^{i\vec p\cdot\vec r}\frac{\paq{\vec p\cdot \pa{p-k}}^2}{p^2\pa{p-k}^2k^2}
\paq{\pa{\vec k\cdot v_1}k^i-k^2v_1^i}k^kS_1^{kj}{\cal E}_{ij}\\
&=&\ds i192\frac{G_N^2m_1m_2}{\Lambda^6r^8}\paq{
  5\pa{\hat n\cdot \vec v_1}n^i\pa{\vec S_1\times \vec r}^j-
  v_1^i\pa{\vec S_1\times\vec r}^j-r^i\pa{\vec S_1\times\vec v_1}^j}
{\cal E}_{ij}\,,\\
\ds A_{fig.\ref{fig:rad_spin_cc_ele}c}&=&\ds i\frac{m_2^2}{m_{Pl}^4\Lambda^6}
\int_{\Pp\,,\K}
\frac{e^{i\vec p\cdot\vec r}}{p^2\pa{p-k}^2k^2}
\pag{\paq{\vec k\cdot\pa{\vec p-\vec k}}^2p^ip^j+
  2\paq{\vec p\cdot\pa{\vec p-\vec k}}^2k^ik^j}\\
&&\ds\times\pa{v_1^mS_1^{mn}+S_1^{0n}}{\cal E}_{ij}\\
&=&\ds -i\pa{1+\kappa}\frac{2688}{\pa{\Lambda r}^6}\frac{G_Nm_2^2}{r^2}
\pag{4n^i n^j\paq{\vec S_1\cdot\pa{\vec r\times \vec v_1}}+
  r^i\pa{\vec S_1\times \vec v_1}^j}{\cal E}_{ij}\,,\\
\ds A_{fig.\ref{fig:rad_spin_cc_ele}d}&=&\ds i\frac{2m_1m_2}{m_{Pl}^4\Lambda^6}
\int_{\Pp\,,\K}
\frac{e^{i\vec p\cdot\vec r}}{p^2\pa{p-k}^2k^2}\Big\{
\paq{\vec p\cdot \vec k}^2\pa{p-k}^i\pa{p-k}^j\\
&&\ds+
    \paq{\vec p\cdot \pa{\vec p-\vec k}}^2k^ik^j+
    \paq{\pa{\vec p-\vec k}\cdot \vec k}^2p^ip^j\Big\}
  k^n\pa{v^m_1S_1^{mn}+S^{0n}}{\cal E}_{ij}\\
&=&\ds -i\pa{1+\kappa}\frac{2688}{\pa{\Lambda r}^6}\frac{G_Nm_1m_2}{r^2}
  \pag{4n^in^j\paq{\vec S_1\cdot\pa{\vec r\times \vec v_1}}
    +r^i\pa{\vec S_1\times \vec v_1}^j}{\cal E}_{ij}\,,
\ea
\ee
whose sum gives eq.~(\ref{eq:rad_ele_cc_spin}).

\subsubsection{$\mathcal{C\tilde C}$}
\label{ssapp:rad_spin_cctilde}
\allowdisplaybreaks
The diagrams in fig.~\ref{fig:rad_spin_cctilde_ele} give the following amplitudes for the electric coupling
\be
\ba{rcl}
\ds A_{fig.\ref{fig:rad_spin_cctilde_ele}{\cal E}a}&=&\ds i\frac{m_2^2}{m_{Pl}^4\Lambda_-^6}\int_{\Pp\,,\K}
\frac{e^{i\vec p\cdot\vec r}\pa{\vec p-\vec k}\cdot \vec k}{p^2\pa{p-k}^2k^2}\\
&&\ds\times\pag{\paq{\pa{\vec p-\vec k}\cdot \vec k}p^i\delta^{jm}
+2k^i\pa{p-k}^jk^m}p^rp^kS_{1lr}\epsilon_{jkl}R_{0i0m}\\
&&\ds=1152\frac{G_N^2m_2^2}{r^2\pa{\Lambda_- r}^6}
\paq{5r^iS_1^j-8n^in^j\pa{\vec r\cdot \vec S_1}}R_{0i0j}\,,\\
\ds A_{fig.\ref{fig:rad_spin_cctilde_ele}{\cal E}b}&=&\ds i2\frac{m_1m_2}{m_{Pl}^4\Lambda_-^6}
\int_{\Pp\,,\K}\frac{e^{i\vec p\cdot\vec r}}{p^2\pa{p-k}^2k^2}\pa{p-k}^k\pa{p-k}^rS_{1lr}\epsilon_{jkl}R_{0i0m}\\
&&\ds
  \times\pag{\pa{\vec p\cdot \vec k}^2\delta^{jm}\pa{p-k}^i+
    \paq{\pa{\vec p-\vec k}\cdot \vec k\,p^ip^mk^j+\pa{\vec p-\vec k}\cdot \vec p\, k^ik^mp^j}}\\

&=&\ds 2688\frac{G_N^2m_1m_2}{r^2\pa{\Lambda_-r}^6}
\paq{r^iS_1^j-4n^in^j\pa{\vec r\cdot \vec S_1}}R_{0i0j}\,,
\ea
\ee
allowing to derive eq.~(\ref{eq:rad_ele_cct_spin_ele}).\\
For the magnetic coupling one has
\be
\ba{rcl}
\ds A_{fig.\ref{fig:rad_spin_cctilde_ele}{\cal B}a}&=&\ds 2i\frac{m_2^2}{m_{Pl}^4\Lambda_-^6}\int_{\Pp\,,\K}\frac{e^{i\vec p\cdot\vec r}\paq{\vec p\cdot\pa{\vec p-\vec k}}}{p^2\pa{p-k}^2k^2}
  \pa{\vec p\cdot\vec v_1}k^ik^jp^kk^lS_{1kl}{\cal B}_{ij}\\
  &=&\ds 384\frac{G_N^2m_2^2}{r^2\pa{\Lambda_-r}^6}
  \paq{\pa{\vec r\times\vec S_1}^iv_1^j+\pa{\vec v_1\times\vec S_1}^ir^j-
    8\frac{\vec r\cdot\vec v_1}{r^2}\pa{\vec r\times\vec S_1}^ir^j}{\cal B}_{ij}\\
\ds A_{fig.\ref{fig:rad_spin_cctilde_ele}{\cal B}b}&=&\ds 2i\frac{m_1m_2}{m_{Pl}^4\Lambda_-^6}
\int_{\Pp\,,\K}\frac{e^{i\vec p\cdot\vec r}\pa{\vec p-\vec k}\cdot \vec v_1}{p^2\pa{p-k}^2k^2}\\
&&\ds\times\paq{\vec k\cdot\pa{\vec p-\vec k}p^ip^jk^m\pa{p-k}^n+
  \vec p\cdot\pa{\vec p-\vec k}k^ik^jp^m\pa{p-k}^n}S_{1mn}{\cal B}_{ij}\\
&=&\ds 576\frac{G_N^2m_1m_2}{r^2\pa{\Lambda_- r}^6}
\paq{2\frac{\vec r\cdot \vec v_1}{r^2}\pa{\vec r\times\vec S_1}^ir^j-2\pa{\vec v_1\times\vec S_1}^ir^j+7n^in^j\pa{\vec r\times\vec v_1}\cdot \vec S_1}{\cal B}_{ij}\,,\\
\ds A_{fig.\ref{fig:rad_spin_cctilde_mag}a}&=&\ds i\frac{m_2^2}{4m_{Pl}^4\Lambda_-^6}\int_{\Pp\,,\K}
\frac{e^{i\vec p\cdot\vec r}p^k}{p^2\pa{p-k}^2k^2}\Big\{2\paq{\vec k\cdot\pa{\vec p-\vec k}}^2\pa{p^2v^i_1\delta^{jl}-v^i_1p^jp^l-\delta^{ik}p^j\pa{\vec p\cdot v_1}}\\
&&\ds+\pa{p-k}^i\pa{p-k}^jp^kk^l\pa{p^2\pa{\vec k\cdot\vec v_1}-\pa{\vec p\cdot\vec k}\pa{\vec p\cdot\vec v_1}}\Big\}S_{1kl}{\cal B}_{ij}\\
&&\ds=48\frac{G_N^2m_2^2}{r^2\pa{\Lambda_- r}^6}
\left[33\pa{\vec r\times \vec S_1}^iv_1^j-66\frac{\vec r\cdot\vec v_1}{r^2}\pa{\vec r\times \vec S_1}^ir^j
  +5n^in^j\pa{\vec r\times \vec v_1}\cdot \vec S_1\right.\\
  &&\ds\qquad\left.+7\pa{\vec v_1\times \vec S_1}^ir^j\right]{\cal B}_{ij}\,,\\
\ds A_{fig.\ref{fig:rad_spin_cctilde_mag}b}&=&\ds i\frac{m_2^2}{2m_{Pl}^4\Lambda_-^6}
\int_{\Pp\,,\K}\frac{e^{i\vec p\cdot\vec r}}{p^2\pa{p-k}^2k^2}\\
&&\ds\pag{p^ip^j\paq{\vec k\cdot\pa{\vec p-\vec k}}^2
  +2k^ik^j\paq{\vec p\cdot\pa{\vec p-\vec k}}^2}\pa{v_{1k}p_lS_1^{kl}-p_iS_1^{i0}}
    {\cal B}_{ij}\\
    &&\ds i1344\pa{1+\kappa}\frac{G_N^2m_2^2}{r^2\pa{\Lambda_- r}^6}
    \paq{4n^in^j\pa{\vec r\times\vec v_1}\cdot\vec S_1-\pa{\vec v_1\times \vec S_1}^ir^j}{\cal B}_{ij}\,.\\
    \ds A_{fig.\ref{fig:rad_spin_cctilde_mag}c}&=&\ds
    384\frac{G_N^2m_2^2}{r^2\pa{\Lambda_-r}^6}\left[-7\pa{\vec r\times \vec S_1}^iv_2^j+32\frac{\vec r\cdot \vec v_2}{r^2}\pa{\vec r\times \vec S_1}^ir^j
    +40n^in^j\pa{\vec r\times\vec v_2}\cdot\vec S_1\right.\\
    &&\ds\qquad\left.-14\pa{\vec v_2\times \vec S_1}^ir^j\right]{\cal B}_{ij}\,,\\
    \ds A_{fig.\ref{fig:rad_spin_cctilde_mag}d}&=&\ds i\frac{m_1m_2}{m_{Pl}^4\Lambda_-^6}
\int_{\Pp\,,\K}\frac{e^{i\vec p\cdot\vec r}}{p^2\pa{p-k}^2k^2}\\
&&\ds\Big\{
\pa{\vec p\cdot\vec k}^2\pa{p-k}^k\left[\pa{\vec p-\vec k}^2v_1^i\delta^{jl}-v_1^i\pa{p-k}^j\pa{p-k}^l-\delta^{jl}\pa{p-k}^i\pa{\vec p-\vec k}\cdot \vec v_1\right]\\
&&\ds\ +
\frac{p^kk^l}4\left[p^ip^j\paq{\vec k\cdot\vec v_1(p-k)^2
    -\pa{\vec p-\vec k}\cdot\vec v_1\vec k\cdot\pa{\vec p-\vec k}}\right.\\
&&\ds \left.\qquad+k^ik^j\paq{\pa{\vec p-\vec k}\cdot\vec v_1\vec p\cdot\pa{\vec p-\vec k}}
-\pa{p-k}^2\pa{\vec p\cdot\vec v_1}\right]\\
&&\ds\quad-
    k^ik^jp^2\pa{p-k}^2\pa{p-k}^kv_1^l\Big\}S_{1kl}{\cal B}_{ij}\\
&=&\ds i24\frac{G_N^2m_1m_2}{r^2\pa{\Lambda_- r}^6}
    \left[16\pa{\vec r\times \vec S_1}^iv_1^j-74\frac{\vec r\cdot\vec v_1}{r^2}\pa{\vec r\times\vec S_1}^ir^j+21\pa{\vec r\times\vec v_1}\cdot \vec S_1n^in^j+\right.\\
      &&\ds\qquad\left.+10\pa{\vec v_1\times\vec S_1}^ir^j\right]{\cal B}_{ij}\,,
    \ea
    \ee
    \be
    \nonumber
    \ba{rcl}
\ds A_{fig.\ref{fig:rad_spin_cctilde_mag}e}&=&\ds i\frac{m_2^2}{m_{Pl}^4\Lambda_-^6}
\int_{\Pp\,,\K}\frac{e^{i\vec p\cdot\vec r}}{p^2\pa{p-k}^2k^2}\\
&&\ds\Big\{\pa{p-k}^i\pa{p-k}^j\pa{\vec k\cdot\vec p}^2
+p^ip^j\pa{\vec k\cdot\pa{\vec p-\vec k}}^2+k^ik^j\pa{\vec p\cdot\pa{\vec p-\vec k}}\Big\}\\
&&\ds\qquad\times\pa{v_1^k\pa{p-k}^lS_1^{kl}+\pa{p-k}^kS_1^{0k}}{\cal B}_{ij}\\
&=&\ds 1344\pa{1+\kappa}\frac{G_Nm_1m_2}{r^2\pa{\Lambda_-r}^6}\paq{4n^in^j\pa{\vec r\times\vec v_1}\cdot \vec S_1+\pa{\vec S_1\times\vec v_1}^ir^j}{\cal B}_{ij}\,,\\
\ds A_{fig.\ref{fig:rad_spin_cctilde_mag}f}&=&\ds 384\frac{G^2_Nm_1m_2}{r^2\pa{\Lambda_-r}^6}
\left[3\pa{\vec r\times\vec S_1}^iv_1^j-6\frac{\vec r\cdot \vec v_1}{r^2}\pa{\vec r\times \vec S_1}^ir^j
  +5n^in^j\pa{\vec r\times\vec v_1}\cdot \vec S_1\right.\\
  &&\ds\qquad\left.-2\pa{\vec v_1\times \vec S_1}^ir^j\right]{\cal B}_{ij}\\
\ds A_{fig.\ref{fig:rad_spin_cctilde_mag}g}&=&\ds 768\frac{G^2_Nm_1m_2}{r^2\pa{\Lambda_-r}^6}
\left[-3\pa{\vec r\times\vec S_1}^iv_2^j+18\frac{\vec r\cdot \vec v_2}{r^2}\pa{\vec r\times \vec S_1}^ir^j
  +23n^in^j\pa{\vec r\times\vec v_2}\cdot \vec S_1\right.\\
  &&\ds\qquad\left.-14\pa{\vec v_2\times \vec S_1}^ir^j\right]{\cal B}_{ij}\,.
\ea
\ee
Expression of amplitudes before momentum integration relative to diagrams in
fig.~\ref{fig:rad_spin_cctilde_mag}c,f,g are very length and have not been
reported here.

\subsubsection{$\tilde{\mathcal{C}}^2$}
\label{ssapp:rad_spin_ctct}

The three diagrams in fig.\ref{fig:rad_spin_ctct_ele} give the amplitudes
\be
\ba{rcl}
\ds A_{fig.\ref{fig:rad_spin_ctct_ele}a}&=&\ds \frac{16m_2^2}{m_{Pl}^4\tilde \Lambda^6}
\int_{\Pp\,,\K}\frac{e^{i\vec p\cdot\vec r}}{p^2\pa{p-k}^2k^2}
\left\{\paq{\pa{\vec p-\vec k}\cdot \vec k}p^ip^k\delta^{lu}\pa{p-k}^nv_2^r\right.\\
&&\left.\qquad
-\pa{\vec p\cdot\vec k}\pa{p-k}^i\pa{p-k}^kv_2^lk^n\delta^{ru}\right\}
k^mp^sS_{1su}\epsilon_{jkl}\epsilon_{mnr}R_{0i0j}\\
&=&\ds i1536 \frac{G_N^2m_2^2}{\tilde \Lambda^6 r^8}\left\{
  16n^in^j\paq{\vec S_1\cdot\pa{\vec r\times\vec v_2}}-
  16n^i\pa{\hat n\cdot\vec v_2}\pa{\vec S_1\times\vec r}^j\right.\\
  &&\ds\left. +6r^i\pa{\vec S_1\times \vec v_2}^j
  +5v_2^i\pa{\vec S_1\times r}^j\right\}R_{0i0j}\,,\\
\ds  A_{fig.\ref{fig:rad_spin_ctct_ele}b}&=&\ds \frac{m_1m_2}{m_{Pl}^4\tilde\Lambda^6}
\int_{\Pp\,,\K}\frac{e^{i\vec p\cdot \vec r}}{p^2\pa{p-k}^2k^2}\left\{\pa{\vec p\cdot \vec k}
  \pa{p-k}^i\pa{p-k}^k\delta^{lu}k^nv_1^r+\right.\\
&&\ds\left.  \paq{\vec p\cdot\pa{\vec p-\vec k}}k^ik^kv_1^l\pa{p-k}^n\delta^{ru}
  \right\}p^m\pa{p-k}^sS_{1su}\epsilon_{jkl}\epsilon_{mnr}R_{0i0j}\\
&=&\ds -i 768\frac{G_N^2 m_1 m_2}{r^2(\tilde \Lambda r)^6}
\left\{11n^in^j\paq{\vec S_1\cdot\pa{\vec r\times\vec v_1}}
-6n^i\pa{\hat n\cdot \vec v_1}\pa{\vec S_1\times\vec r}^j\right.\\
&&\ds\qquad\left.+2r^i\pa{\vec S_1\times \vec v_1}^j
+3v_1^i\pa{\vec S_1\times \vec r}^j\right\} R_{0i0j}\,,\\
\ds  A_{fig.\ref{fig:rad_spin_ctct_ele}c}&=&\ds \frac{m_1m_2}{m_{Pl}^4\tilde\Lambda^6}
\int_{\Pp\,,\K}\frac{e^{i\vec p\cdot\vec r}}{p^2\pa{p-k}^2k^2}
\left\{p^ip^kv_2^l\delta^{mu}k^n\pa{p-k}^r\paq{\vec k\cdot\pa{\vec p-\vec k}}
\right.\\
&&\ds\left. \pa{p-k}^i\pa{p-k}^k\delta^{lu}v_2^mk^np^r\right\}
\pa{p-k}^sS_{1su}\epsilon_{jkl}\epsilon_{mnr}R_{0i0j}\\
&=&\ds i 1536\frac{G_N^2m_1m_2}{r^2(\tilde \Lambda r)^6}
\left\{47n^in^j\paq{\vec S_1\cdot\pa{\vec r\times\vec v_2}}
-42n^i\pa{\hat n\cdot \vec v_2}\pa{\vec S_1\times\vec r}^j\right.\\
&&\ds\qquad\left.+32r^i\pa{\vec S_1\times \vec v_2}-
6v_2^i\pa{\vec S_1\times \vec r}^j\right\} R_{0i0j}\,,
\ea
\ee
and their sum gives eq.~(\ref{eq:rad_ele_ctct_spin}).

The two diagrams in fig.\ref{fig:rad_spin_ctct_mag} give the amplitudes
\be
\ba{rcl}
\ds A_{fig.\ref{fig:rad_spin_ctct_mag}a}&=&\ds -\frac{4m_2^2}{m_{Pl}^4\tilde \Lambda^6}\int_{\Pp\,,\K}\frac{e^{i\vec p\cdot\vec r}\pa{\vec p\cdot\vec k}}{p^2k^2\pa{p-k}^2}
\pa{p-k}^i\pa{p-k}^jk^mp^np^sS_1^{sr}\epsilon_{mnr}\mathcal{B}_{ij}\\
&=&\ds i768\frac{G_N^2m_2^2}{\tilde \Lambda^6 r^8}
\paq{-5S_1^ir^j+8n^i n^j\pa{\vec S_1\cdot \vec r}}{\cal B}_{ij}\\
  \ds A_{fig.\ref{fig:rad_spin_ctct_mag}b}&=&\ds -\frac{4m_1m_2}{m_{Pl}^4\tilde \Lambda^6}\int_{\Pp\,,\K}\frac{e^{i\vec p\cdot\vec r}}{p^2\pa{p-k}^2k^2}\left\{
  \paq{\pa{\vec p-\vec k}\cdot\vec k}p^ip^jk^mp^n \right.\\
  &&\ds\left.\qquad -\paq{\pa{\vec p-\vec k}\cdot \vec p}k^ik^jp^m
  k^n \right\}\pa{p-k}^sS_{1sr}\epsilon_{mnr}{\cal B}_{ij}\\
  &=&\ds i2304\frac{G_Nm_1m_2}{\tilde \Lambda^6 r^8}
  \paq{-S_1^ir^j+6n^in^j\pa{\vec S_1\cdot \vec r}}{\cal B}_{ij}\,,
\ea
\ee
which allow the derivation of eq.~(\ref{eq:rad_mag_ctct_spin}).


\begin{thebibliography}{38}
\expandafter\ifx\csname natexlab\endcsname\relax\def\natexlab#1{#1}\fi
\expandafter\ifx\csname bibnamefont\endcsname\relax
  \def\bibnamefont#1{#1}\fi
\expandafter\ifx\csname bibfnamefont\endcsname\relax
  \def\bibfnamefont#1{#1}\fi
\expandafter\ifx\csname citenamefont\endcsname\relax
  \def\citenamefont#1{#1}\fi
\expandafter\ifx\csname url\endcsname\relax
  \def\url#1{\texttt{#1}}\fi
\expandafter\ifx\csname urlprefix\endcsname\relax\def\urlprefix{URL }\fi
\providecommand{\bibinfo}[2]{#2}
\providecommand{\eprint}[2][]{\url{#2}}

\bibitem[{\citenamefont{Abbott
  et~al.}(2019{\natexlab{a}})}]{LIGOScientific:2018mvr}
\bibinfo{author}{\bibfnamefont{B.~P.} \bibnamefont{Abbott}}
  \bibnamefont{et~al.} (\bibinfo{collaboration}{LIGO Scientific, Virgo}),
  \bibinfo{journal}{Phys. Rev.} \textbf{\bibinfo{volume}{X9}},
  \bibinfo{pages}{031040} (\bibinfo{year}{2019}{\natexlab{a}}),
  \eprint{1811.12907}.

\bibitem[{\citenamefont{Abbott et~al.}(2020{\natexlab{a}})}]{Abbott:2020niy}
\bibinfo{author}{\bibfnamefont{R.}~\bibnamefont{Abbott}} \bibnamefont{et~al.}
  (\bibinfo{collaboration}{LIGO Scientific, Virgo})
  (\bibinfo{year}{2020}{\natexlab{a}}), \eprint{2010.14527}.

\bibitem[{\citenamefont{Aasi et~al.}(2015)}]{TheLIGOScientific:2014jea}
\bibinfo{author}{\bibfnamefont{J.}~\bibnamefont{Aasi}} \bibnamefont{et~al.}
  (\bibinfo{collaboration}{LIGO Scientific}), \bibinfo{journal}{Class.\ Quant.\
  Grav.} \textbf{\bibinfo{volume}{32}}, \bibinfo{pages}{074001}
  (\bibinfo{year}{2015}), \eprint{1411.4547}.

\bibitem[{\citenamefont{Acernese et~al.}(2015)}]{TheVirgo:2014hva}
\bibinfo{author}{\bibfnamefont{F.}~\bibnamefont{Acernese}} \bibnamefont{et~al.}
  (\bibinfo{collaboration}{VIRGO}), \bibinfo{journal}{Class.\ Quant.\ Grav.}
  \textbf{\bibinfo{volume}{32}}, \bibinfo{pages}{024001}
  (\bibinfo{year}{2015}), \eprint{1408.3978}.

\bibitem[{\citenamefont{Allen et~al.}(2012)\citenamefont{Allen, Anderson,
  Brady, Brown, and Creighton}}]{Allen:2005fk}
\bibinfo{author}{\bibfnamefont{B.}~\bibnamefont{Allen}},
  \bibinfo{author}{\bibfnamefont{W.~G.} \bibnamefont{Anderson}},
  \bibinfo{author}{\bibfnamefont{P.~R.} \bibnamefont{Brady}},
  \bibinfo{author}{\bibfnamefont{D.~A.} \bibnamefont{Brown}}, \bibnamefont{and}
  \bibinfo{author}{\bibfnamefont{J.~D.~E.} \bibnamefont{Creighton}},
  \bibinfo{journal}{Phys.\ Rev.\ D} \textbf{\bibinfo{volume}{85}},
  \bibinfo{pages}{122006} (\bibinfo{year}{2012}), \eprint{gr-qc/0509116}.

\bibitem[{\citenamefont{Endlich et~al.}(2017)\citenamefont{Endlich, Gorbenko,
  Huang, and Senatore}}]{Endlich:2017tqa}
\bibinfo{author}{\bibfnamefont{S.}~\bibnamefont{Endlich}},
  \bibinfo{author}{\bibfnamefont{V.}~\bibnamefont{Gorbenko}},
  \bibinfo{author}{\bibfnamefont{J.}~\bibnamefont{Huang}}, \bibnamefont{and}
  \bibinfo{author}{\bibfnamefont{L.}~\bibnamefont{Senatore}},
  \bibinfo{journal}{JHEP} \textbf{\bibinfo{volume}{09}}, \bibinfo{pages}{122}
  (\bibinfo{year}{2017}), \eprint{1704.01590}.

\bibitem[{\citenamefont{Goldberger and Rothstein}(2006)}]{Goldberger:2004jt}
\bibinfo{author}{\bibfnamefont{W.~D.} \bibnamefont{Goldberger}}
  \bibnamefont{and} \bibinfo{author}{\bibfnamefont{I.~Z.}
  \bibnamefont{Rothstein}}, \bibinfo{journal}{Phys.\ Rev.\ D}
  \textbf{\bibinfo{volume}{73}}, \bibinfo{pages}{104029}
  (\bibinfo{year}{2006}), \eprint{hep-th/0409156}.

\bibitem[{\citenamefont{Porto and Rothstein}(2006)}]{Porto:2006bt}
\bibinfo{author}{\bibfnamefont{R.~A.} \bibnamefont{Porto}} \bibnamefont{and}
  \bibinfo{author}{\bibfnamefont{I.~Z.} \bibnamefont{Rothstein}},
  \bibinfo{journal}{Phys.\ Rev.\ Lett.} \textbf{\bibinfo{volume}{97}},
  \bibinfo{pages}{021101} (\bibinfo{year}{2006}), \eprint{gr-qc/0604099}.

\bibitem[{\citenamefont{Porto}(2016)}]{Porto:2016pyg}
\bibinfo{author}{\bibfnamefont{R.~A.} \bibnamefont{Porto}},
  \bibinfo{journal}{Phys.\ Rept.} \textbf{\bibinfo{volume}{633}},
  \bibinfo{pages}{1} (\bibinfo{year}{2016}), \eprint{1601.04914}.

\bibitem[{\citenamefont{Levi and Steinhoff}(2015)}]{Levi:2015msa}
\bibinfo{author}{\bibfnamefont{M.}~\bibnamefont{Levi}} \bibnamefont{and}
  \bibinfo{author}{\bibfnamefont{J.}~\bibnamefont{Steinhoff}},
  \bibinfo{journal}{JHEP} \textbf{\bibinfo{volume}{09}}, \bibinfo{pages}{219}
  (\bibinfo{year}{2015}), \eprint{1501.04956}.

\bibitem[{\citenamefont{Camanho et~al.}(2016)\citenamefont{Camanho, Edelstein,
  Maldacena, and Zhiboedov}}]{Camanho:2014apa}
\bibinfo{author}{\bibfnamefont{X.~O.} \bibnamefont{Camanho}},
  \bibinfo{author}{\bibfnamefont{J.~D.} \bibnamefont{Edelstein}},
  \bibinfo{author}{\bibfnamefont{J.}~\bibnamefont{Maldacena}},
  \bibnamefont{and}
  \bibinfo{author}{\bibfnamefont{A.}~\bibnamefont{Zhiboedov}},
  \bibinfo{journal}{JHEP} \textbf{\bibinfo{volume}{02}}, \bibinfo{pages}{020}
  (\bibinfo{year}{2016}), \eprint{1407.5597}.

\bibitem[{\citenamefont{Kol and Smolkin}(2008)}]{Kol:2007bc}
\bibinfo{author}{\bibfnamefont{B.}~\bibnamefont{Kol}} \bibnamefont{and}
  \bibinfo{author}{\bibfnamefont{M.}~\bibnamefont{Smolkin}},
  \bibinfo{journal}{Class.\ Quant.\ Grav.} \textbf{\bibinfo{volume}{25}},
  \bibinfo{pages}{145011} (\bibinfo{year}{2008}), \eprint{0712.4116}.

\bibitem[{\citenamefont{Foffa and Sturani}(2011)}]{Foffa:2011ub}
\bibinfo{author}{\bibfnamefont{S.}~\bibnamefont{Foffa}} \bibnamefont{and}
  \bibinfo{author}{\bibfnamefont{R.}~\bibnamefont{Sturani}},
  \bibinfo{journal}{Phys. Rev. D} \textbf{\bibinfo{volume}{84}},
  \bibinfo{pages}{044031} (\bibinfo{year}{2011}), \eprint{1104.1122}.

\bibitem[{\citenamefont{Foffa and Sturani}(2013)}]{Foffa:2012rn}
\bibinfo{author}{\bibfnamefont{S.}~\bibnamefont{Foffa}} \bibnamefont{and}
  \bibinfo{author}{\bibfnamefont{R.}~\bibnamefont{Sturani}},
  \bibinfo{journal}{Phys. Rev. D} \textbf{\bibinfo{volume}{87}},
  \bibinfo{pages}{064011} (\bibinfo{year}{2013}), \eprint{1206.7087}.

\bibitem[{\citenamefont{Thorne}(1980)}]{Thorne:1980ru}
\bibinfo{author}{\bibfnamefont{K.~S.} \bibnamefont{Thorne}},
  \bibinfo{journal}{Rev. Mod. Phys.} \textbf{\bibinfo{volume}{52}},
  \bibinfo{pages}{299} (\bibinfo{year}{1980}).

\bibitem[{\citenamefont{Hanson and Regge}(1974)}]{Hanson:1974qy}
\bibinfo{author}{\bibfnamefont{A.~J.} \bibnamefont{Hanson}} \bibnamefont{and}
  \bibinfo{author}{\bibfnamefont{T.}~\bibnamefont{Regge}},
  \bibinfo{journal}{Annals Phys.} \textbf{\bibinfo{volume}{87}},
  \bibinfo{pages}{498} (\bibinfo{year}{1974}).

\bibitem[{\citenamefont{Porto}(2006)}]{Porto:2005ac}
\bibinfo{author}{\bibfnamefont{R.~A.} \bibnamefont{Porto}},
  \bibinfo{journal}{Phys.\ Rev.\ D} \textbf{\bibinfo{volume}{73}},
  \bibinfo{pages}{104031} (\bibinfo{year}{2006}), \eprint{gr-qc/0511061}.

\bibitem[{\citenamefont{Mathisson}(1937)}]{Mathisson:1937zz}
\bibinfo{author}{\bibfnamefont{M.}~\bibnamefont{Mathisson}},
  \bibinfo{journal}{Acta Phys.\ Polon.} \textbf{\bibinfo{volume}{6}},
  \bibinfo{pages}{163} (\bibinfo{year}{1937}).

\bibitem[{\citenamefont{Papapetrou}(1951)}]{Papapetrou:1951pa}
\bibinfo{author}{\bibfnamefont{A.}~\bibnamefont{Papapetrou}},
  \bibinfo{journal}{Proc.\ Roy.\ Soc.\ Lond.\ A}
  \textbf{\bibinfo{volume}{A209}}, \bibinfo{pages}{248} (\bibinfo{year}{1951}).

\bibitem[{\citenamefont{Dixon}(1970)}]{Dixon:1970zza}
\bibinfo{author}{\bibfnamefont{W.}~\bibnamefont{Dixon}},
  \bibinfo{journal}{Proc.\ Roy.\ Soc.\ Lond.\ A}
  \textbf{\bibinfo{volume}{A314}}, \bibinfo{pages}{499} (\bibinfo{year}{1970}).

\bibitem[{\citenamefont{Goldstein}(2002)}]{goldstein}
\bibinfo{author}{\bibfnamefont{H.}~\bibnamefont{Goldstein}},
  \emph{\bibinfo{title}{Classical Mechanics}}
  (\bibinfo{publisher}{Addison-Wesley}, \bibinfo{year}{2002}).

\bibitem[{\citenamefont{Barker and O'Connell}(1975)}]{Barker:1975ae}
\bibinfo{author}{\bibfnamefont{B.}~\bibnamefont{Barker}} \bibnamefont{and}
  \bibinfo{author}{\bibfnamefont{R.}~\bibnamefont{O'Connell}},
  \bibinfo{journal}{Phys.\ Rev.\ D} \textbf{\bibinfo{volume}{12}},
  \bibinfo{pages}{329} (\bibinfo{year}{1975}).

\bibitem[{\citenamefont{Foffa and Sturani}(2014)}]{Foffa:2013qca}
\bibinfo{author}{\bibfnamefont{S.}~\bibnamefont{Foffa}} \bibnamefont{and}
  \bibinfo{author}{\bibfnamefont{R.}~\bibnamefont{Sturani}},
  \bibinfo{journal}{Class. Quant. Grav.} \textbf{\bibinfo{volume}{31}},
  \bibinfo{pages}{043001} (\bibinfo{year}{2014}), \eprint{1309.3474}.

\bibitem[{\citenamefont{Barker and O'Connell}(1970)}]{Barker:1970zr}
\bibinfo{author}{\bibfnamefont{B.}~\bibnamefont{Barker}} \bibnamefont{and}
  \bibinfo{author}{\bibfnamefont{R.}~\bibnamefont{O'Connell}},
  \bibinfo{journal}{Phys.\ Rev.\ D} \textbf{\bibinfo{volume}{2}},
  \bibinfo{pages}{1428} (\bibinfo{year}{1970}).

\bibitem[{\citenamefont{Thorne and Hartle}(1984)}]{Thorne:1984mz}
\bibinfo{author}{\bibfnamefont{K.~S.} \bibnamefont{Thorne}} \bibnamefont{and}
  \bibinfo{author}{\bibfnamefont{J.~B.} \bibnamefont{Hartle}},
  \bibinfo{journal}{Phys. Rev. D} \textbf{\bibinfo{volume}{31}},
  \bibinfo{pages}{1815} (\bibinfo{year}{1985}).

\bibitem[{\citenamefont{Poisson and Sasaki}(1995)}]{Poisson:1994yf}
\bibinfo{author}{\bibfnamefont{E.}~\bibnamefont{Poisson}} \bibnamefont{and}
  \bibinfo{author}{\bibfnamefont{M.}~\bibnamefont{Sasaki}},
  \bibinfo{journal}{Phys.\ Rev.\ D} \textbf{\bibinfo{volume}{51}},
  \bibinfo{pages}{5753} (\bibinfo{year}{1995}), \eprint{gr-qc/9412027}.

\bibitem[{\citenamefont{Kidder et~al.}(1993)\citenamefont{Kidder, Will, and
  Wiseman}}]{Kidder:1992fr}
\bibinfo{author}{\bibfnamefont{L.~E.} \bibnamefont{Kidder}},
  \bibinfo{author}{\bibfnamefont{C.~M.} \bibnamefont{Will}}, \bibnamefont{and}
  \bibinfo{author}{\bibfnamefont{A.~G.} \bibnamefont{Wiseman}},
  \bibinfo{journal}{Phys. Rev. D} \textbf{\bibinfo{volume}{47}},
  \bibinfo{pages}{R4183} (\bibinfo{year}{1993}), \eprint{gr-qc/9211025}.

\bibitem[{\citenamefont{Kidder}(1995)}]{Kidder:1995zr}
\bibinfo{author}{\bibfnamefont{L.~E.} \bibnamefont{Kidder}},
  \bibinfo{journal}{Phys.\ Rev.\ D} \textbf{\bibinfo{volume}{52}},
  \bibinfo{pages}{821} (\bibinfo{year}{1995}), \eprint{gr-qc/9506022}.

\bibitem[{\citenamefont{Abbott
  et~al.}(2019{\natexlab{b}})}]{LIGOScientific:2019fpa}
\bibinfo{author}{\bibfnamefont{B.}~\bibnamefont{Abbott}} \bibnamefont{et~al.}
  (\bibinfo{collaboration}{LIGO Scientific, Virgo}), \bibinfo{journal}{Phys.
  Rev. D} \textbf{\bibinfo{volume}{100}}, \bibinfo{pages}{104036}
  (\bibinfo{year}{2019}{\natexlab{b}}), \eprint{1903.04467}.

\bibitem[{\citenamefont{Abbott et~al.}(2020{\natexlab{b}})}]{Abbott:2020jks}
\bibinfo{author}{\bibfnamefont{R.}~\bibnamefont{Abbott}} \bibnamefont{et~al.}
  (\bibinfo{collaboration}{LIGO Scientific, Virgo})
  (\bibinfo{year}{2020}{\natexlab{b}}), \eprint{2010.14529}.

\bibitem[{\citenamefont{Sennett et~al.}(2020)\citenamefont{Sennett, Brito,
  Buonanno, Gorbenko, and Senatore}}]{Sennett:2019bpc}
\bibinfo{author}{\bibfnamefont{N.}~\bibnamefont{Sennett}},
  \bibinfo{author}{\bibfnamefont{R.}~\bibnamefont{Brito}},
  \bibinfo{author}{\bibfnamefont{A.}~\bibnamefont{Buonanno}},
  \bibinfo{author}{\bibfnamefont{V.}~\bibnamefont{Gorbenko}}, \bibnamefont{and}
  \bibinfo{author}{\bibfnamefont{L.}~\bibnamefont{Senatore}},
  \bibinfo{journal}{Phys. Rev. D} \textbf{\bibinfo{volume}{102}},
  \bibinfo{pages}{044056} (\bibinfo{year}{2020}), \eprint{1912.09917}.

\bibitem[{\citenamefont{Cardoso et~al.}(2018)\citenamefont{Cardoso, Kimura,
  Maselli, and Senatore}}]{Cardoso:2018ptl}
\bibinfo{author}{\bibfnamefont{V.}~\bibnamefont{Cardoso}},
  \bibinfo{author}{\bibfnamefont{M.}~\bibnamefont{Kimura}},
  \bibinfo{author}{\bibfnamefont{A.}~\bibnamefont{Maselli}}, \bibnamefont{and}
  \bibinfo{author}{\bibfnamefont{L.}~\bibnamefont{Senatore}},
  \bibinfo{journal}{Phys. Rev. Lett.} \textbf{\bibinfo{volume}{121}},
  \bibinfo{pages}{251105} (\bibinfo{year}{2018}), \eprint{1808.08962}.

\bibitem[{\citenamefont{Breton et~al.}(2008)\citenamefont{Breton, Kaspi,
  Kramer, McLaughlin, Lyutikov, Ransom, Stairs, Ferdman, Camilo, and
  Possenti}}]{Breton:2008xy}
\bibinfo{author}{\bibfnamefont{R.~P.} \bibnamefont{Breton}},
  \bibinfo{author}{\bibfnamefont{V.~M.} \bibnamefont{Kaspi}},
  \bibinfo{author}{\bibfnamefont{M.}~\bibnamefont{Kramer}},
  \bibinfo{author}{\bibfnamefont{M.~A.} \bibnamefont{McLaughlin}},
  \bibinfo{author}{\bibfnamefont{M.}~\bibnamefont{Lyutikov}},
  \bibinfo{author}{\bibfnamefont{S.~M.} \bibnamefont{Ransom}},
  \bibinfo{author}{\bibfnamefont{I.~H.} \bibnamefont{Stairs}},
  \bibinfo{author}{\bibfnamefont{R.~D.} \bibnamefont{Ferdman}},
  \bibinfo{author}{\bibfnamefont{F.}~\bibnamefont{Camilo}}, \bibnamefont{and}
  \bibinfo{author}{\bibfnamefont{A.}~\bibnamefont{Possenti}},
  \bibinfo{journal}{Science} \textbf{\bibinfo{volume}{321}},
  \bibinfo{pages}{104} (\bibinfo{year}{2008}), \eprint{0807.2644}.

\bibitem[{\citenamefont{Perera et~al.}(2010)}]{Perera:2010sp}
\bibinfo{author}{\bibfnamefont{B.}~\bibnamefont{Perera}} \bibnamefont{et~al.},
  \bibinfo{journal}{Astrophys. J.} \textbf{\bibinfo{volume}{721}},
  \bibinfo{pages}{1193} (\bibinfo{year}{2010}), \eprint{1008.1097}.

\bibitem[{\citenamefont{Perrodin and Sesana}(2018)}]{Perrodin:2017bxr}
\bibinfo{author}{\bibfnamefont{D.}~\bibnamefont{Perrodin}} \bibnamefont{and}
  \bibinfo{author}{\bibfnamefont{A.}~\bibnamefont{Sesana}},
  \bibinfo{journal}{Astrophys. Space Sci. Libr.}
  \textbf{\bibinfo{volume}{457}}, \bibinfo{pages}{95} (\bibinfo{year}{2018}),
  \eprint{1709.02816}.

\bibitem[{\citenamefont{Everitt et~al.}(2011)\citenamefont{Everitt, DeBra,
  Parkinson, Turneaure, Conklin, Heifetz, Keiser, Silbergleit, Holmes,
  Kolodziejczak et~al.}}]{PhysRevLett.106.221101}
\bibinfo{author}{\bibfnamefont{C.~W.~F.} \bibnamefont{Everitt}},
  \bibinfo{author}{\bibfnamefont{D.~B.} \bibnamefont{DeBra}},
  \bibinfo{author}{\bibfnamefont{B.~W.} \bibnamefont{Parkinson}},
  \bibinfo{author}{\bibfnamefont{J.~P.} \bibnamefont{Turneaure}},
  \bibinfo{author}{\bibfnamefont{J.~W.} \bibnamefont{Conklin}},
  \bibinfo{author}{\bibfnamefont{M.~I.} \bibnamefont{Heifetz}},
  \bibinfo{author}{\bibfnamefont{G.~M.} \bibnamefont{Keiser}},
  \bibinfo{author}{\bibfnamefont{A.~S.} \bibnamefont{Silbergleit}},
  \bibinfo{author}{\bibfnamefont{T.}~\bibnamefont{Holmes}},
  \bibinfo{author}{\bibfnamefont{J.}~\bibnamefont{Kolodziejczak}},
  \bibnamefont{et~al.}, \bibinfo{journal}{Phys. Rev. Lett.}
  \textbf{\bibinfo{volume}{106}}, \bibinfo{pages}{221101}
  (\bibinfo{year}{2011}), \eprint{1105.3456},
  \urlprefix\url{https://link.aps.org/doi/10.1103/PhysRevLett.106.221101}.

\bibitem[{\citenamefont{Racine et~al.}(2009)\citenamefont{Racine, Buonanno, and
  Kidder}}]{Racine:2008kj}
\bibinfo{author}{\bibfnamefont{E.}~\bibnamefont{Racine}},
  \bibinfo{author}{\bibfnamefont{A.}~\bibnamefont{Buonanno}}, \bibnamefont{and}
  \bibinfo{author}{\bibfnamefont{L.~E.} \bibnamefont{Kidder}},
  \bibinfo{journal}{Phys. Rev. D} \textbf{\bibinfo{volume}{80}},
  \bibinfo{pages}{044010} (\bibinfo{year}{2009}), \eprint{0812.4413}.

\bibitem[{\citenamefont{Ross}(2012)}]{Ross:2012fc}
\bibinfo{author}{\bibfnamefont{A.}~\bibnamefont{Ross}}, \bibinfo{journal}{Phys.
  Rev. D} \textbf{\bibinfo{volume}{85}}, \bibinfo{pages}{125033}
  (\bibinfo{year}{2012}), \eprint{1202.4750}.

\end{thebibliography}
\end{document}